\documentclass[showpacs,preprintnumbers,amsmath,amssymb,nofootinbib]{revtex4}
\usepackage{setspace}
\usepackage{graphicx}  % Include figure files
\usepackage{dcolumn}   % Align table columns on decimal point
\usepackage{bm}        % bold math

\newcommand{\met}{\ensuremath{\,/\!\!\!\!E_{T}}}
\newcommand{\et}{\ensuremath{E_T}}
\newcommand{\pt}{\ensuremath{p_T}}
\newcommand{\sht}{\ensuremath{H_T}}
\newcommand{\ttbar}{\ensuremath{{t\bar{t}}}}
\newcommand{\ppbar}{{\ensuremath{p\bar{p}}}}

\newcommand{\pbi}{pb$^{-1}$}

\newcommand{\gevcii}{GeV/$c^{2}$}

\newcommand{\cms}{cm$^{-2}$s$^{-1}$}

\begin{document}

\preprint{hep-ex/05xxxix}
\date{\today}

\begin{center}
\begin{large}
\bf{Measurement of the Cross Section for $\ttbar$ Production in $\ppbar$\ Collisions using the Kinematics of Lepton+Jets Events}
\end{large}
\end{center}

\font\eightit=cmti8
\def\r#1{\ignorespaces $^{#1}$}
\hfilneg
\begin{sloppypar}
\noindent 
D.~Acosta,\r {16} J.~Adelman,\r {12} T.~Affolder,\r 9 T.~Akimoto,\r {54}
M.G.~Albrow,\r {15} D.~Ambrose,\r {15} S.~Amerio,\r {42}
D.~Amidei,\r {33} A.~Anastassov,\r {50} K.~Anikeev,\r {15}
% ...
\end{sloppypar}
\vskip .026in
\begin{center}
(CDF Collaboration)
\end{center}

\vskip .026in
\begin{center}
\r 1  {\eightit Institute of Physics, Academia Sinica, Taipei, Taiwan 11529,
Republic of China} \\
\r 2  {\eightit Argonne National Laboratory, Argonne, Illinois 60439} \\
\r 3  {\eightit Institut de Fisica d'Altes Energies, Universitat Autonoma
de Barcelona, E-08193, Bellaterra (Barcelona), Spain} \\
\r 4  {\eightit Istituto Nazionale di Fisica Nucleare, University of Bologna,
I-40127 Bologna, Italy} \\
\r 5  {\eightit Brandeis University, Waltham, Massachusetts 02254} \\
\r 6  {\eightit University of California, Davis, Davis, California  95616} \\
\r 7  {\eightit University of California, Los Angeles, Los
Angeles, California  90024} \\
\r 8  {\eightit University of California, San Diego, La Jolla, California  92093} \\
\r 9  {\eightit University of California, Santa Barbara, Santa Barbara, California
93106} \\
\r {10} {\eightit Instituto de Fisica de Cantabria, CSIC-University of Cantabria,
39005 Santander, Spain} \\
\r {11} {\eightit Carnegie Mellon University, Pittsburgh, PA  15213} \\
\r {12} {\eightit Enrico Fermi Institute, University of Chicago, Chicago,
Illinois 60637} \\
\r {13}  {\eightit Joint Institute for Nuclear Research, RU-141980 Dubna, Russia}
\\
\r {14} {\eightit Duke University, Durham, North Carolina  27708} \\
\r {15} {\eightit Fermi National Accelerator Laboratory, Batavia, Illinois
60510} \\
\r {16} {\eightit University of Florida, Gainesville, Florida  32611} \\
\r {17} {\eightit Laboratori Nazionali di Frascati, Istituto Nazionale di Fisica
               Nucleare, I-00044 Frascati, Italy} \\
\r {18} {\eightit University of Geneva, CH-1211 Geneva 4, Switzerland} \\
\r {19} {\eightit Glasgow University, Glasgow G12 8QQ, United Kingdom}\\
\r {20} {\eightit Harvard University, Cambridge, Massachusetts 02138} \\
\r {21} {\eightit Division of High Energy Physics, Department of
Physics, University of Helsinki and Helsinki Institute of Physics,
FIN-00014, Helsinki, Finland}\\
\r {22} {\eightit Hiroshima University, Higashi-Hiroshima 724, Japan} \\
\r {23} {\eightit University of Illinois, Urbana, Illinois 61801} \\
\r {24} {\eightit The Johns Hopkins University, Baltimore, Maryland 21218} \\
\r {25} {\eightit Institut f\"{u}r Experimentelle Kernphysik,
Universit\"{a}t Karlsruhe, 76128 Karlsruhe, Germany} \\
\r {26} {\eightit High Energy Accelerator Research Organization (KEK), Tsukuba,
Ibaraki 305, Japan} \\
\r {27} {\eightit Center for High Energy Physics: Kyungpook National
University, Taegu 702-701; Seoul National University, Seoul 151-742; and
SungKyunKwan University, Suwon 440-746; Korea} \\
\r {28} {\eightit Ernest Orlando Lawrence Berkeley National Laboratory,
Berkeley, California 94720} \\
\r {29} {\eightit University of Liverpool, Liverpool L69 7ZE, United Kingdom} \\
\r {30} {\eightit University College London, London WC1E 6BT, United Kingdom} \\
\r {31} {\eightit Massachusetts Institute of Technology, Cambridge,
Massachusetts  02139} \\
\r {32} {\eightit Institute of Particle Physics: McGill University,
Montr\'{e}al, Canada H3A~2T8; and University of Toronto, Toronto, Canada
M5S~1A7} \\
\r {33} {\eightit University of Michigan, Ann Arbor, Michigan 48109} \\
\r {34} {\eightit Michigan State University, East Lansing, Michigan  48824} \\
\r {35} {\eightit Institution for Theoretical and Experimental Physics, ITEP,
Moscow 117259, Russia} \\
\r {36} {\eightit University of New Mexico, Albuquerque, New Mexico 87131} \\
\r {37} {\eightit Northwestern University, Evanston, Illinois  60208} \\
\r {38} {\eightit The Ohio State University, Columbus, Ohio  43210} \\
\r {39} {\eightit Okayama University, Okayama 700-8530, Japan}\\
\r {40} {\eightit Osaka City University, Osaka 588, Japan} \\
\r {41} {\eightit University of Oxford, Oxford OX1 3RH, United Kingdom} \\
\r {42} {\eightit University of Padova, Istituto Nazionale di Fisica
          Nucleare, Sezione di Padova-Trento, I-35131 Padova, Italy} \\
\r {43} {\eightit University of Pennsylvania, Philadelphia,
        Pennsylvania 19104} \\
\r {44} {\eightit Istituto Nazionale di Fisica Nucleare Pisa, Universities 
of Pisa, Siena and Scuola Normale Superiore, I-56127 Pisa, Italy} \\
\r {45} {\eightit University of Pittsburgh, Pittsburgh, Pennsylvania 15260} \\
\r {46} {\eightit Purdue University, West Lafayette, Indiana 47907} \\
\r {47} {\eightit University of Rochester, Rochester, New York 14627} \\
\r {48} {\eightit The Rockefeller University, New York, New York 10021} \\
\r {49} {\eightit Istituto Nazionale di Fisica Nucleare, Sezione di Roma 1,
University di Roma ``La Sapienza," I-00185 Roma, Italy}\\
\r {50} {\eightit Rutgers University, Piscataway, New Jersey 08855} \\
\r {51} {\eightit Texas A\&M University, College Station, Texas 77843} \\
\r {52} {\eightit Texas Tech University, Lubbock, Texas 79409} \\
\r {53} {\eightit Istituto Nazionale di Fisica Nucleare, University of Trieste/\
Udine, Italy} \\
\r {54} {\eightit University of Tsukuba, Tsukuba, Ibaraki 305, Japan} \\
\r {55} {\eightit Tufts University, Medford, Massachusetts 02155} \\
\r {56} {\eightit Waseda University, Tokyo 169, Japan} \\
\r {57} {\eightit Wayne State University, Detroit, Michigan  48201} \\
\r {58} {\eightit University of Wisconsin, Madison, Wisconsin 53706} \\
\r {59} {\eightit Yale University, New Haven, Connecticut 06520} \\
\end{center}

\begin{abstract}

We present a measurement of the top pair production cross section in $\ppbar$ collisions at 
$\sqrt{s}$=1.96~TeV. We collect a data sample with an integrated luminosity of 
194$\pm$11~\pbi\ with the CDF II detector at the Fermilab Tevatron. We use an artificial 
neural network technique to discriminate between top pair production and background
processes in a sample of 519 lepton+jets events, which have one isolated energetic charged lepton, large 
missing transverse energy and at least three energetic jets. We measure the top pair production
 cross section to be  $\sigma_{t\overline{t}}= 6.6$$\pm 1.1 \pm 1.5$~pb, where the first uncertainty
is statistical and the second is systematic. 

\end{abstract}

  % PACS, the Physics and Astronomy
                               % Classification Scheme.
%\keywords{Suggested keywords} % Use showkeys class option if keyword
                               % display desired

\maketitle

%------------------------------------------------------------------
\section{\label{sec:int} Introduction}

%Since the discovery of the top quark~\cite{discoverycdf, discoveryd0} in 1995, experimental attention 
%has turned to the question of whether the properties of this most massive fundamental particle are 
%in good agreement with the predictions of the standard model.  

We report on a measurement with the Collider Detector at the Fermilab Tevatron of the rate of pair production of 
top quarks in the lepton+jets channel, $\ppbar \rightarrow \ttbar \rightarrow W^{+}W^{-} b\bar{b} \rightarrow \ell\bar{\nu_{\ell}}q\bar{q'}b\bar{b}$.  
Recent theoretical calculations predict the cross section for top pair production~\cite{mlm,kidonakis}
with an uncertainty of less than 15\%. 
The increase in the Fermilab 
Tevatron center-of-mass energy to 1.96~TeV from 1.80~TeV is expected to have enhanced the top pair production cross section by 30\%.  
Each top quark is expected to decay into a $W$\ boson and a $b$\ quark, with a branching fraction of almost 100\%.  
A significant deviation of the observed rate of top pair production from the standard model prediction 
could indicate either a novel top quark production mechanism, {\it e.g.} the 
production and decay of a heavy resonance into \ttbar\ pairs~\cite{heavyresonance}, or a novel top quark decay mechanism, 
{\it e.g.} a decay into supersymmetric particles~\cite{supersymmetry}, or a similar final state
signature from a top-like particle~\cite{tprime,tprime1,beautmirrors,littlehier}. 
Previous measurements of the properties of the top quark~\cite{topProperties}
are consistent with expectations from the standard model but suffer from large statistical uncertainties. 

We first show that it is feasible to measure the top pair production cross section with a single kinematic event property,  which may be used to discriminate between the signal from top pair production and 
the dominant background from $W$ boson production with associated jets~\cite{KinEvidence}.  
This property is the total transverse energy in the event~\cite{Htfirst}, which has been used as a discriminant 
by several recent top analyses~\cite{ttbardilepton,svxruniipaper,sltpaper}.
In addition, we test the modeling of the kinematics of top pair and $W+$jets production.  
A good understanding of the kinematics of these processes will be required to discover single top 
quark production and will benefit searches for the Higgs boson and physics beyond the standard model at both the Tevatron and the future Large Hadron Collider, where techniques using kinematic discrimination have been proposed.

We then develop an artificial neural network technique in order to maximize 
the discriminating power available from kinematic and topological properties~\cite{KinStudiesRun1}. 
Throughout this paper, we quantify the gain of our neural network approach
relative to the single event property of total transverse energy.
The statistical sensitivity of our neural network technique is comparable to that of methods employing 
secondary vertex $b$-tagging~\cite{svxruniipaper,Taka+Mel}, and is 
independent of the assumptions and systematic uncertainties specific to $b$-tagging. 

%------------------------------------------------------------------
\section{\label{sec:cdf} Experimental Apparatus}

The Collider Detector at Fermilab (CDF) has been substantially upgraded for the current Tevatron 
collider run, which began in 2001.  The major upgrades include new charged particle tracking detectors, 
forward calorimetry, trigger and data acquisition electronics
and infrastructure as well as extended muon coverage. 
A thorough description of the detector is provided elsewhere~\cite{cdfdetector}. The  
essential components of the detector for this analysis are briefly described here.  

The reconstruction of charged particles with high transverse momentum is essential to the electron and muon triggers that 
collect our data sample, the identification of electrons and muons, and the measurement of the muon momentum.
The charged particle tracking detectors are immersed in a 1.4~T magnetic field from a superconducting solenoid, 
which is oriented parallel to the proton beam direction~\cite{itsthegeometrystupid}. The Central Outer Tracker~\cite{cotref} (COT) 
has eight super-layers of 310~cm long wires covering radii from 40 to 137~cm. Each super-layer consists of planes of 12 sense wires.
The super-layers alternate between having wires parallel to the cylinder axis and wires displaced by a $2^{\circ}$\ stereo angle.
This provides three dimensional charged particle track reconstruction, with 
up to 96 position measurements with a spatial resolution of about 180 $\mathrm{\mu}$m in the transverse plane.
The COT transverse momentum resolution is $\sigma_{p_{T}}/p^{2}_{T} \approx 0.0017$ $\mathrm{[GeV/c]^{-1}}$.     
The inner tracking detector is a silicon strip detector~\cite{L00,svxref,ISL} that
provides up to eight position measurements with a spatial resolution of about 15 $\mathrm{\mu}$m. 

%consists of three sub-detectors.  
%A single-sided layer of silicon 
%sensors (L00)~\cite{L00} is installed directly onto the beryllium beam pipe, at a radius of 1.35~cm. 
%It is followed by five concentric layers of double-sided silicon sensors (SVX)~\cite{svxref} 
%located at radii between 2.5 and 10.6~cm.  The Intermediate Silicon Layers (ISL)~\cite{ISL} 
%consist of one double-sided layer at a radius of 23~cm in the central region, and two double-sided 
%layers at radii 20 and 29~cm in the forward regions.

Calorimetry is used to measure the transverse energy of electrons and jets, as
well as to infer the presence of neutrinos from a significant imbalance in the observed transverse energy.
The calorimeters lie outside the solenoid and are physically 
divided into a central region~\cite{CEM,CHA} covering pseudo-rapidity $|\eta|<1.1$ and an upgraded plug region~\cite{PEM} covering
$1.1<|\eta|<3.6$. The electromagnetic calorimeter is a lead-scintillator sandwich, which is 
18 radiation lengths deep in the central region (CEM), with energy resolution of $14\%/\sqrt{E_{T}}$. 
The hadronic calorimeter is an iron-scintillator sandwich, which is 4.5 nuclear interaction lengths 
deep in the central region (CHA), with energy resolution of $50\%/\sqrt{E}$. 
The calorimeters are segmented into a projective ``tower'' geometry, where each tower subtends an area 
of 0.11 in $\eta$\ and 15$^\circ$ in azimuth in the central region. 
Finer position resolution for electron and photon identification is provided by proportional chambers (CES), located 
at the approximate electromagnetic shower maximum depth in each tower.

Muons are identified in drift chambers which surround the calorimeters up to $|\eta|<1.0$. 
The Central Muon Detector (CMU)~\cite{CMU} consists of a set of drift 
chambers located outside the central hadronic calorimeters and covers $|\eta|<0.6$. 
An additional 60~cm thick layer of steel shields the four layers of single wire drift tubes that comprise
 the Central Muon Upgrade detector (CMP).
The Central Muon Extension detector (CMX) consists of drift tubes, located at each end of the central detector between
$42^{\circ}-55^{\circ}$\ in polar angle, that extend the coverage to muons between 
$0.6< \eta <1.0$. 
  
Gas Cerenkov light detectors~\cite{CLC} located in the $3.7 < |\eta| < 4.7$\ region measure 
the number of inelastic \ppbar\ collisions per bunch crossing and thereby the luminosity delivered to CDF by the Tevatron.  The total uncertainty on the luminosity is 5.9\%, 
where 4.4\% comes from the acceptance  and operation of the luminosity monitor and 4.0\% from the calculation of the total \ppbar\ cross section~\cite{cdflumi}.

%------------------------------------------------------------------
\section{\label{sec:sel} Selection of data sample}

Top quark events in the lepton+jets channel\footnote{For the rest of this paper, lepton and the 
symbol $\ell$\ imply electron or muon of either charge.}, 
$\ppbar \rightarrow \ttbar \rightarrow \ell\bar{\nu_{\ell}}q\bar{q'}b\bar{b}$, are characterized by 
a high transverse momentum lepton and substantial missing transverse energy due to the leptonic $W$ decay along with several 
hadronic jets with high transverse energy. Two jets are expected from the hadronic $W$ decay, two more are expected from
the $b$ and $\overline{b}$ quarks originating from the respective $t$\ and $\bar{t}$\ decays. 
In practice, not all of these jets may be reconstructed due to kinematic requirements and 
limitations of the detector geometry, while other jets may arise from initial and final state hard radiation 
effects. 

The data sample in this paper is collected by a trigger based solely on the presence
of a high transverse momentum lepton.  In this section, we discuss the trigger and lepton identification requirements,
the reconstruction of the jets and the missing transverse energy, and further requirements we impose to reduce specific backgrounds.  The same criteria are applied to both data and Monte Carlo simulation.

\subsection{\label{subsec:data} Data}

This analysis uses data from $p\bar{p}$\ collisions at a center-of-mass energy of
$\sqrt{s}=1.96$~TeV collected with CDF between March 2002 and
September 2003.  All of the detector subsystems important for lepton 
identification and kinematic reconstruction, namely the central outer tracker, calorimeters and muon chambers, 
were carefully monitored over this period and any
segment of data with a problem in any of these systems was excluded from consideration.
No requirement was made on the silicon detectors for this analysis.  
The integrated luminosity of this data sample was measured to be 194$\pm$11~\pbi~\cite{cdflumi}. 
%While the ongoing collider run is expected to collect over 4000~\pbi by 2009, this sample is already 
%twice as large as and independent of the sample, collected in the previous collider run between 1992 and 1995, 
%used for previous publications on top quark properties.

\subsection{\label{subsec:trigger} Trigger}

CDF uses a three-level trigger and data acquisition system to filter interesting events from the 1.7~MHz 
beam crossing rate and write them to permanent storage at an average rate of 60~Hz.
We describe here only the triggers important for this analysis, which select events containing an
electron or muon with high transverse momentum (\pt).  
The efficiencies of these triggers have been measured directly from the data~\cite{wzprd} and are listed 
in Table~\ref{tab:acc}.

At the first level (L1), charged particle tracks reconstructed in the  COT $r-\phi$\ projection by
a hardware track processor, the eXtremely Fast Tracker (XFT)~\cite{xftref}, are required
to point to a cluster of energy in the electromagnetic calorimeter or to a track segment in the muon chambers.
The L1 electron trigger requires an XFT track with $\pt> 8$~{GeV/$c$}, matched to a single trigger tower
in the central electromagnetic calorimeter having transverse energy $\et>8$~GeV and with a
ratio of hadronic to electromagnetic energy less than 0.125.  The L1 muon trigger requires that either
an XFT track with $\pt>4$~{GeV/$c$}\ be matched to a muon track segment with $\pt>6$~{GeV/$c$} from the CMU and the CMP, or that an XFT track with $\pt>8$~{GeV/$c$}\ be matched to a muon track segment with $\pt>6$~{GeV/$c$} from the CMX.  

The second level (L2) electron trigger requires the XFT track matched to a cluster of energy 
in the central electromagnetic calorimeter with $\et > 16$~GeV and with a ratio of hadronic 
to electromagnetic energy less than 0.125.  The calorimeter cluster is formed by adding 
the energy in neighboring trigger towers with $\et>7.5$~GeV to the original L1 trigger tower.    
For this data set, the L2 muon trigger automatically accepts events passing the L1 muon trigger.

At the third level (L3), a farm of Linux computers performs on-line a complete event 
reconstruction, including three-dimensional charged particle track reconstruction.  
The L3 electron trigger requires: a track with  $\pt > 9$~{GeV/$c$} matched 
 to a cluster of energy in three adjacent towers in pseudo-rapidity in the central electromagnetic calorimeter 
with $\et >18$~GeV; the ratio of hadronic to electromagnetic
  energy less than 0.125; a lateral shower profile\footnote{See 
  Section~\ref{subsec:eleid} on electron identification.} of the calorimeter cluster less than 0.4;
  and the distance between the extrapolated track position and the CES measurement in the $z$\ 
  view less than 10~cm. The L3 muon trigger requires a track with $\pt > 18$~{GeV/$c$} matched to a 
 track segment in the muon chambers within $10$~cm in the $r-\phi$\ view and, for CMU and CMP 
 muons only, within $20$~cm in the $z$\ view.

\subsection{\label{subsec:eleid} Electron identification}

Electron candidates are required to have a COT track with $\pt>9$~{GeV/$c$}\ 
that extrapolates to a cluster of energy with $\et>20$~GeV formed by three adjacent towers in pseudo-rapidity 
in the central electromagnetic calorimeter.  The electron energy is corrected by less than 
5\% for the non-uniform response across each calorimeter tower by using the CES measurement of the shower position.
The shower position is required to be away from the calorimeter tower boundaries to ensure high quality 
discrimination between electrons and charged hadrons. 
This fiducial volume for electrons covers 84\% of the solid angle in 
the central $|\eta|<1.0$ region.  The selection requirements are defined below and listed in Table~\ref{tab:elesel}:

\begin{itemize}

\item Ratio of hadronic energy in the cluster, $E_{\textrm had}$, to the 
 electromagnetic energy in the cluster, $E_{\textrm em}$.

\item Comparison of the lateral shower profile~\cite{lshr}, 
the distribution of adjacent CEM tower energies as a function of the
seed tower energy in the calorimeter, with that expected from test beam electrons, $L_{\textrm{shr}}$.

\item $\chi^{2}$\ comparison of the CES shower profiles with those of test beam electrons in the $z$\ view, $\chi^{2}_{\textrm{strip}}$.

\item Distance between the position of the extrapolated track and the CES 
shower profiles measured in the $r-\phi$\ and $z$\ views, $\Delta x$\ and $\Delta z$.
The limits on $\Delta x$\ are asymmetric and signed by electric charge $Q$\ to allow 
for energy deposition from bremsstrahlung photons emitted as the electron/positron passes through the detector material.

\item Ratio of cluster energy to track momentum, $E/P$.

\item Isolation, $I$, defined as the ratio between any 
additional transverse energy in a cone of radius $R=\sqrt{(\delta \eta)^{2}+(\delta \phi)^{2}}=0.4$\ 
around the cluster and the transverse energy of the cluster. 

\end{itemize}

\begin{table}[!hbtp]
\caption{\label{tab:elesel}Selection requirements for electron candidates from $W$ boson decay. }
\begin{center}
\begin{ruledtabular}
\begin{tabular}{ll}
Property & Requirement \\
\hline
$E_{T}$ 		& $\geq$ 20 GeV 			\\
$E_{\textrm{had}}/E_{\textrm{em}}$ 	& $\leq$ 0.055+0.00045*E (GeV)		\\
$L_{\textrm{shr}}$ 		& $\leq$ 0.2 				\\
$\chi^{2}_{\textrm{strip}}$	& $\leq$ 10.0 				\\
$|\Delta z|$		& $\leq$ 3.0 cm 			\\
$Q*\Delta x$		& $\geq$ -3.0 cm, $\leq$ 1.5 cm  	\\
$E/P$			& $\leq$ 2.0 or $p_T > 50$ {GeV/$c$} 	\\
Isolation		& $\leq$ 0.1				\\
Conversion	        & Veto					\\
\end{tabular}
\end{ruledtabular}
\end{center}
\end{table}

For electrons in the fiducial volume, the identification efficiency is determined from a data sample of $Z\to e^{+}e^{-}$\ events and
is found to be 82.5 $\pm$ 0.5\%, where the uncertainty is statistical only.  In our estimate
of the selection efficiency for top pair events, we are sensitive to systematic differences
in electron identification between data and simulation.  We use $Z\to e^{+}e^{-}$\ data and simulation samples
to measure a correction factor of 0.965$\pm$0.014 for the electron identification efficiency, where the uncertainty is statistical only.
We discuss systematic uncertainties and differences between the electron environment in $Z\to e^{+}e^{-}$\ events 
and \ttbar\ events further in section~\ref{sec:sys}.

Photon conversions occur throughout the detector material and are a major source of electrons and positrons that pass the above
selection criteria.  We identify photon conversions by the characteristic small opening angle between two oppositely charged tracks 
that are parallel at their distance of closest approach to each other.  Specifically, we require the distance between 
the tracks in the $r-\phi$\ plane at the radius where the tracks are parallel to be less than 0.2~cm, and the 
difference between the cotangent of polar angles to be less than 0.04.   We reject electron candidates with an oppositely
charged partner track meeting these requirements. In this analysis, we are sensitive to any loss in efficiency 
from the mis-identification of an electron from $W$\ boson decay as a photon conversion. 
We measure the loss in efficiency  with a $Z\rightarrow e^{+}e^{-}$\ data sample.  
We find that we can halve the loss in efficiency to $2.3\pm0.04$\% by not rejecting electrons accompanied by a converted 
bremsstrahlung photon.  Specifically, we
do not reject electron candidates where the nearby oppositely charged particle track itself has an additional conversion partner.
For completeness, we note here that the performance of this algorithm to identify electrons from photon conversions is 
estimated~\cite{svxruniipaper} at $72.6 \pm 0.07$\%, where the error covers both statistical and systematic uncertainties.
  
\subsection{\label{subsec:muid} Muon identification}

Muon candidates are required to have a COT track with $\pt>20$~{GeV/$c$} that 
extrapolates to a track segment in the muon chambers. 
The muon COT track curvature, and thus the muon transverse momentum, 
is corrected in order to remove a small azimuthal dependence from residual detector alignment effects~\cite{wzprd}.
The selection requirements used to separate muons from products of hadrons that interact in the calorimeters 
and from cosmic rays are defined below and listed in Table~\ref{tab:musel}:

\begin{itemize}
\item Energy deposition in the electromagnetic and hadronic calorimeter 
expected to be characteristic of minimum ionizing particles, $E_{\textrm{em}}$\ and $E_{\textrm{had}}$.

\item Distance between the extrapolated track and the track segment in the muon chamber, 
$\Delta x$. A track matched to a segment in the CMU muon chambers is required to have a matched track
segment in the CMP chambers as well, and vice versa.

\item Distance of closest approach of the reconstructed track to the beam-line in the transverse plane, $d_{0}$. If available, information
from the silicon tracking detector is included to increase precision and improve rejection of muons from 
cosmic rays and decays-in-flight of charged hadrons.

\item Cosmic ray muons that pass through the detector close to the beam-line may be reconstructed as a 
pair of charged particles.  We use the timing capabilities of the COT to reject events where one 
of the tracks from a charged particle appears to travel toward instead of away from the center of the 
detector.

\item Isolation, $I$, defined as the ratio between any additional transverse energy 
in a cone of radius $R=0.4$\ around the track direction and the muon transverse momentum.

\end{itemize}

\begin{table}[!hbtp]
\caption{\label{tab:musel}
Selection requirements for muon candidates from $W$ boson decay.}
\begin{ruledtabular}
\begin{tabular}{ll}
Property & Requirement 				\\
\hline
$p_{T}$			& $\geq$ 20 GeV \\
$E_{\textrm had}$ 		& $\leq  \max (6, 6+0.0280(p-100))$\ GeV  \\
$E_{\textrm em}$ 		& $\leq  \max (2, 2+0.0115(p-100))$\ GeV \\
${\textrm CMU} |\Delta x|$	& $\leq$ 3.0 cm 			\\
${\textrm CMP} |\Delta x|$	& $\leq$ 5.0 cm 			\\
${\textrm CMX} |\Delta x|$	& $\leq$ 6.0 cm 			\\
$|d_{0}|$		& $\leq$ 0.02 cm (0.2 cm) with (without) silicon tracking		\\
Isolation		& $\leq$ 0.1				\\
Cosmic ray		& Veto 		\\
\end{tabular}
\end{ruledtabular}
\end{table}

%Depending on the azimuthal position and
%the sign of the charge of the particle producing the  track, a correction term ranging from 0.00147 to 0.00073 is added to or 
%subtracted from the COT track curvature: $1/p_{T}$ (GeV/$c$)$^{-1}$, in order to remove an 
%azimuthal dependence from  residual detector alignment effects. 

The identification efficiency is determined from a data sample of $Z\to \mu^{+}\mu^{-}$\ events and
is found to be 85.1 $\pm$ 0.7\% for muons fiducial to CMU/CMP and 90.1 $\pm$ 0.8\% for muons fiducial to CMX,
where the uncertainty is statistical only.  In our estimate of the selection efficiency for top pair events, 
we are sensitive to systematic differences in muon identification between data and simulation.  
We use $Z\to \mu^{+}\mu^{-}$\ data and simulation samples
to measure correction factors of 0.887$\pm$0.014 for CMU/CMP and 1.001$\pm$0.017 for CMX muon
identification efficiencies, where the uncertainty is statistical only.
We discuss systematic uncertainties and differences between the muon environment in $Z\to \mu^{+}\mu^{-}$\ events 
and \ttbar\ events further in section~\ref{sec:sys}.

\subsection{\label{subsec:track} Track quality and primary vertex reconstruction}

For both electron and muon candidates, the charged particle track is required to have at least 3 axial and 3 
stereo COT super-layer track segments, with each segment  having at least 7 hits attached out of a possible total 
of 12 hits. We constrain the COT track fit to be consistent with the beam position in the transverse plane. 
We use an unbiased data sample collected by a calorimeter-only trigger to calibrate 
the track reconstruction efficiency for isolated leptons and 
we find a correction factor of 1.009 $\pm$ 0.002 to the simulation efficiency. 

We reconstruct the $z$ position of each primary interaction using an
algorithm based on COT and silicon tracking information.  Since there may be multiple $p\overline{p}$
interactions, we identify the $z$ coordinate of the event with the $z$ position of the reconstructed primary vertex 
closest to the lepton COT track $z$ position, $z_{0}$, at its point of closest approach to the beam-line in the transverse plane.   
In less than 1\% of the cases the separation is greater than 5 cm, so 
we use instead the $z_{0}$ of the lepton COT track as the event $z$ position.

We require the $z$\ position of the event to be within $60$~cm of the center of the detector, 
in order to ensure good event reconstruction in the projective tower geometry of the CDF calorimeter.
However, the integrated luminosity of the data sample is measured for the full \ppbar\ luminous region,
which extends beyond this range. 
Our simulation attempts to model the $z$ profile of the \ppbar\ luminous region but may not be correct on average.
Therefore, we estimate the selection efficiency for top pair events in simulation
with respect to events that have a $z$ position in this range.  We use minimum bias data to find 
that this range covers $94.8 \pm 0.3$\% of the full \ppbar\ luminous region. 
We then apply this number as a correction factor to our estimate of the selection efficiency for top pair events.

\subsection{\label{subsec:jet} Jet reconstruction and systematic uncertainties}

This analysis is heavily dependent on jet-based kinematic properties to  
discriminate between signal and background processes.   Therefore we discuss here
the reconstruction of jets and the uncertainties related to the jet energy scale~\cite{ARun1Paper}.        

The jets used in this analysis are reconstructed from calorimeter towers 
using a cone algorithm~\cite{JETCLUref} with a 
radius $R=\sqrt{(\Delta \phi)^{2}+(\Delta \eta)^{2}} \leq$0.4, where the \et\ of each tower is calculated with 
respect to the $z$\ coordinate of the event, as defined in the previous section.   
The calorimeter towers belonging to any electron candidate are not used by the 
jet clustering algorithm. We require three or more jets with $\et \geq 15$~GeV and 
$|\eta| < 2.0$, where we have corrected for the pseudo-rapidity dependence of the calorimeter 
response, the calibration of the calorimeter energy scale, and extra \et\ from any multiple \ppbar\ interactions.
 
The response of the calorimeter relative to the central region, $0.2 < |\eta| < 0.6$, 
is calibrated using a di-jet data sample.  For a 2 $\to$\ 2 process like di-jet production, the transverse energy of
the two jets should balance on average.  This property is
used to determine corrections as a function of jet pseudo-rapidity. 
The correction is largest (1.15) in the overlap region, $1.0<|\eta| < 1.4$, between the central and 
plug calorimeters.  In the region $|\eta|>1.0$, we find the simulation response differs from the data response by more than 2\%.
Therefore for this region, we derive a separate correction function for the simulation by applying the same 
technique to di-jet PYTHIA~\cite{PYTHIA} Monte Carlo. 
We take half of the difference between data and simulation as a systematic uncertainty. 
The systematic uncertainty on the relative calorimeter response is summarized in Table~\ref{tab:jesrel},
and includes additional contributions from the stability of the calibration in the central region
and variations in the parametrization function. 

The response of the central electromagnetic calorimeter is well understood ($<$1\%) from the 
position of the invariant mass peak in $Z \to e^{+}e^{-}$\ data. Therefore, with a sample of photon-jet events,
the well-measured energy of the photon can be used to check the calibration of the jet energy scale and to assess the modeling of the 
calorimeter response to jets.  We correct the simulation jet energy scale by a factor of 1.05, and 
assign a systematic uncertainty of 5\% based on comparison of photon-jet data to PYTHIA and HERWIG~\cite{HERWIG} Monte Carlo.   
A systematic uncertainty 
in the 3\% to 2\% range for jets with \et\ between 15 and 100 GeV is derived from the convolution of
the uncertainty on the simulation of the non-linear calorimeter response to low-energy particles with the \pt\ spectrum of particles
from the jet fragmentation.  

We use a jet cone size of $R=0.4$\ to separately reconstruct the many jets in \ttbar\ events.
However, a significant fraction of the particles from relatively broad low energy jets will lie outside this jet cone.
Checks of the modeling of the energy outside the jet cone introduce an additional systematic uncertainty in 
the 5\% to 1.5\% range for jets with \et\ between 15 and 100 GeV.

Particles from additional soft \ppbar\ interactions may deposit energy in the calorimeter that falls inside the jet cone.
For the highest instantaneous luminosity of $50 \times 10^{30}$~\cms\ in this dataset,
the mean number of \ppbar\ interactions per bunch crossing is about 1.8.
A good indicator of the number of \ppbar\ interactions in the same bunch crossing is the number of reconstructed primary vertices in the event.
We measure the amount of transverse energy inside a randomly chosen cone as a function of the number of reconstructed primary vertices
in an independent data sample collected with a minimum bias trigger. 
We subtract 260$\pm$100~MeV from the observed jet \et\ for each additional reconstructed primary vertex in the event.

The systematic uncertainties on the jet energy scale are summarized in Table~\ref{tab:jesrel}.
The total uncertainty is their sum in quadrature,
which gives a total systematic uncertainty of 11-12\% for jets with \et\ of
15~GeV and 5-8\% for jets with \et\ of 100~GeV.  Future improvements, including improved
simulation of the forward calorimeter response to low-energy particles, 
are expected to substantially reduce these rather large uncertainties.

\begin{table}[!hbtp]
\begin{ruledtabular}
\caption{\label{tab:jesrel}
Systematic uncertainties on the calorimeter response for a jet with \et\ of 15 (100) GeV.}
\begin{tabular}{lc}
Source					& Jet Energy Scale Uncertainty (\%) \\
\hline
Relative     $ |\eta| <$ 0.2    	& 3.2 (3.2) \\ %%	& 3 	& 1 \\
Relative 0.2 $<|\eta| <$ 0.6    	& 1.1 (1.1) \\ %%	& 0.5 	& 1 \\
Relative 0.6 $<|\eta| <$ 1.0    	& 2.2 (2.2) \\ %%	& 2 	& 1 \\
Relative 1.0 $<|\eta| <$ 1.4    	& 8.1 (8.1) \\ %%	& 4 	& 7 \\
Relative 1.4 $<|\eta| <$ 2.0    	& 6.3 (6.3) \\ %%	& 2 	& 6 \\
Relative    $ |\eta| >$ 2.0    		& 9.9 (9.9) \\ %%	& 7 	& 7 \\
Photon-jet balance			& 5.0 (5.0) \\
Single particle response		& 3.0 (2.0) \\
Out-of-cone energy			& 5.0 (1.5) \\
Multiple \ppbar\ interactions  		& 0.7 (0.1)\\ 
\end{tabular}
\end{ruledtabular}
\end{table}

\subsection{\label{subsec:met} Missing transverse energy reconstruction}

The presence of neutrinos in an event is inferred from an observed imbalance of transverse energy  
in the detector. The missing transverse energy, \met, is defined as 
the magnitude of the vector $-\sum_{i} (E_{T,i} \cos \phi_{i}, E_{T,i} \sin \phi_{i} )$, 
where $E_{T,i}$\ is the transverse energy, calculated with respect to the $z$\ coordinate of the event, 
in calorimeter tower $i$\ with azimuthal angle $\phi_{i}$. 
In the presence of any muon candidates, the \met\ vector is recalculated by subtracting
the transverse momentum of the muon COT track and adding back in the small amounts of 
transverse energy in the calorimeter towers traversed by the muon.  
For all jets with $\et \geq 8$~GeV and $|\eta| < 2.5$, the \met\ vector is adjusted for the
effect of the jet corrections discussed in the previous section.
In this analysis, we require $\met \geq 20$~GeV. 

\subsection{\label{subsec:furtherbkg} Multi-jet and multi-lepton rejection}

Multi-jet background events can pass the selection criteria and enter the data sample in several ways including:
semi-leptonic decay of a $b$ or $c$ quark producing both a charged lepton and missing transverse energy
from the neutrino; an electron from a photon conversion;
jet fragmentation with a charged pion and a neutral pion that mimics the signature of an electron;
jet fragmentation with decay-in-flight of a charged kaon that mimics the signature of a muon; and, in combination with the above,
mis-measurement of jet energies causing significant missing transverse energy.
However, in contrast to the isolated lepton from $W$ boson decay,
these lepton candidates tend to be surrounded by other particles from the parent jet.
Furthermore, the direction of the \met\ tends to be parallel or anti-parallel with the most
energetic jet in the event.  

\begin{figure}[!btp]
  \begin{center}
\resizebox{3.2in}{!}{ \includegraphics{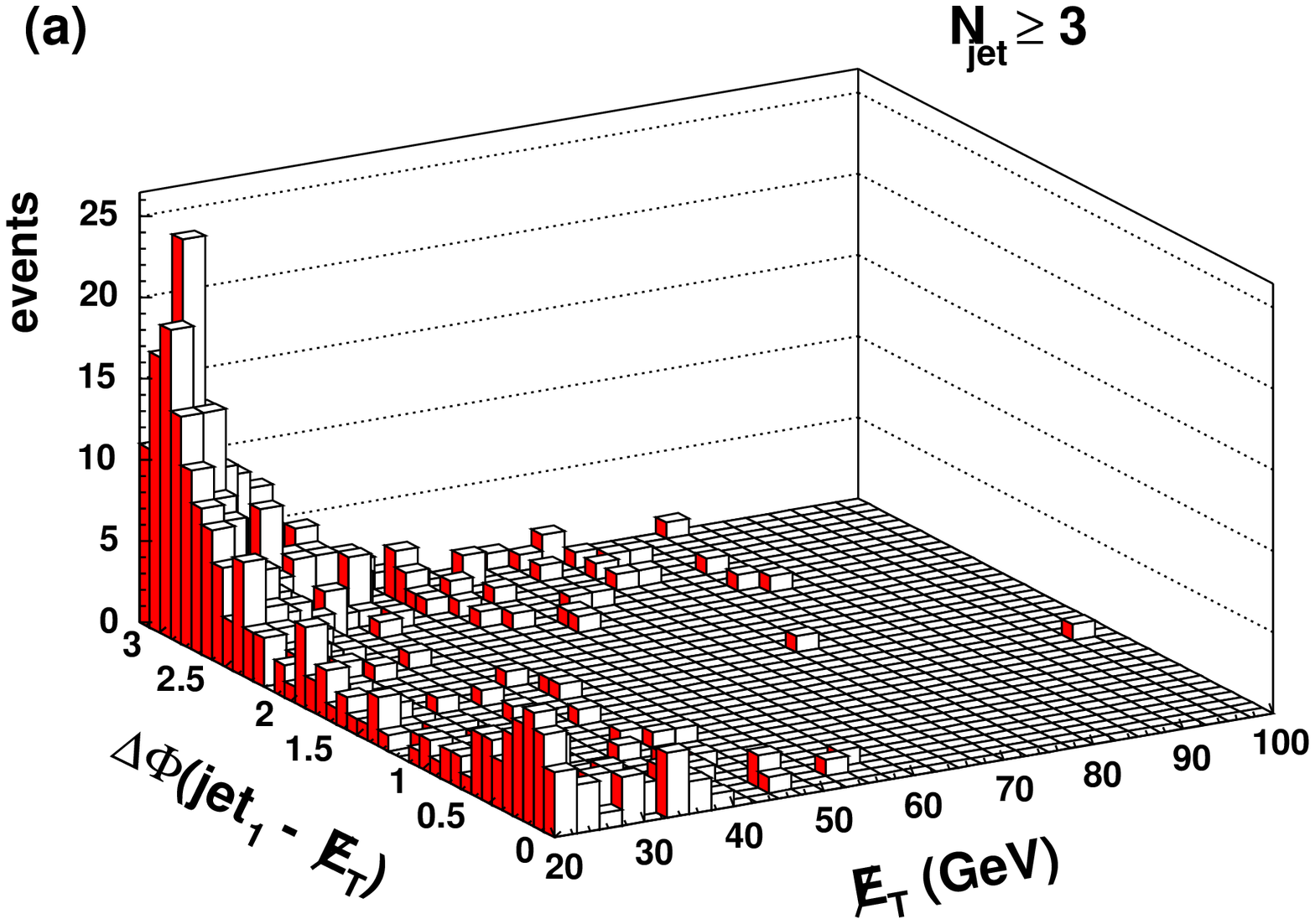}}
\resizebox{3.2in}{!}{ \includegraphics{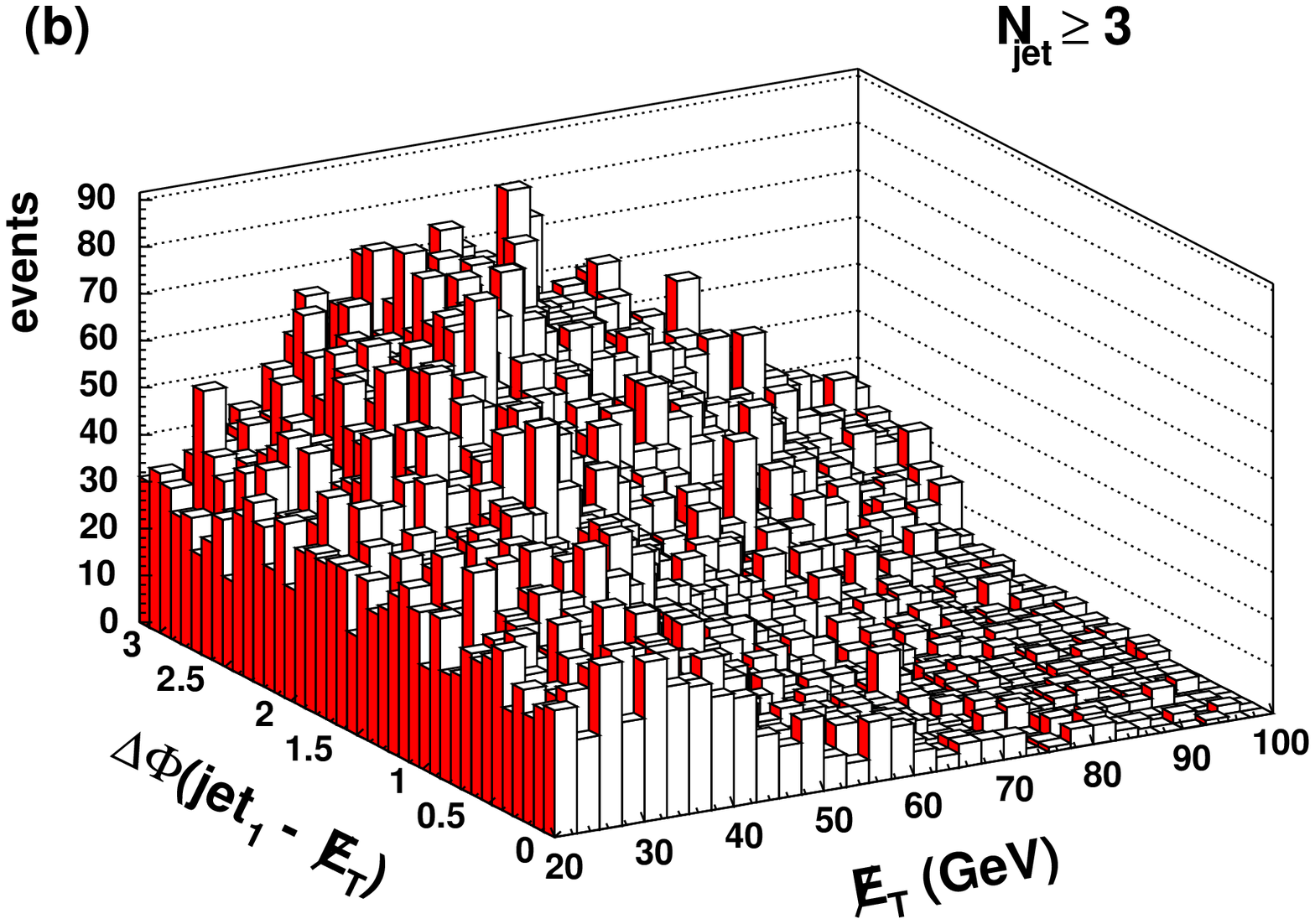}}
%%\resizebox{3.2in}{!}{ \includegraphics{fig0_wjets.eps}}
  \end{center}
  \caption{\label{fig:dphi_legooo} The angle in the transverse plane between the direction of the \met\ and the leading jet 
   versus the \met\ for (a) our model of the multi-jet background from 
 the non-isolated lepton data sample, and (b) PYTHIA \ttbar\ Monte Carlo. %%, and (c) ALPGEN+HERWIG $W$+3 partons Monte Carlo.  
}
\end{figure}

Due to the high purity of the lepton identification criteria, 
it is difficult to create a high statistics model of this background by using Monte Carlo simulations.
Therefore, we model the kinematics of the multi-jet background using data events that 
pass all of our selection requirements except lepton isolation, where instead we require poor isolation, $I>0.2$.
The \met\ distribution versus the azimuthal angle, $\Delta \phi$, 
between the direction of the \met\ and the highest \et\ jet is shown in 
Fig.~\ref{fig:dphi_legooo}(a) for our model of the multi-jet background derived from non-isolated lepton data 
and in Fig.~\ref{fig:dphi_legooo}(b) for the PYTHIA Monte Carlo simulation of the \ttbar\ signal. 
We find that we can reduce the multi-jet background by 50\%
by requiring that $0.5 < \Delta\phi < 2.5$ radians for events with $\met <$ 30~GeV.  
This multi-jet veto is 95\% efficient for \ttbar\ events passing the previous requirements.

Backgrounds from processes with two or more high \pt\ leptons include single top production and $Z$ boson, 
$WW$, $WZ$\ and $ZZ$\ diboson  production with associated jets. 
We remove all events with two or more leptons satisfying the usual identification 
criteria in Tables~\ref{tab:elesel} and~\ref{tab:musel}. To avoid overlap with the 
\ttbar\ dilepton analysis~\cite{ttbardilepton}, we also remove events that contain an additional 
lepton identified either as an electron in the plug calorimeter or as a muon with a track segment in 
CMU but outside the fiducial volume of CMP and vice versa. 
To further reduce the residual background from processes with leptonic $Z$ decays, 
we remove events where the primary lepton and a second object form an invariant mass within the 
76-106 ~{GeV/$c^{2}$} window containing the $Z$ boson mass.  The criteria for this 
second object are designed to remove events where the second lepton is outside the fiducial volume of a calorimeter tower 
or muon chamber:

\begin{itemize}
\item The second object may be a lepton with relaxed identification requirements as listed in Table~\ref{tab:zveto}.

\item  The second object may be an isolated oppositely charged particle track with 
$\pt \geq 10$~{GeV/$c$} that extrapolates back to within 10~cm of the z position of the event. 
In this case, isolated means that any additional tracks within a cone of
radius $R=0.4$\ have transverse momentum sum below 4.0~{GeV/$c$}.

\item If the primary lepton is an electron, the second object may also be a jet with 
$\et \geq 15$~GeV, $|\eta| \leq 2.0$,  less than three tracks inside a cone of radius
$R=0.4$, and over 95\% of the total energy in the electromagnetic calorimeter.

\end{itemize}

The multi-lepton veto removes about 90\% of $Z \rightarrow e^{+}e^{-}$\ events and 
about 50\% of $Z \rightarrow \mu^{+}\mu^{-}$\ events, where the difference is due to the larger geometrical coverage
of the calorimeter for electrons compared to that of the tracking system for muons. 
This multi-lepton veto is 96\% efficient for \ttbar\ events passing the previous requirements.

\begin{table}[!hbtp]
\caption{\label{tab:zveto}
Selection requirements for second lepton used in the $Z$ boson veto.}
\begin{ruledtabular}
\begin{tabular}{ll}
Property & Requirement \\
\hline
\multicolumn{2}{c}{Electron} \\
$E_{T}$ 		& $\geq$ 10.0 GeV 			\\
$E_{\textrm had}/E_{\textrm em}$ 	& $\leq$ 0.12 		\\
Isolation		& $\leq$ 0.15				\\
\hline
\multicolumn{2}{c}{Muon with a track segment in the muon chambers} \\
$p_{T}$			& $\geq$ 10.0 GeV/$c$ \\
$E_{\textrm had}$ 		& $\leq$ 10.0 GeV  \\
$E_{\textrm em}$ 		& $\leq$ 5.0 GeV \\
$|\Delta x|$		& $\leq$ 10.0 cm 		\\
$|d_{0}|$		& $\leq$ 0.5 cm 		\\
Isolation		& $\leq$ 0.15			\\
\hline
\multicolumn{2}{c}{Muon without a track segment in the muon chambers} \\
$p_{T}$			& $\geq$ 10.0 GeV/$c$ \\
$E_{\textrm had}$ 		& $\leq$ 6.0 GeV  \\
$E_{\textrm em}$ 		& $\leq$ 2.0 GeV \\
$E_{\textrm em}+E_{\textrm had}$ 	& $\leq$ 10.0 GeV \\
$|d_{0}|$		& $\leq$ 0.5 cm 		\\
Isolation		& $\leq$ 0.15			\\
\end{tabular}
\end{ruledtabular}
\end{table}

\subsection{\label{subsec:summary} Observed data events}

In summary, our selection of $\ttbar \rightarrow \ell\bar{\nu_{\ell}}q\bar{q'}b\bar{b}$\ decays
requires a $W \to \ell\nu$\ candidate and at least three jets, which we will refer
to as $W+\geq3$jets. The $W$ boson candidate is one isolated lepton with $\et \geq 20$~GeV and missing 
transverse energy $\met \geq 20$~GeV.  Jets are reconstructed with a cone algorithm of 
radius $R=0.4$\ and are required to have $\et \geq 15$~GeV and $|\eta| < 2.0$.  
In order to reduce the background from multi-jet processes, we require the directions of 
the \met\ and the most energetic jet to be well-separated in the transverse plane if $\met < 30$~GeV.

Table~\ref{tab:ndata} lists the number of observed events in 194~\pbi\ of data, 
for the electron and muon channels separately and combined, as a function of the jet multiplicity. 
We also show our expectation for the number of \ttbar\ events, where we use our estimate from the next section of the acceptance
for a top mass of 175~GeV/$c^{2}$\ and assume the NLO production cross section of 6.7~pb~\cite{mlm,kidonakis}. 

\begin{table}[!hbtp]
\caption{\label{tab:ndata} The observed number of $W \to \ell\nu$\ candidates as a function of the jet multiplicity, compared to
the expectation from PYTHIA \ttbar\ Monte Carlo simulation, where we assume a top mass of 175~\gevcii.  
We require at least 3 jets. 
}
  \begin{ruledtabular}
  \begin{tabular}{rrrr|r}
Jet multiplicity & Electron & Muon  & Total & Expected \ttbar\\
\hline 
%\multicolumn{5}{c}{Before $\Delta \phi$} \\
%\hline
%0   		& 99454  & 76203 & 175657 & 0.2  \\ 
%1  		& 10683  & 7744  &  18427 & 4.5 \\ 
%2 	  	& 1736   & 1180  &   2916 & 23.5 \\ 
%3   		& 304    & 178   &    482 & 44.5\\ 
%$\geq$ 4   	& 86     & 44  	 &    130 & 52.8 \\ 
%\hline
%\multicolumn{5}{c}{After $\Delta \phi$} \\
%\hline
0   		& 99454 & 76203  & 175657 & 0.2  \\
1   		& 9407  & 6982   & 16389  & 4.4  \\
2 	  	& 1442  & 1054   & 2496   & 22.6 \\
\hline
3 	  	& 254   & 147    & 401    & 42.3 \\
$\geq$ 4   	& 78    & 40     & 118    & 49.9 \\
  \end{tabular}
  \end{ruledtabular}
  \end{table}

%------------------------------------------------------------------
\section{\label{sec:ttbaracc} Signal Acceptance}

We measure the fraction of \ttbar\ events accepted by our event selection requirements using a 
combination of Monte Carlo simulation and data. We generate \ttbar\ events with the PYTHIA Monte Carlo program, which 
has a leading order matrix element for the parton hard scattering convoluted with the CTEQ5L 
parton distribution functions~\cite{CTEQ}.  The acceptances from PYTHIA for each type of 
identified lepton are shown in the top line of Table~\ref{tab:acc}.
We correct these raw fractions for several effects, described in the previous section,
that are not sufficiently well-modeled in our simulation: 
the lepton trigger efficiencies, measured from data; 
the fraction of the \ppbar\ luminous region well-contained in the CDF detector, measured from data;
the difference between the track reconstruction efficiency measured in data and simulation;
and the difference between lepton identification efficiencies measured in 
$Z \rightarrow \ell^{+}\ell^{-}$\ data and PYTHIA Monte Carlo.  
All of the correction factors for each type of identified lepton are shown in Table~\ref{tab:acc}.

The total acceptance of our event selection for \ttbar\ is $7.11 \pm 0.56$\%, given by 
the sum of the corrected acceptance weighted by the integrated luminosity of the data sample for each type of identified lepton.
The uncertainty includes the systematic uncertainties discussed later in Section~\ref{sec:sys}.
We assume a top mass of 175~GeV/$c^{2}$\ and the  PYTHIA branching fraction for $W\rightarrow \ell \nu$\ of 10.8\%.
The acceptance is mostly from the $\ttbar \rightarrow \ell \nu q\bar{q} b\bar{b}$\ channel, 
but also contains small contributions from other \ttbar\ decay modes, as shown in Table~\ref{tab:acccomp}.

\begin{table*}[!btp]
\caption{\label{tab:acc} \ttbar\ acceptance and correction factors.  We assume a top mass of 175~\gevcii.  }
\begin{ruledtabular}
\begin{tabular}{lccc}
Quantity & CEM Electron  & CMU/CMP Muon & CMX Muon\\
\hline
PYTHIA acceptance & 0.0462 $\pm$ 0.0004  &  0.0283 $\pm$ 0.0003  & 0.0104 $\pm$ 0.002   \\
\hline
Efficiency:  Trigger 	  	    &  0.962 $\pm$ 0.006   & 0.887 $\pm$ 0.007   & 0.954 $\pm$ 0.004\\
Efficiency:  Luminous region        & {0.948 $\pm$ 0.003}  & {0.948 $\pm$ 0.003} & {0.948 $\pm$ 0.003} \\
Correction:  Track reconstruction   &  {1.009 $\pm$ 0.002} & {1.009 $\pm$ 0.002} & {1.009 $\pm$ 0.002}\\
Correction:  Lepton identification  &  0.965 $\pm$ 0.014   & 0.887 $\pm$ 0.014   & 1.001 $\pm$ 0.017\\
\hline
Corrected acceptance & \bf{0.0412 $\pm$ 0.0033} & \bf{0.0213 $\pm$ 0.0017} & \bf{0.0095 $\pm$ 0.0008}\\ 
\hline
Integrated luminosity (pb$^{-1}$) & 194 & 194 & 175 \\
\end{tabular}
\end{ruledtabular}
\end{table*}

\begin{table*}[!htbp]
\caption{\label{tab:acccomp} Expected composition of selected \ttbar\ events in terms of the various \ttbar\ decay modes, as determined from PYTHIA \ttbar\ Monte Carlo simulation.}
\begin{ruledtabular}
\begin{tabular}{cccc}
\ttbar\ decay mode				     & Signal composition(\%) \\
\hline
$\ttbar \rightarrow \ell \nu q\bar{q} b\bar{b}$      & 82  \\
$\ttbar \rightarrow \tau \nu \ell \nu b\bar{b}$      &  7  \\
$\ttbar \rightarrow \tau \nu q\bar{q} b\bar{b}$      &  6  \\
$\ttbar \rightarrow \ell \nu \ell \nu b\bar{b}$      &  5  \\
\end{tabular}
\end{ruledtabular}
\end{table*}

%The composition of the acceptance is 82\% from $\ttbar \rightarrow \ell \nu q\bar{q} b\bar{b}$, 
%7\% from $\ttbar \rightarrow \tau \nu \ell \nu b\bar{b}$, 6\% from $\ttbar \rightarrow \tau \nu q\bar{q} b\bar{b}$\ and
% 5\% from $\ttbar \rightarrow \ell \nu \ell \nu b\bar{b}$. The systematic uncertainties relevant
%to the calculation of the \ttbar\ acceptance are calculated separately for each effect:

% This table has been killed, the discussion of acceptance systematics was moved to the 
% systematics section
%\begin{table}[!htbp]
%\caption{\label{tab:accsys} Relative systematics uncertainties on the acceptance.}
%\begin{ruledtabular}
%\begin{tabular}{lc}
%Source	& Systematic\\
%\hline
%Lepton Identification & 2.0\%\\
%Lepton Isolation & 5.0\%\\
%Jet Energy Scale & 4.7\%\\
%ISR/FSR          & 3.0\%\\
%PDF              & 1.5\%\\
%Generator        & 1.4\%\\
%\end{tabular}
%\end{ruledtabular}
%\end{table}

%------------------------------------------------------------------
\section{\label{sec:bkg} Backgrounds}

A variety of non-$\ttbar$ processes can also produce events that pass
our $W+\geq$3 jets selection requirements.  These backgrounds can be grouped into three categories: 
production of a $W$ boson with associated jets, $W$+jets; other electroweak processes resulting 
in at least one high $p_{T}$ lepton and jets; and generic QCD multi-jet processes. 
However, theoretical predictions for the total rate of these processes only exist at leading-order, with associated uncertainties of 50\%
from the choice of scale used to evaluate the strong coupling constant $\alpha_{s}(Q^{2})$.
Instead, we estimate their contribution to the data sample by exploiting the difference between the kinematics of these
background processes and \ttbar\ production.  In this section, we discuss the Monte Carlo model we use to describe the kinematics 
of the $W$+jets and other electroweak processes.  For the multi-jet events,  we model their kinematics
from an independent data sample and derive an estimate for their contribution.

Much theoretical progress has been made recently to improve the description of the kinematics of the $W$+jets process, 
with leading-order matrix-element generators now available to describe the parton hard
 scattering for processes with a $W$\ boson and up to six well-separated partons in the final state.  
We use the ALPGEN~\cite{ALPGEN} matrix element generator, convoluted with the
CTEQ5L parton distribution functions. We require parton $|\eta|\leq 3.0$, $\pt \geq 8$~{GeV/$c$}  
and a minimum separation  $\Delta R\geq0.2$ between {\it u, d, s} and {\it g} partons at the 
generation level.  We have verified that the shapes of the kinematic distributions used in our
 analysis  are not sensitive to these values. We choose a default  
momentum transfer scale of $Q^{2}= M_{W}^{2}+\sum_{i} p_{T,i}^{2}$\
for the parton distribution functions and the evaluation of $\alpha_{s}$, where $p_{T,i}$\ 
is the transverse momentum of the $i$-th parton. We use the HERWIG parton shower 
algorithm to evolve the final state partons to colorless hadrons.  
Note that the addition of all of the $W$+n parton ALPGEN+HERWIG samples does not give a 
good model of the kinematics of the entire $W$+jets sample. For instance, for a given $W$+1 parton matrix element, 
the parton shower may radiate a gluon with large enough $\pt$\ such that this final state would also 
be covered by the $W$+2 parton matrix element. 
We note that there has been significant recent theoretical and phenomenological progress here: 
an approach developed to solve this double-counting problem~\cite{CKKW} at $e^{+}e^{-}$\ colliders 
has been adapted to the more complicated environment of hadron colliders and
implemented in the PYTHIA and HERWIG Monte Carlo generators~\cite{mrenna}. 

We use the $W$+n parton ALPGEN+HERWIG Monte Carlo to model the $W+\geq$n jet final state,
where we rely on gluon radiation in the parton shower algorithm to adequately model the larger
jet multiplicities.  We also use the ALPGEN+HERWIG Monte Carlo to model $Z$\ boson and diboson ($WW$, $WZ$, $ZZ$) production 
with associated jets. PYTHIA is used to simulate single top production. 
We show the composition of the background from electroweak processes in Table~\ref{tab:wbkg}, where we use the leading order
cross section from ALPGEN to normalize the contributions from different processes.  
We use the term ``W-like'' to refer collectively to all of these electroweak background processes.

\begin{table*}[!htbp]
  \caption{\label{tab:wbkg} 
Expected composition of the $W$-like electroweak background in the electron and muon channels. 
} 
  \begin{ruledtabular}
  \begin{tabular}{lc|cc}
   Process & Generator $\sigma$ (pb) & Electron (\%) & Muon (\%)\\
    \hline 
    $W\rightarrow \ell \nu+$3 parton 		& 179.8 & 87.3 & 84.8 \\	
    $W\rightarrow \tau\nu+$3 parton 		& 89.9  &  4.6 &  4.6 \\	
    $Z \rightarrow \ell^{+}\ell^{-} +2$ parton 	& 46.6  &  1.5 &  4.2 \\	
    $Z \rightarrow \tau^{+}\tau^{-} +2$ parton 	& 23.3  &  1.3  & 1.3 \\	
    $WW$+1 parton				& 4.38  &  3.8  & 3.7 \\	
    $WZ$+1 parton 				& 2.37  &  0.4  & 0.4 \\	
    single top 					& 3.0   &  1.0  & 1.0 \\
  \end{tabular}
  \end{ruledtabular}
  \end{table*}

\begin{figure}[!htbp]
  \begin{center}
\resizebox{3.2in}{!}{ \includegraphics{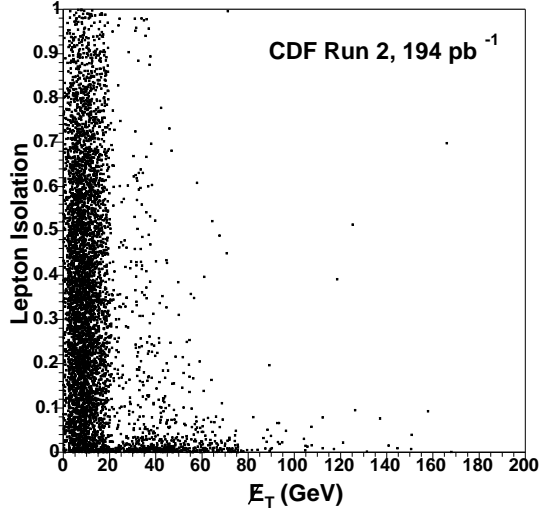}}
  \end{center}
  \caption{\label{fig:met_vs_iso} The \met\ versus isolation distributions for events with
   a lepton and 3 or more jets. The structure apparent between 20 and 30 GeV comes from removing events
    where missing transverse energy lies close to the direction of the highest \et\ jet.
}
\end{figure}

As discussed previously in Section~\ref{subsec:furtherbkg},
multi-jet background events are often characterized by significant additional energy in the cone around the lepton and
low missing transverse energy.  We model the kinematics of the multi-jet background using data 
events that pass all of the selection requirements except for a lepton isolation requirement of  $\geq$\ 0.2.  
To estimate the rate of this background, we assume that there is no correlation between the
\met\ and the isolation of the identified lepton, shown in Fig.~\ref{fig:met_vs_iso}. 
The number of background events passing the selection requirements can then be estimated by comparing 
the number of events in various control regions:

\begin{itemize}
\item $n_{A}$: lepton isolation $I>0.2$ and $\met<10$~GeV 
\item $n_{B}$: lepton isolation $I<0.1$ and $\met<10$~GeV 
\item $n_{C}$: lepton isolation $I>0.2$ and $\met>20$~GeV. 
\end{itemize}

\noindent Since the above numbers should reflect only the multi-jet process, corrections are made
for the  expected contribution from $W+$jets and $\ttbar$\ events. In our signal region, defined by 
$\met>20$~GeV and lepton isolation $I<0.1$, the number of multi-jet events is estimated as 
$n_{C} \times n_{B}/n_{A}$.  Table~\ref{tab:qcdbkg} lists the fraction of events in the signal 
region from QCD multi-jet processes as a function of jet multiplicity. 
We check the assumption of no correlation between \met\ and isolation by 
variation of the requirements that define the control regions: this changes the estimates by $\pm$50\%.  
We discuss the systematic uncertainty on this estimate further in Section~\ref{sec:sys}.

The larger multi-jet background in the electron data sample is partly due to electrons from 
unidentified photon conversions in detector material. 
The number of events identified by the photon conversion algorithm described in Section~\ref{subsec:eleid} 
can be written as $N_{i} = \epsilon \times N_{c} + m \times (N + N_{i} - N_{c})$.
The first term is the number of events with a photon conversion $N_{c}$\ multiplied by the efficiency of the conversion algorithm, $\epsilon=72.6\pm 0.7\%$.  
The second term is the number of events without a photon conversion that are mis-identified by the 
conversion algorithm, where the mis-identification rate is $m=2.3\pm0.04\%$\ and 
$N$\ is the number of events in the electron data sample.
Therefore, the number of events remaining in the electron data sample 
with an unidentified photon conversion is $N_{u}= N_{c} \times (1-\epsilon)$.  This estimate is shown in 
Table~\ref{tab:conv} and demonstrates that the majority of the QCD multi-jet background in the electron data sample 
comes from unidentified photon conversions.

\begin{table}[!htpb]
\caption{\label{tab:qcdbkg} 
The estimated fraction of the QCD multi-jet background in the $W$+jets data sample as a 
function of jet multiplicity. The uncertainty is statistical only.}
  \begin{ruledtabular}
  \begin{tabular}{cccc}
   Jet multiplicity & Electron & Muon & Total \\
    \hline 
        1 jet        	 & 3.8 $\pm$ 0.2\% & 2.9 $\pm$ 0.2\% & 3.4 $\pm$ 0.3\% \\
        2 jets        	 & 6.1 $\pm$ 0.5\% & 2.0 $\pm$ 0.2\% & 4.3 $\pm$ 0.5\% \\
	$\ge$3 jets	 & 7.7 $\pm$ 1.3\% & 3.1 $\pm$ 0.9\% & 6.3 $\pm$ 1.6\% \\ 
  \end{tabular}
  \end{ruledtabular}
   \end{table}

\begin{table}[!htpb]
\caption{\label{tab:conv} 
Estimate of the contribution to the electron data sample from unidentified conversions.}
\begin{ruledtabular}
  \begin{tabular}{lrrr|c}
   Jet  		& $N_{i}$\ 		   & $N$     	      &  $N_{u}$    		    & $N_{u}/N$ \\
    multiplicity        & Identified conversion    & Electron Data    &  Unidentified conversion    &  (\%)         \\
    \hline 
        1 jet        	& 791 		& 9407   &  217 $\pm$ 13   & 2.3 $\pm$ 0.2\% \\
        2 jets        	& 296 		& 1442   &  100 $\pm$ 8    & 6.9 $\pm$ 0.5\% \\
	$\ge$3 jets	&  81 		&  332   &  28 $\pm$  4    & 8.4 $\pm$ 1.2\% \\ 
  \end{tabular}
  \end{ruledtabular}
   \end{table}

%//Better version
%
%{
%  //  float Ni = 791.0; //id conversions
%  //float N = 9407.0; //data - id conversions
%  
%  //float Ni = 296.0; //id conversions
%  //float N = 1442.0; //data - id conversions
%  
%  float Ni = 81; //id conversions
%  float N = 332; //data - id conversions
%  
%  
%  float e = 0.726;  // eff conversion finder
%  float m = 0.023;  // mis-id rate of Z electrons
%  float erre= 0.007;
%  float errm= 0.0004;
%  
%  float Nc=0.0;
%  float errNc=0.0;
%  float Nu=0.0;
%  float errNu=0.0;
% 
%  Nc = (Ni*(1.0-m) - m*N)/(e-m);
%  Nu = Nc*(1.0-e);
%  
%  float ddm = -(Ni+N)*(1.0-e)/(e-m) - Nu/(e-m);
%  float dde = - Nc - Nu/(e-m);
%  
%  float ddNi = (1.0-m)*(1.0-e)/(e-m);
%  float ddN =  m*(1.0-e)/(e-m);
%  
%  errNu = sqrt(ddm*ddm *errm*errm 
%	       + dde*dde *erre*erre 
%	       + ddNi*ddNi * Ni 
%	       + ddN*ddN * N); 
%  
%  float frac = 100.0 * Nu/N;
%  float errf = 100.0 * errNu/N;
%  
%  std::cout << " Data " << N
%	    << " id " << Ni
%	    << " conv " << Nc 
%	    << " unid " << Nu << "+-" << errNu
%	    << " fraction " << frac << "+-" << errf << std::endl;
% 
%}

%------------------------------------------------------------------
\section{\label{sec:kinvar} Cross Section Measurement Method}

%A comparison of the observed data events with the expected number of \ttbar\ signal
%events in Table~\ref{tab:ndata} indicates the expected signal to background ratio 
%is about 1:4  and 1:1 in the $W+\geq$3 jets and $W+\geq$4 jets samples respectively.
%At such low signal purities, 

A comparison of the observed number of data events with the expected number of signal
for a \ttbar\ cross section in the range predicted by theory is shown in Table~\ref{tab:ndata}.
The sensitivity to top pair production from counting the observed
number of events alone is overwhelmed by the 50\% uncertainty on the leading-order theoretical
 prediction for the $W$+jets background.  Previous CDF measurements of the top pair production 
cross section in the lepton+jets channel~\cite{svxruniipaper} have used $b$-tagging, at the cost of about 45\% loss in signal 
acceptance, in order to improve the signal-to-background ratio 
and also use the more accurate prediction for the fraction of $W$+jets containing heavy flavor,
where the leading-order scale dependence of the absolute cross sections largely cancels.

This analysis instead exploits the discrimination available from kinematic and topological 
properties to distinguish \ttbar\ from background processes.  Due to the large mass of the top 
quark, top pair production is associated with central, spherical events with large total \et,
 unlike most of the background processes. We model the kinematics of \ttbar\ and $W$-like background processes
with Monte Carlo simulation. For the QCD multi-jet background, we model the kinematics with a non-isolated lepton data sample.
We use these models to describe the data distribution of a suitably discriminating property.  We  
extract the most likely number of events from \ttbar\ production, $\mu_{\ttbar}$, from a binned maximum likelihood fit:
\begin{equation}
 L(\mu_{\ttbar},\mu_{w},\mu_{q}) = \prod^{N_{bins}}_{i=1} \frac{e^{-\mu_{i}} \mu_{i}^{d_{i}}}{d_{i}!},
\label{eqn:lik}
\end{equation}
\noindent where $\mu_{\ttbar}$, $\mu_{w}$, $\mu_{q}$\ are the parameters of the fit, 
representing Poisson means for the number of \ttbar, W-like, and multi-jet events 
in our data sample. The expected number of events in the $i$-th bin is 
$\mu_{i} = (\mu_{\ttbar}P_{\ttbar,i} + \mu_{w}P_{w,i} + \mu_{q}P_{q,i})$, where $P_{\ttbar,i}$, 
$P_{w,i}$, $P_{q,i}$\ is the probability for observing an event in the  $i$-th bin from \ttbar, 
W-like and  multi-jet processes respectively.  The variable $d_{i}$\ is the number of observed 
data events that populate the $i$-th bin.  The number of multi-jet background events, $\mu_{q}$, 
 is fixed to that expected from Table~\ref{tab:qcdbkg}. Note that the uncertainty on our estimate of the
 number of multi-jet background events is included in the systematic uncertainties discussed in Section~\ref{sec:sys}.
 
We convert the fitted number of \ttbar\ events into the top pair 
production cross section, $\sigma_{\ttbar}$, using the acceptance estimate, $\epsilon_{\ttbar}$, from Section~\ref{sec:ttbaracc}, 
including the branching  ratio for $W\rightarrow \ell\nu$, and the luminosity measurement 
$\mathcal{L}$:
\begin{equation}
\sigma_{\ttbar} = \frac{\mu_{\ttbar}}{\epsilon_{\ttbar} \mathcal{L}}.
\label{eqn:llk}
\end{equation}

In the rest of this section, we first describe our choice of a single kinematic discriminant, then 
how we maximize our discriminating power by developing an optimal variable with an 
Artificial Neural Network (ANN) technique.  ANN's employ information from several 
properties while accounting for the correlations among them~\cite{NN}. 

\subsection{\label{subsec:1var} Single discriminant}

\begin{figure*}[!hbtp]
  \begin{center}
  %1.734
\resizebox{1.7in}{!}{ \includegraphics{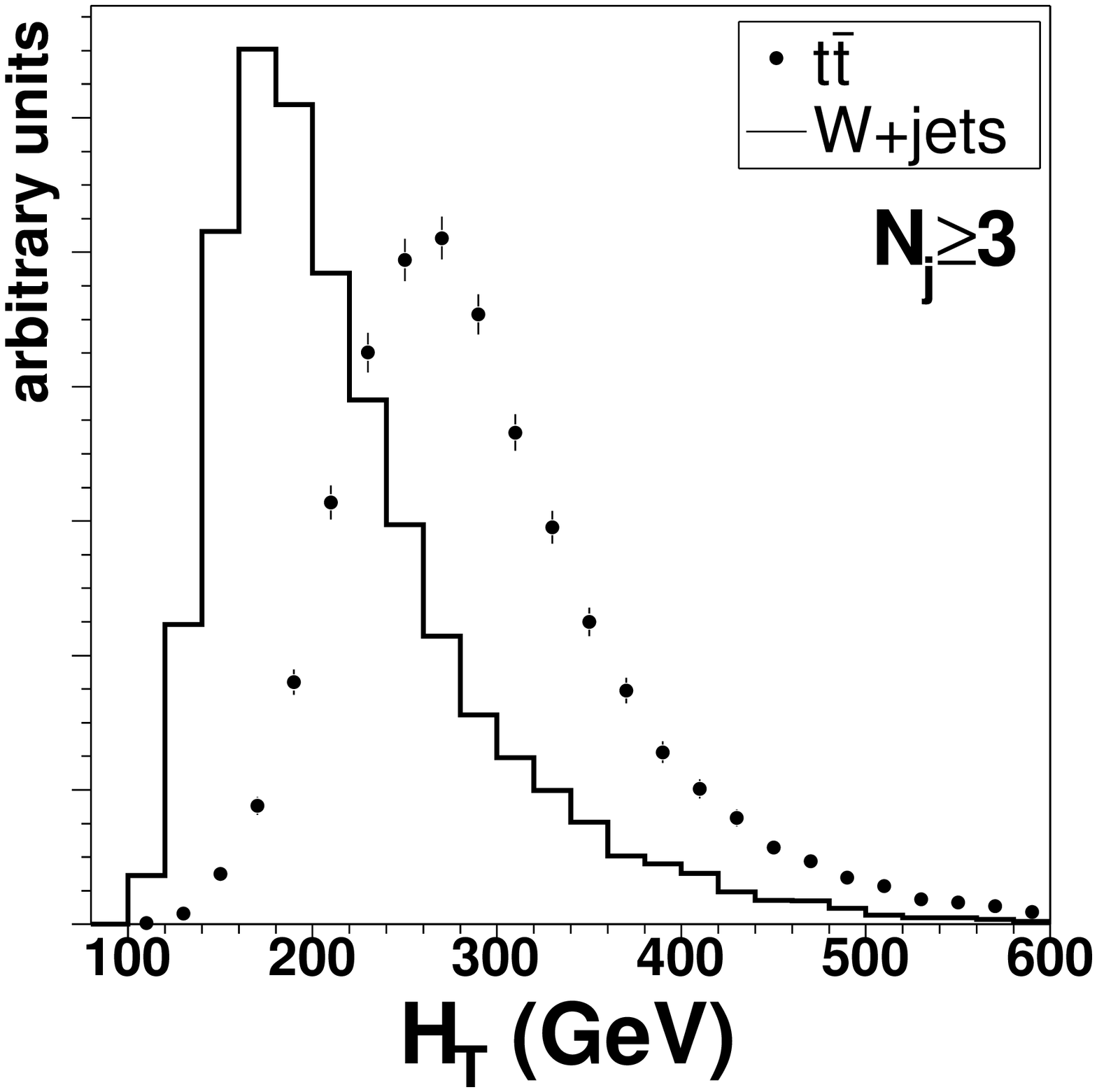}}
\resizebox{1.7in}{!}{ \includegraphics{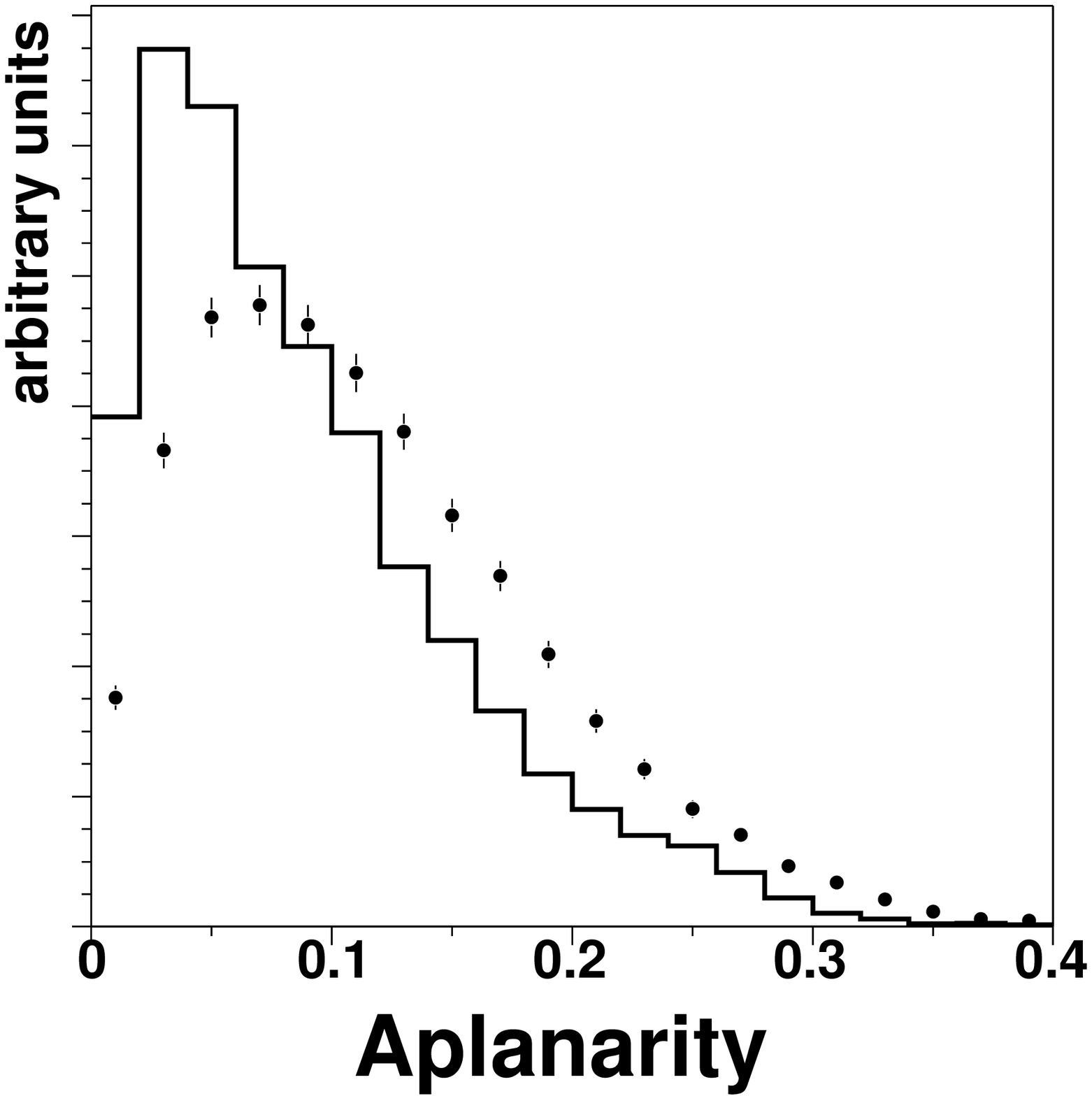}}
\resizebox{1.7in}{!}{ \includegraphics{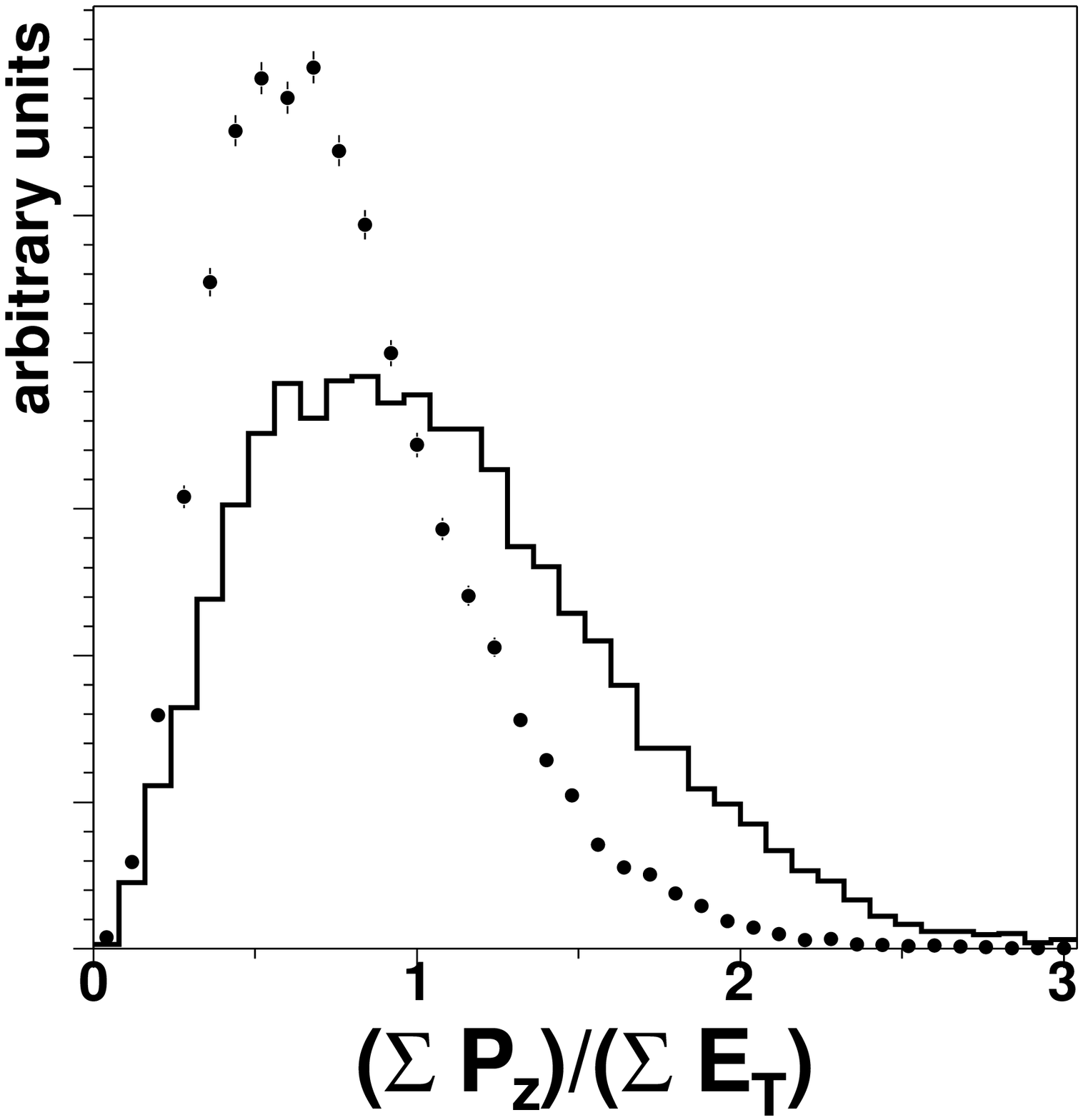}}
\resizebox{1.7in}{!}{ \includegraphics{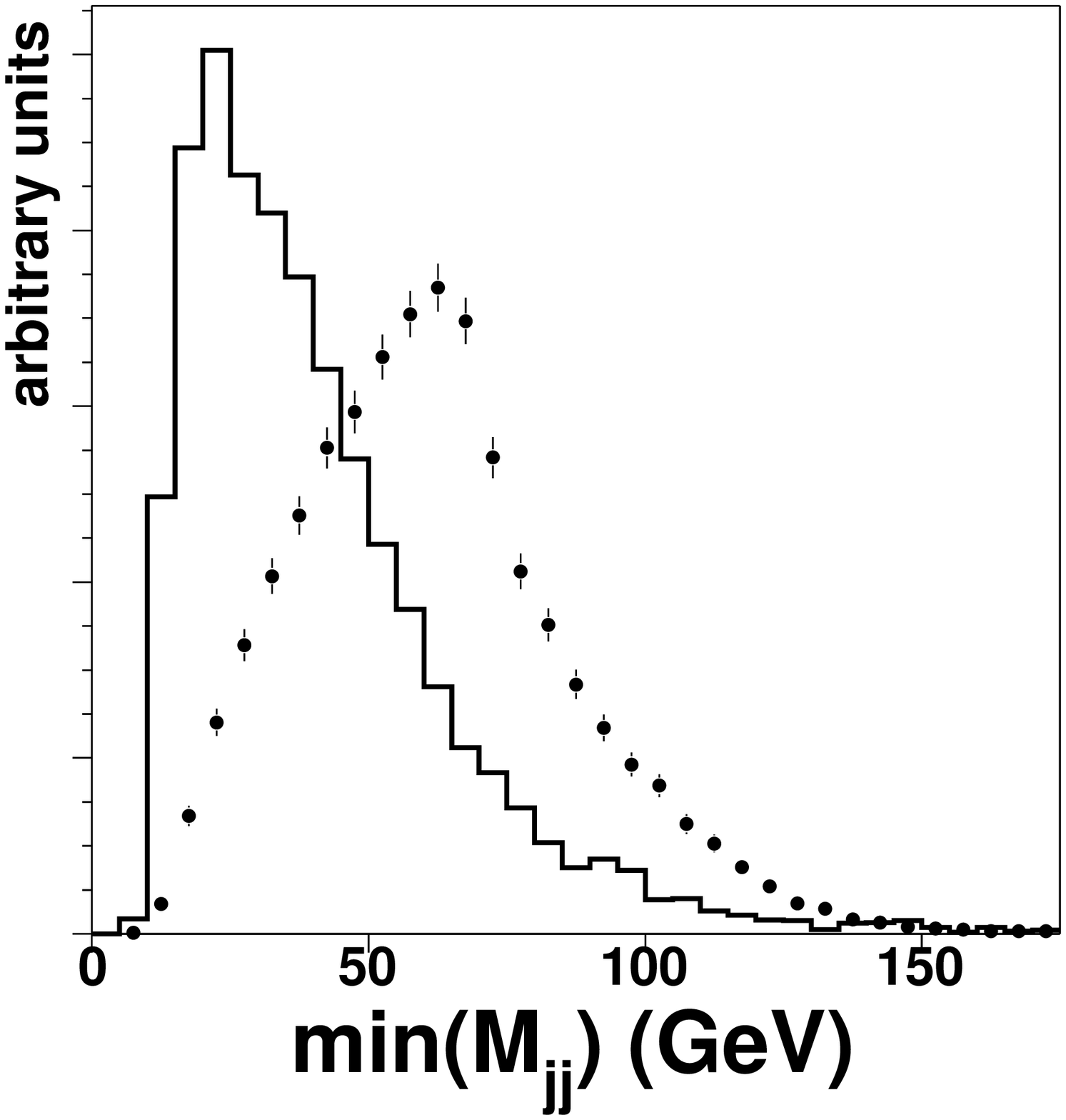}}
\resizebox{1.7in}{!}{ \includegraphics{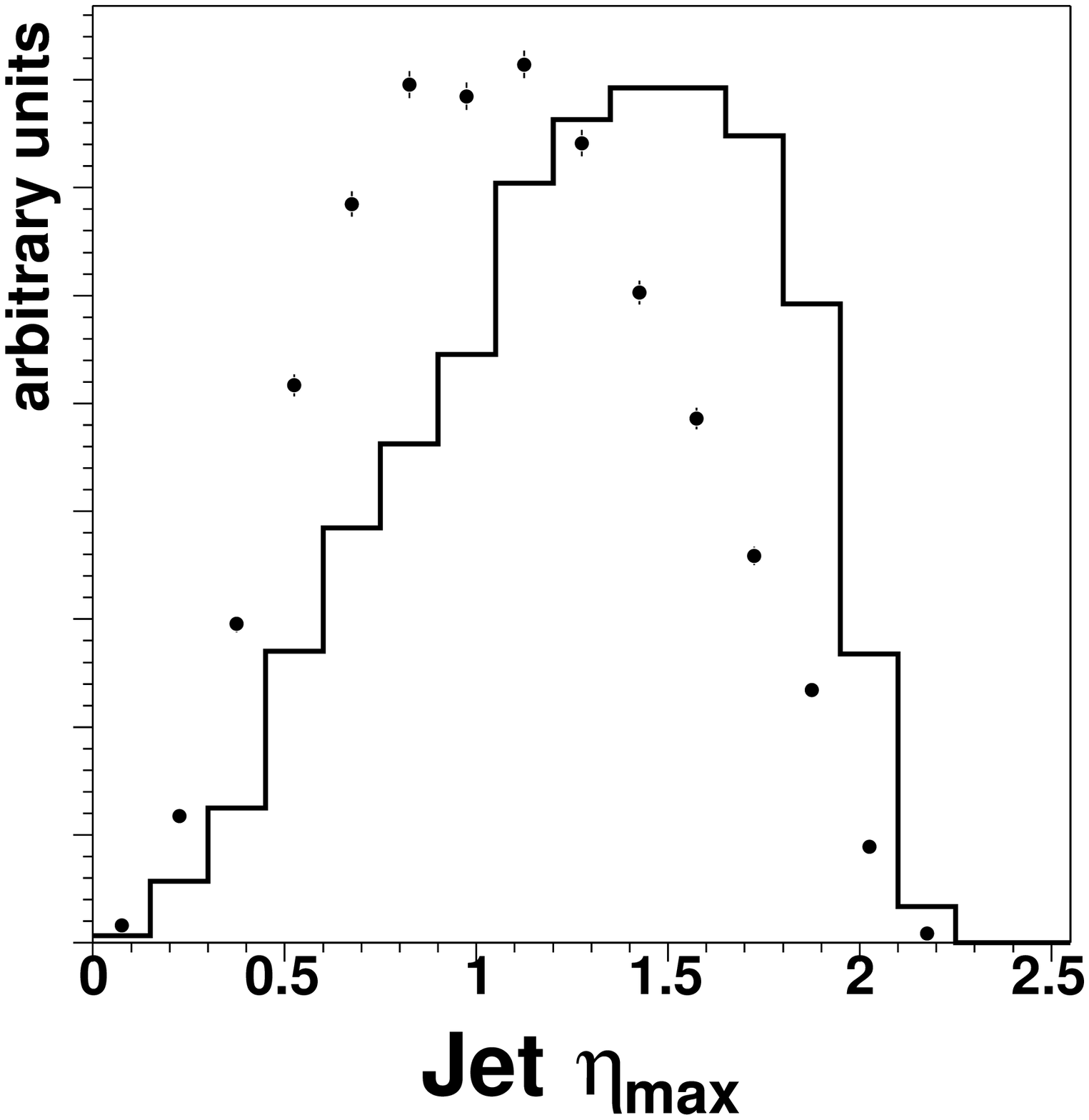}}
\resizebox{1.7in}{!}{ \includegraphics{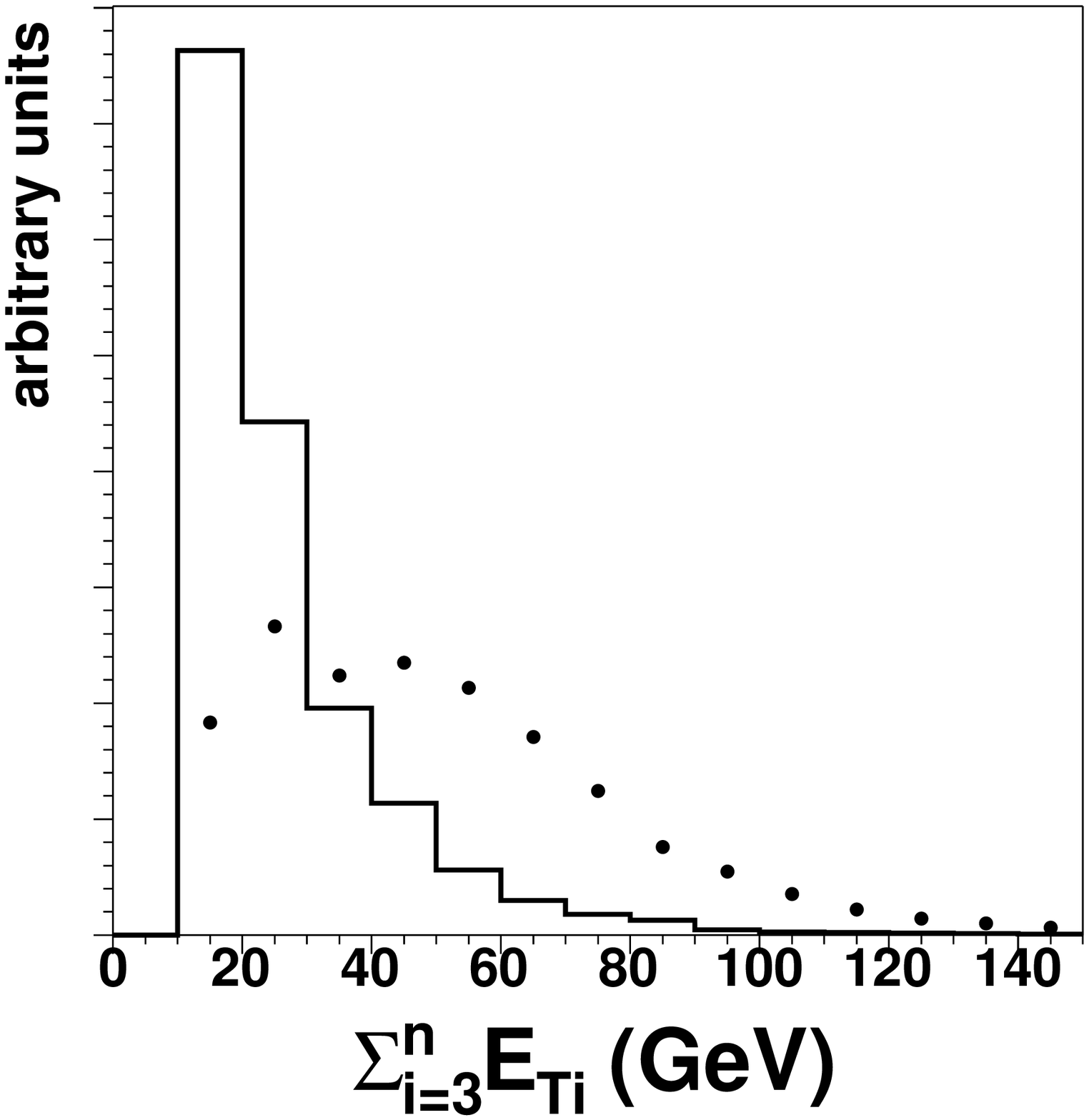}}
\resizebox{1.7in}{!}{ \includegraphics{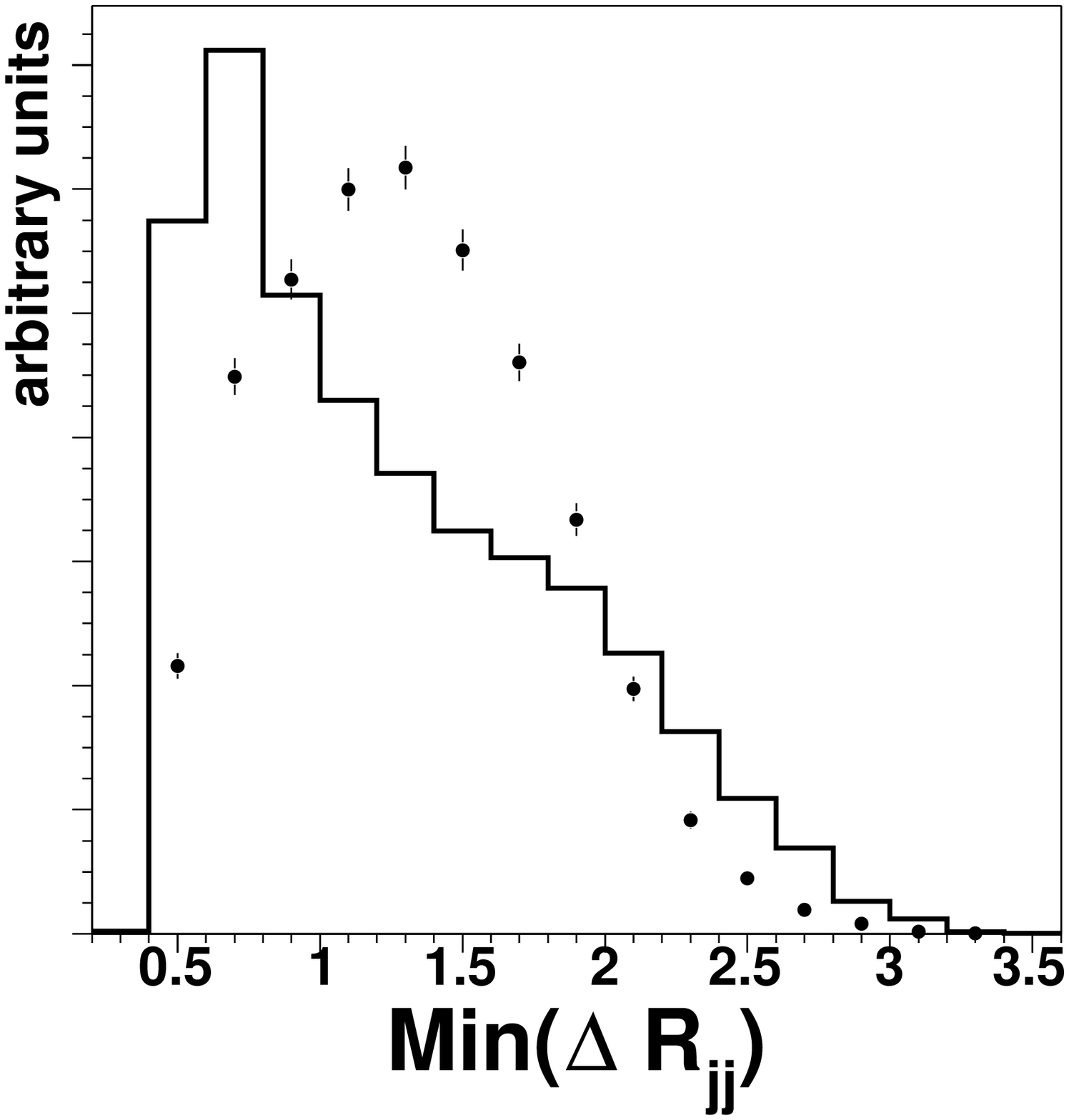}}
\resizebox{1.7in}{!}{ \includegraphics{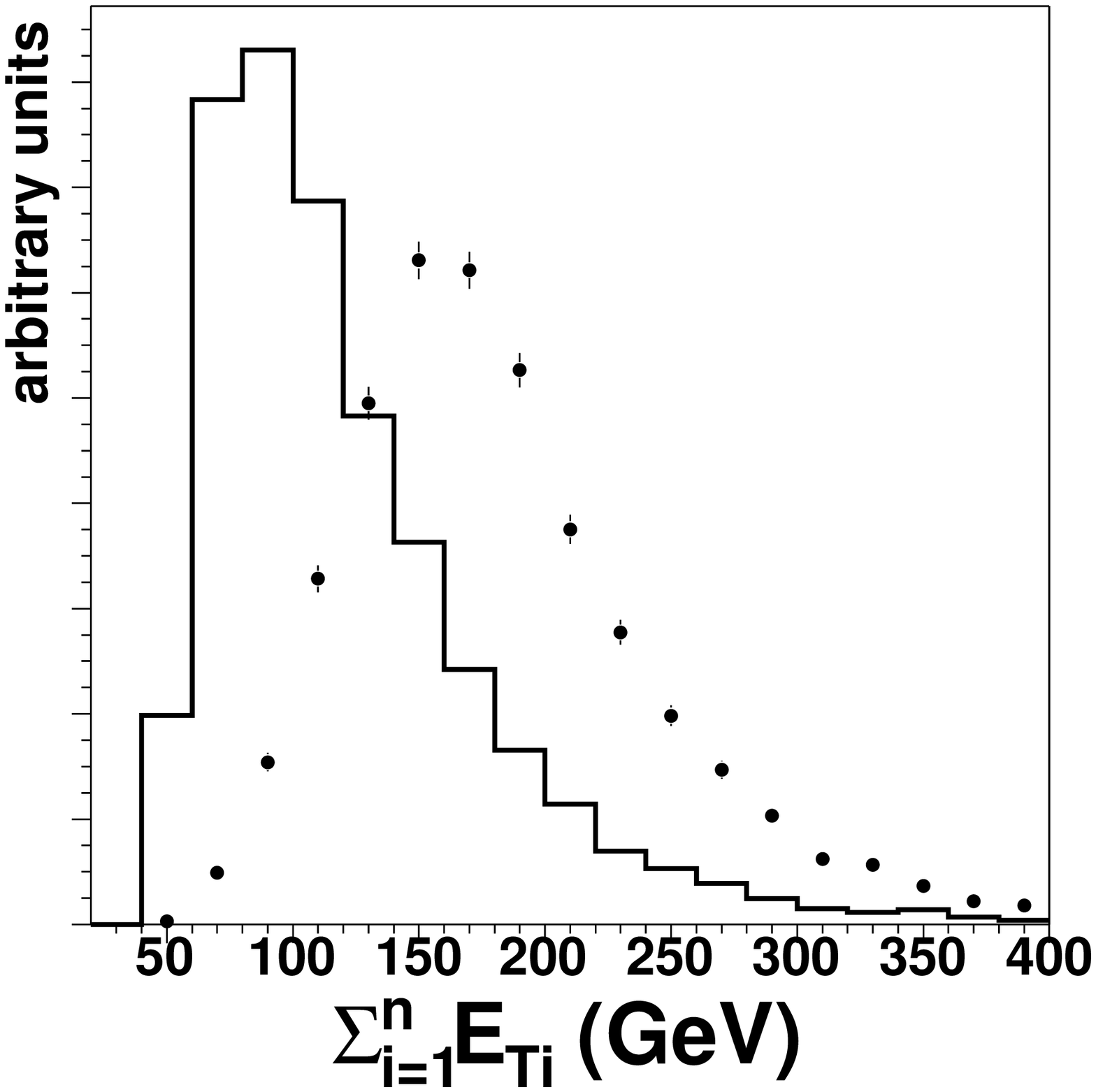}}
\resizebox{1.7in}{!}{ \includegraphics{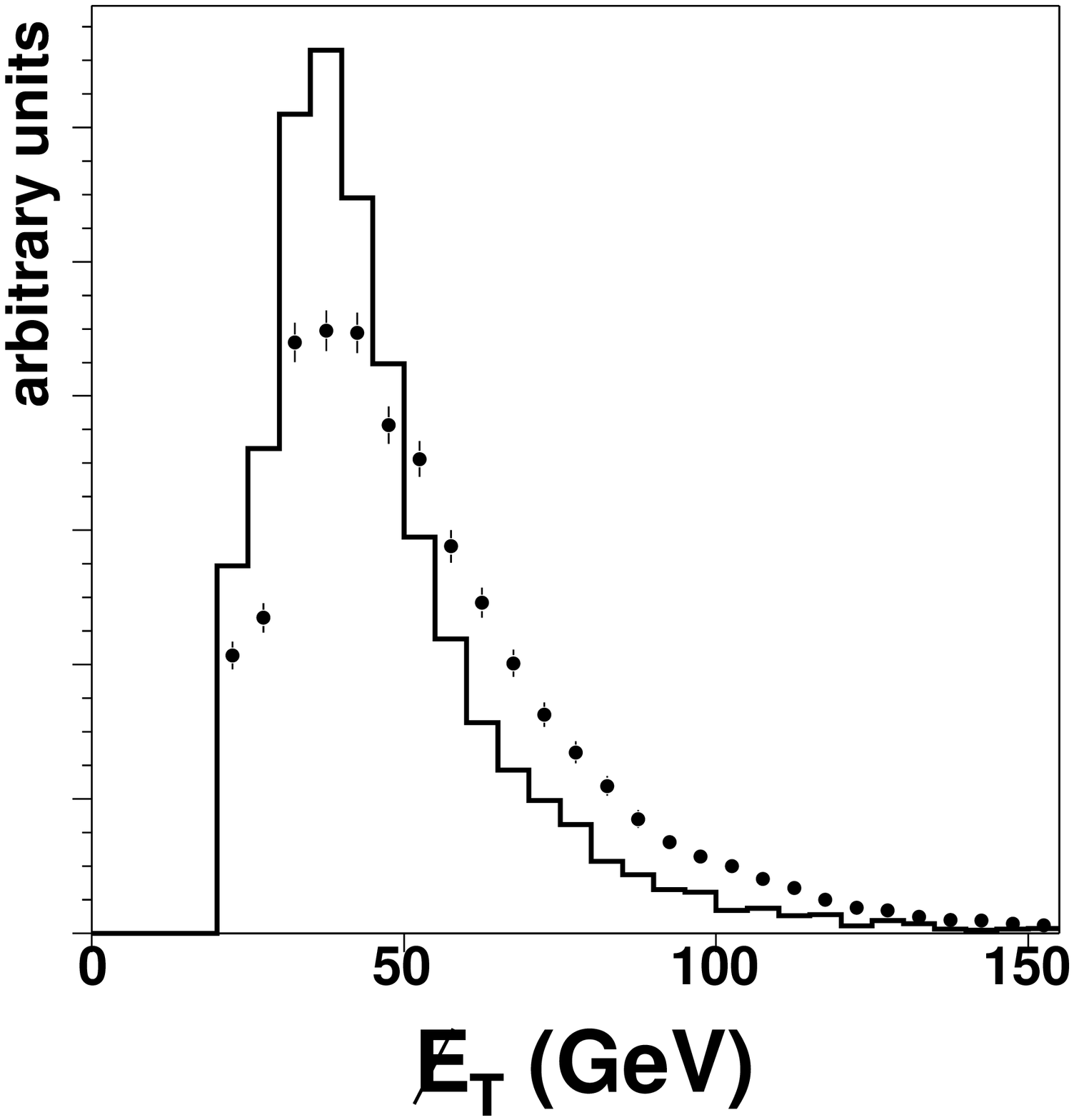}}
\resizebox{1.7in}{!}{ \includegraphics{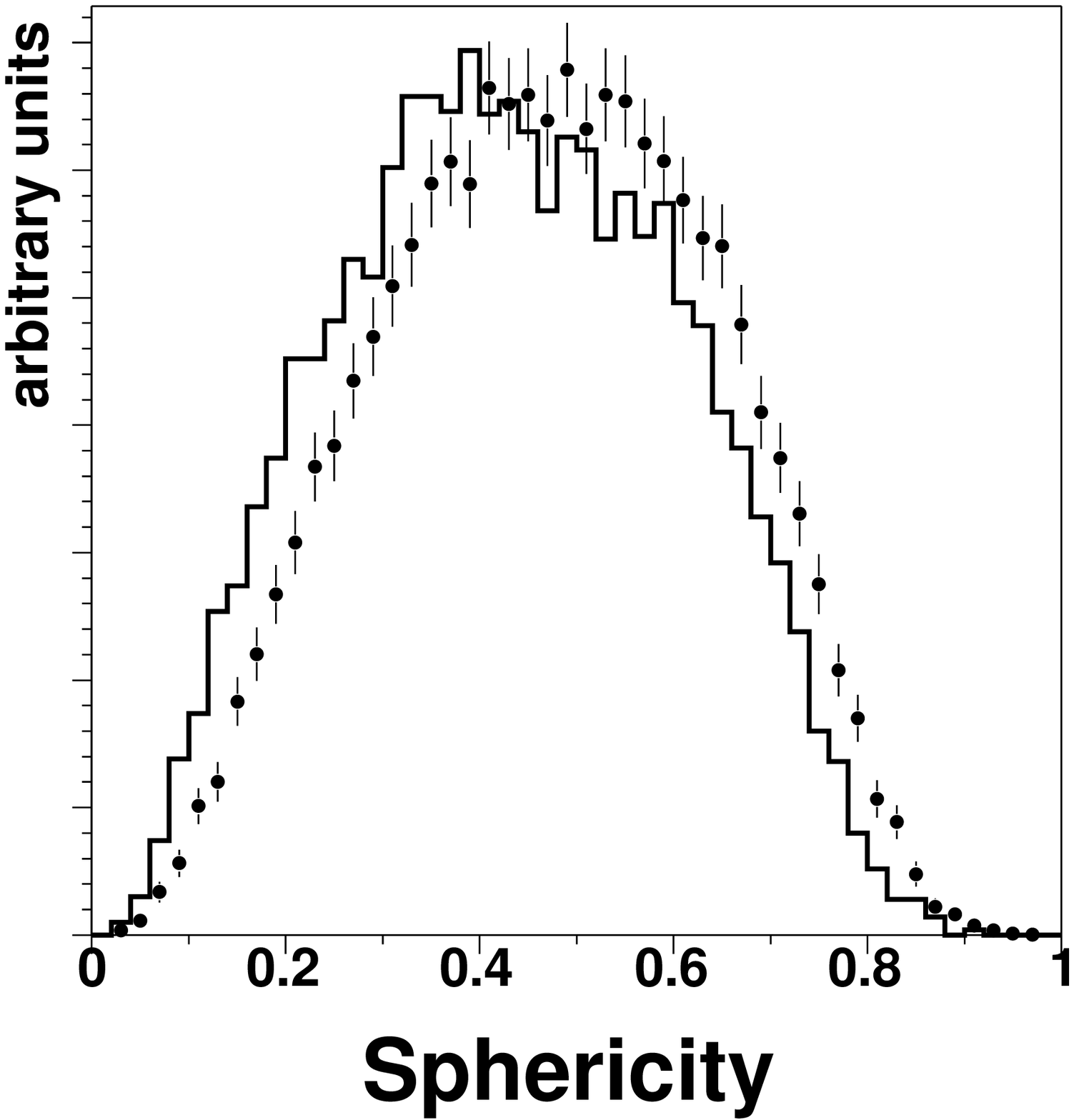}}
\resizebox{1.7in}{!}{ \includegraphics{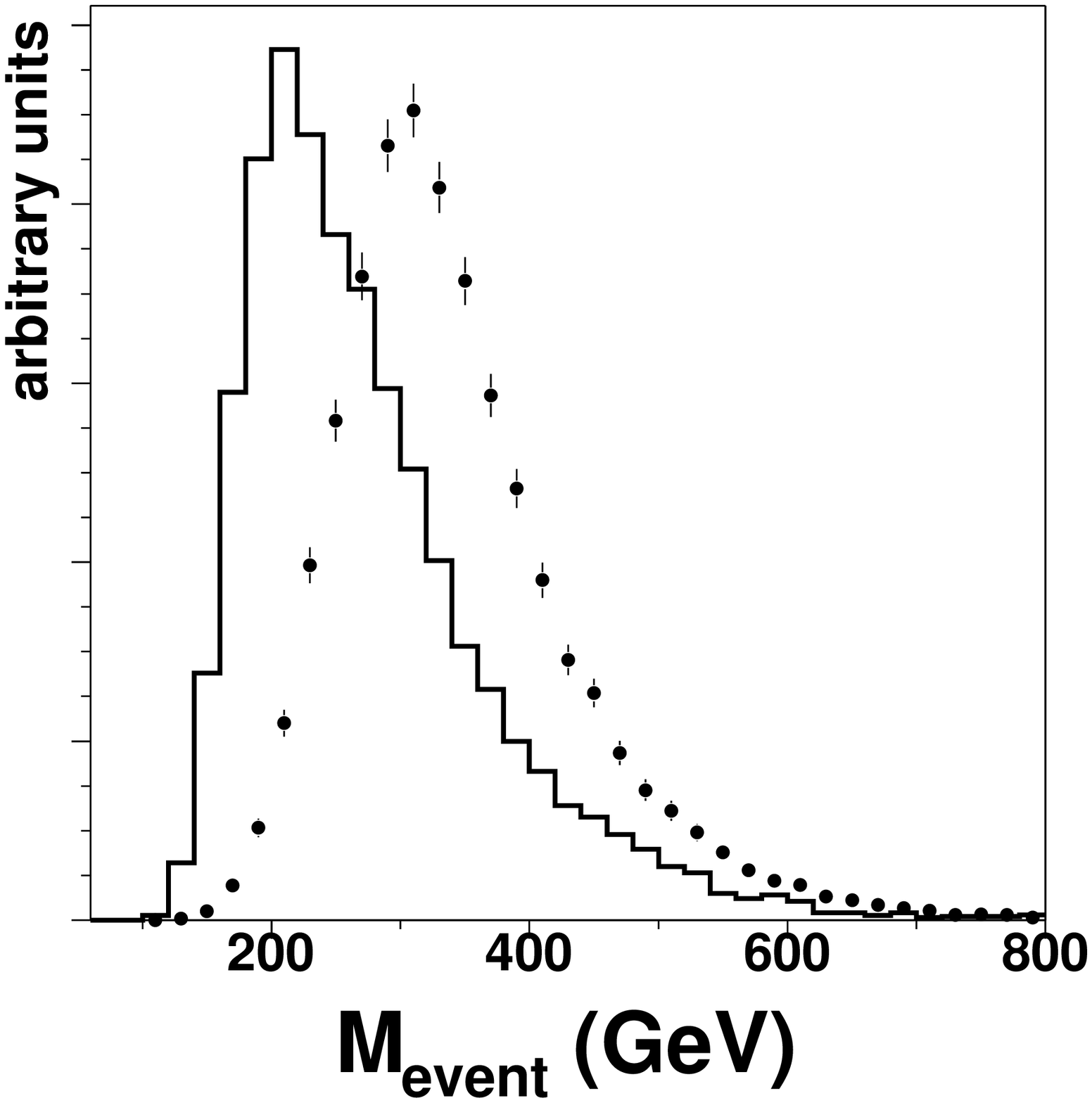}}
\resizebox{1.7in}{!}{ \includegraphics{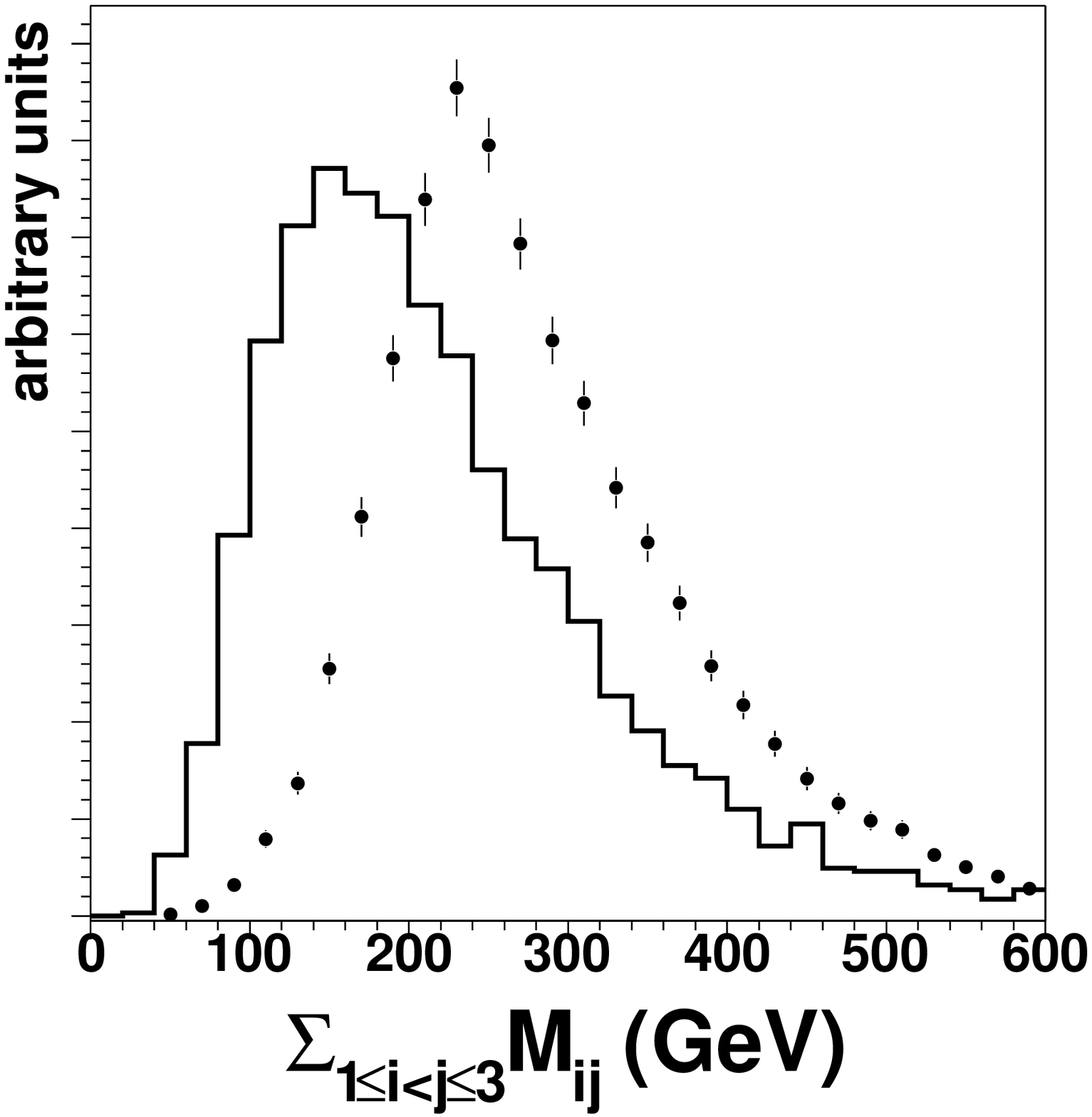}}
\resizebox{1.7in}{!}{ \includegraphics{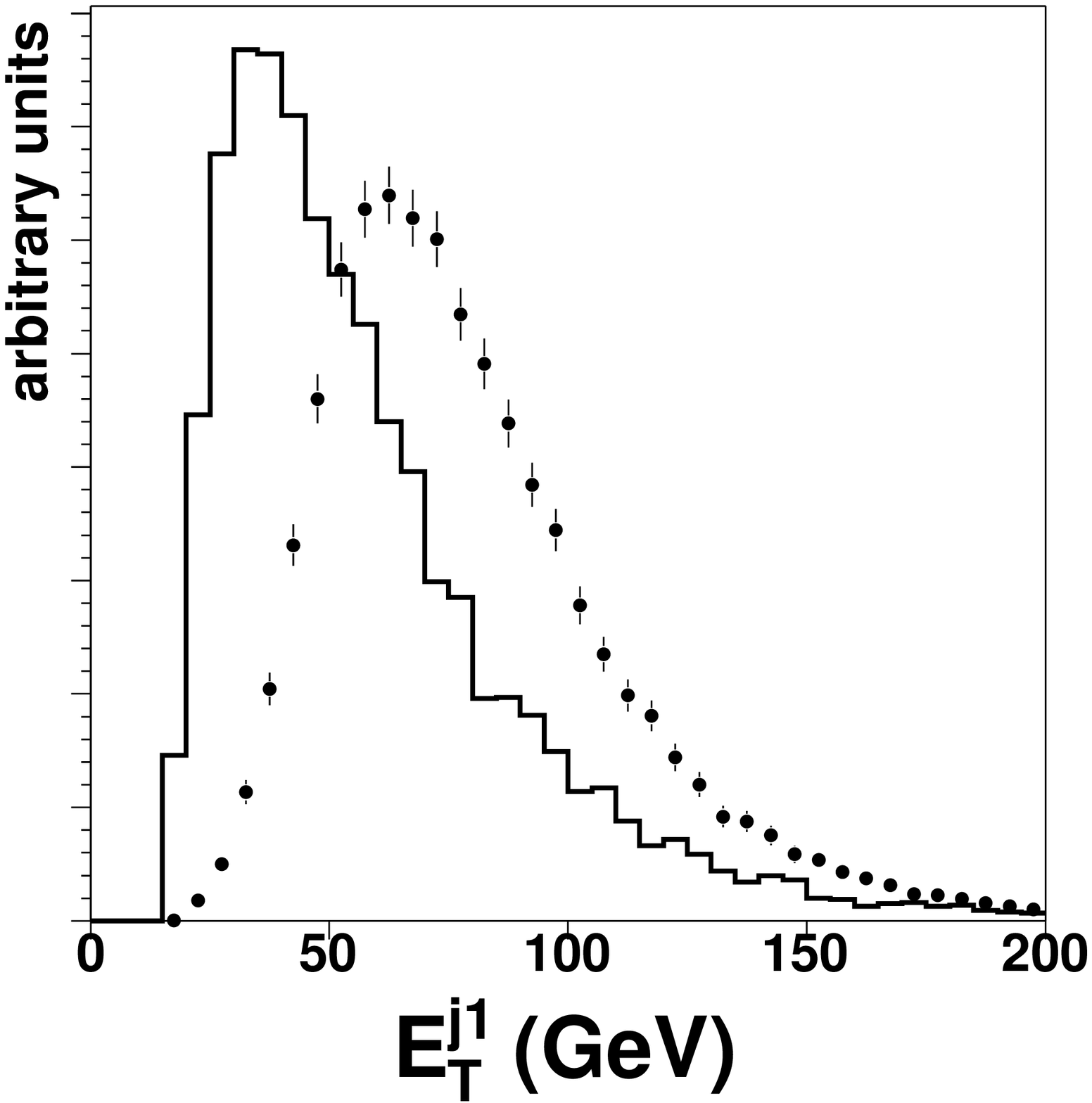}}
\resizebox{1.7in}{!}{ \includegraphics{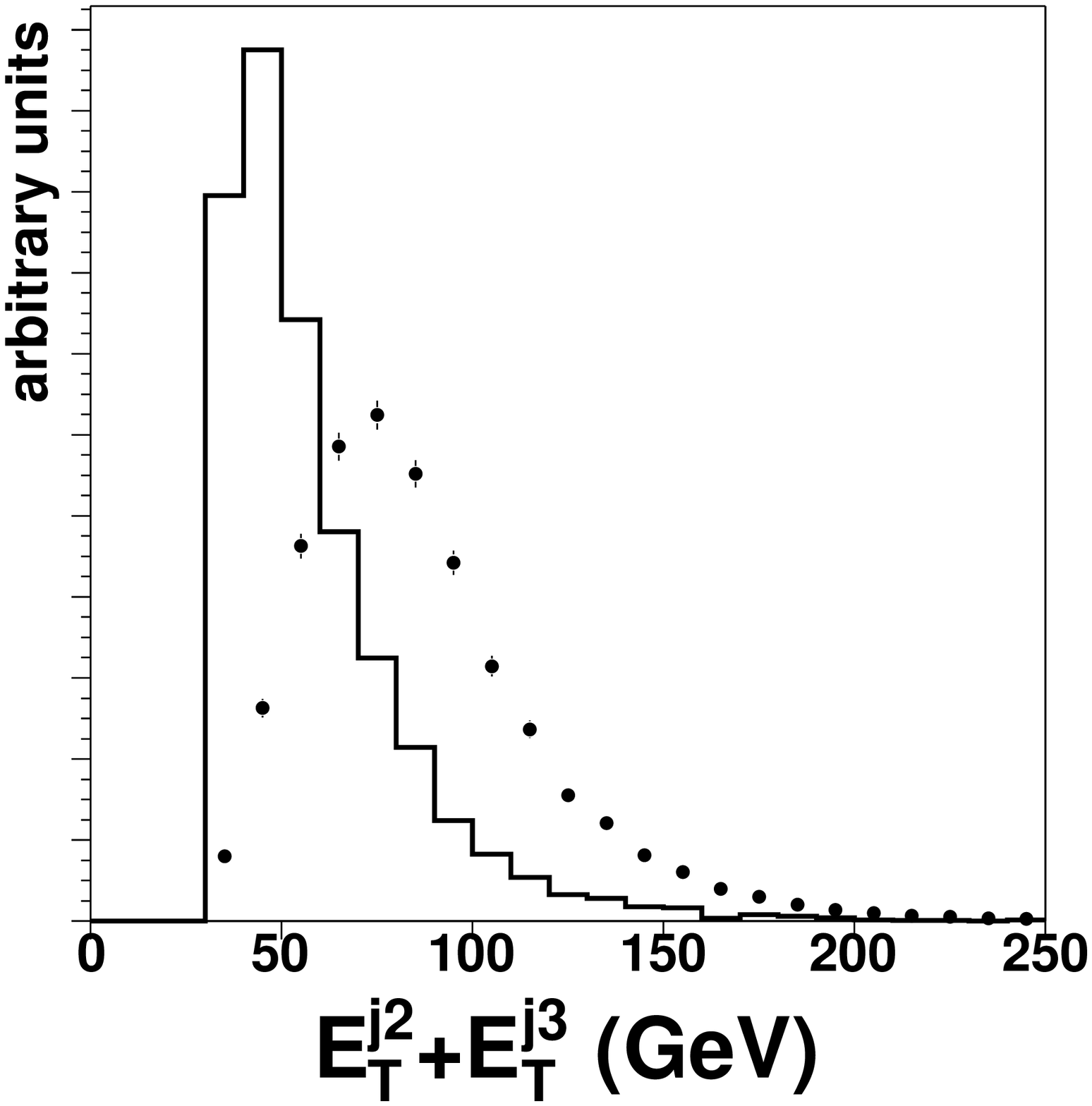}}
\resizebox{1.7in}{!}{ \includegraphics{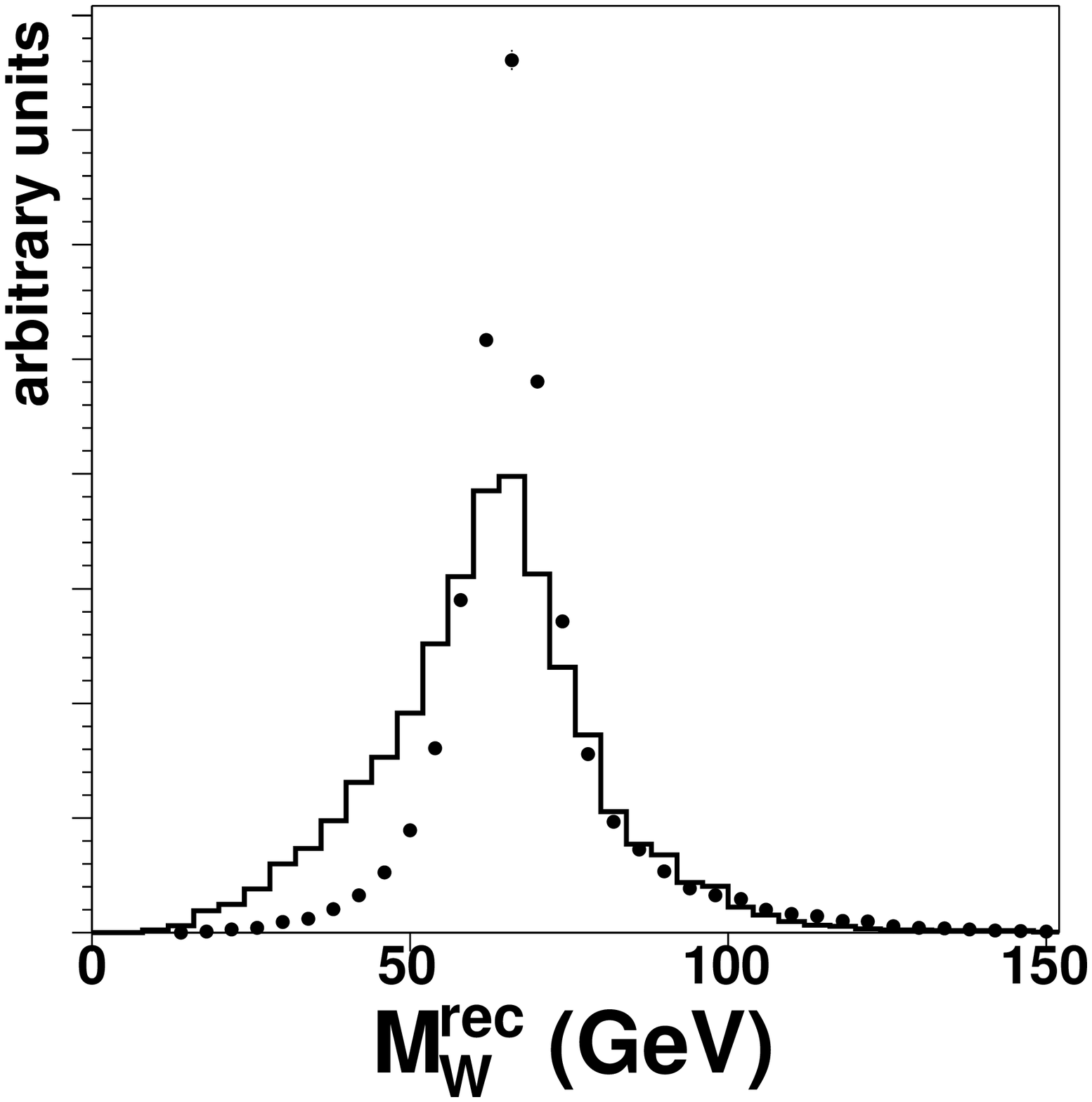}}
\resizebox{1.7in}{!}{ \includegraphics{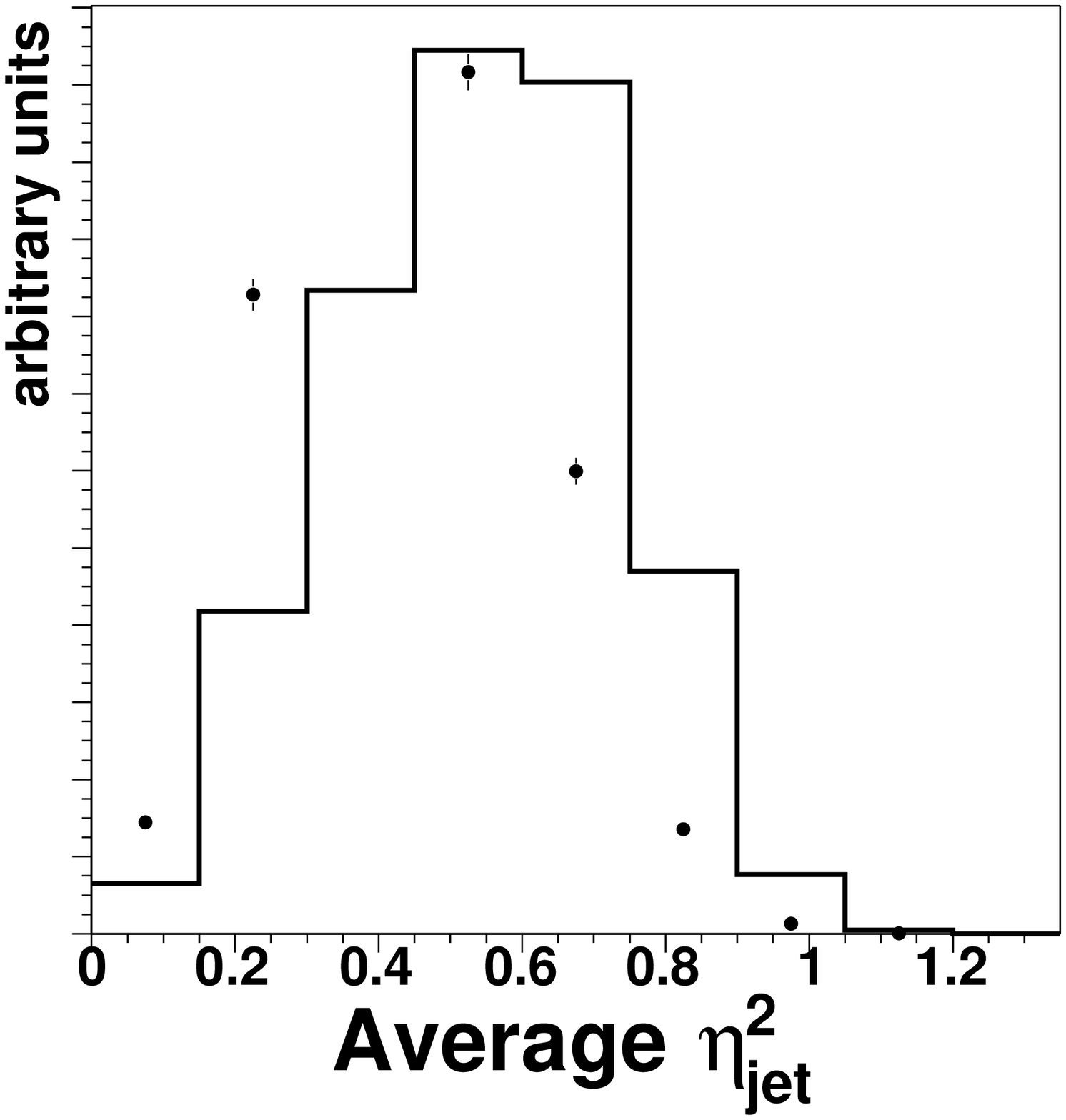}}
\resizebox{1.7in}{!}{ \includegraphics{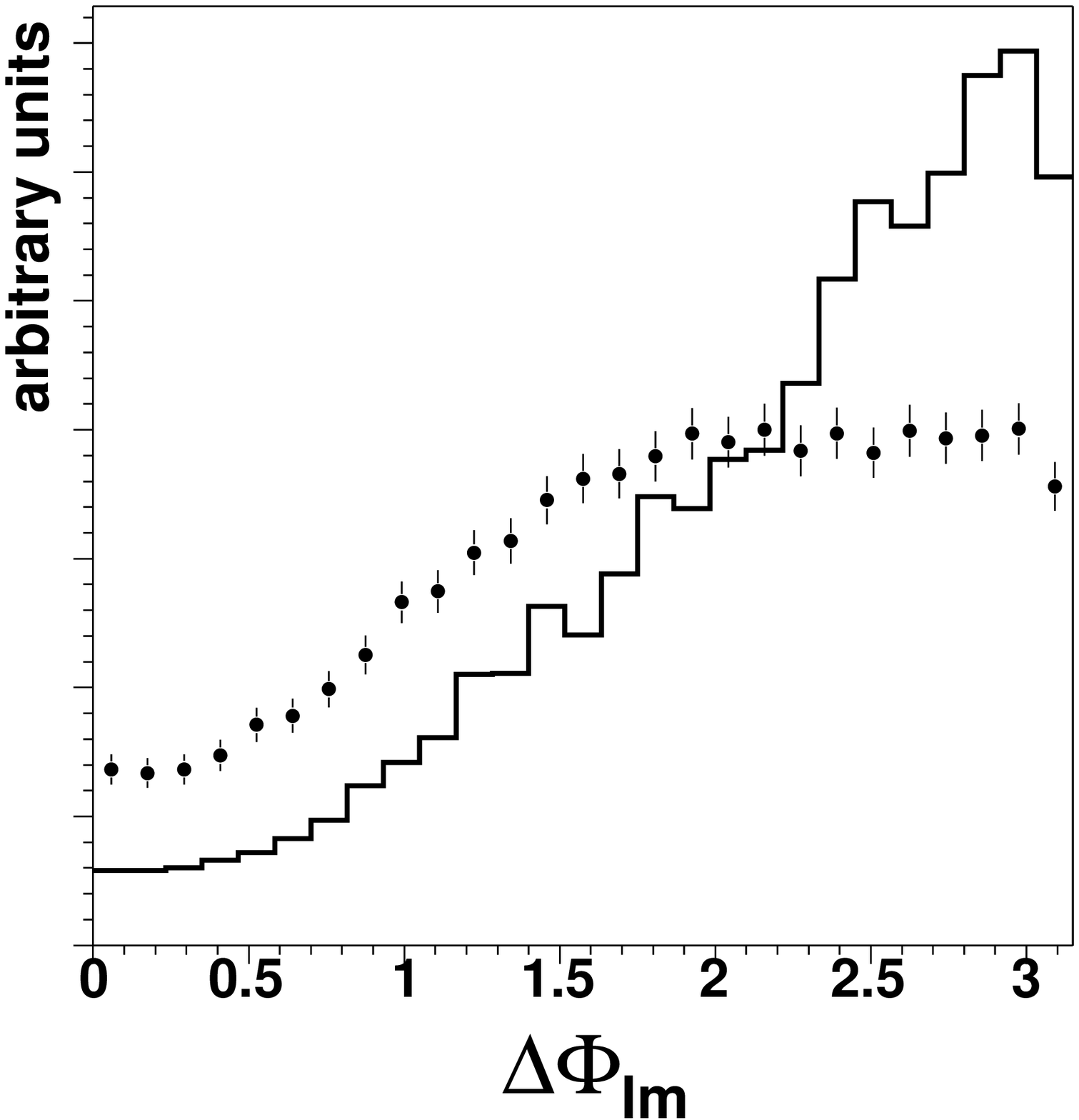}}
\resizebox{1.7in}{!}{ \includegraphics{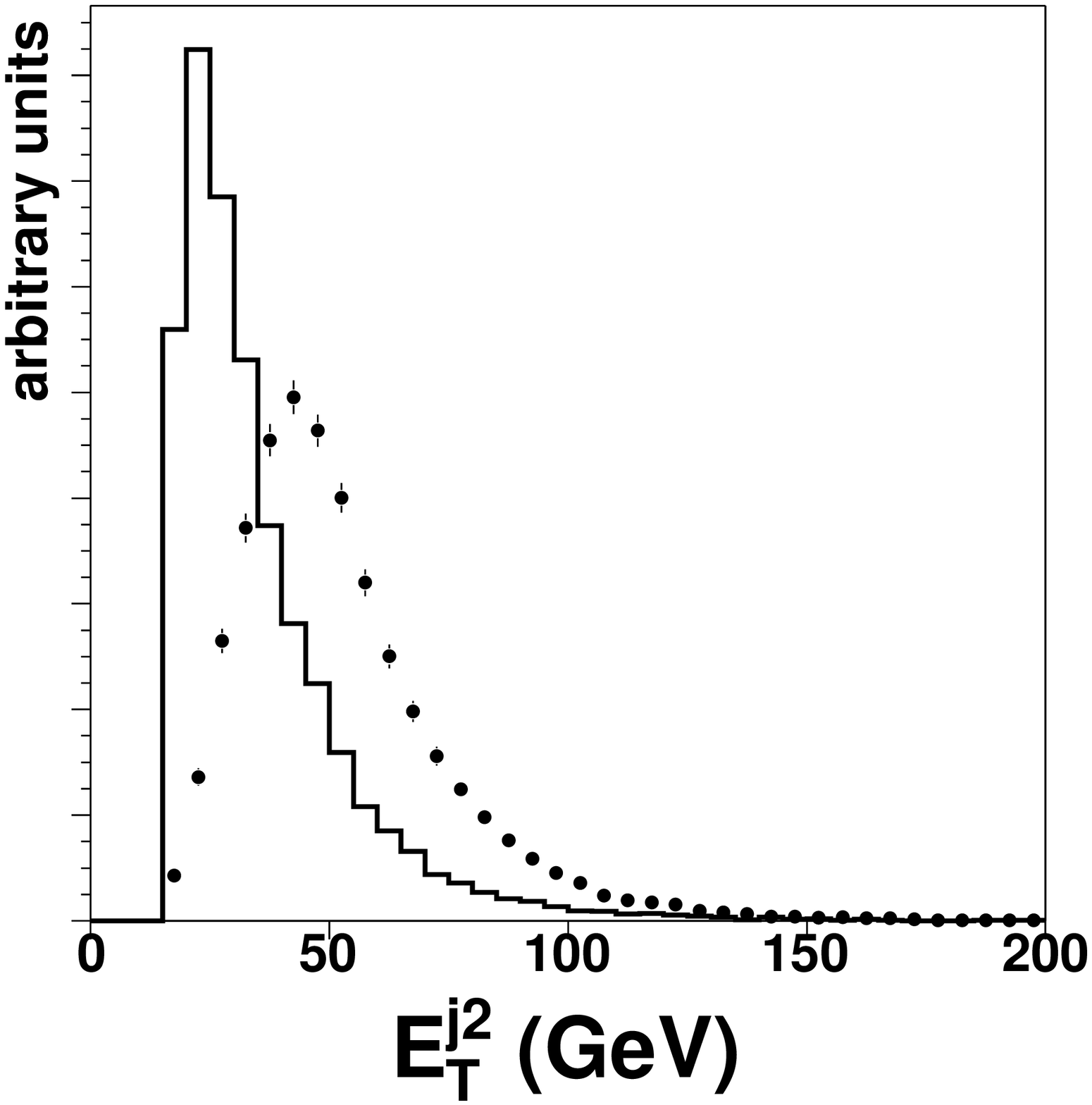}}
\resizebox{1.7in}{!}{ \includegraphics{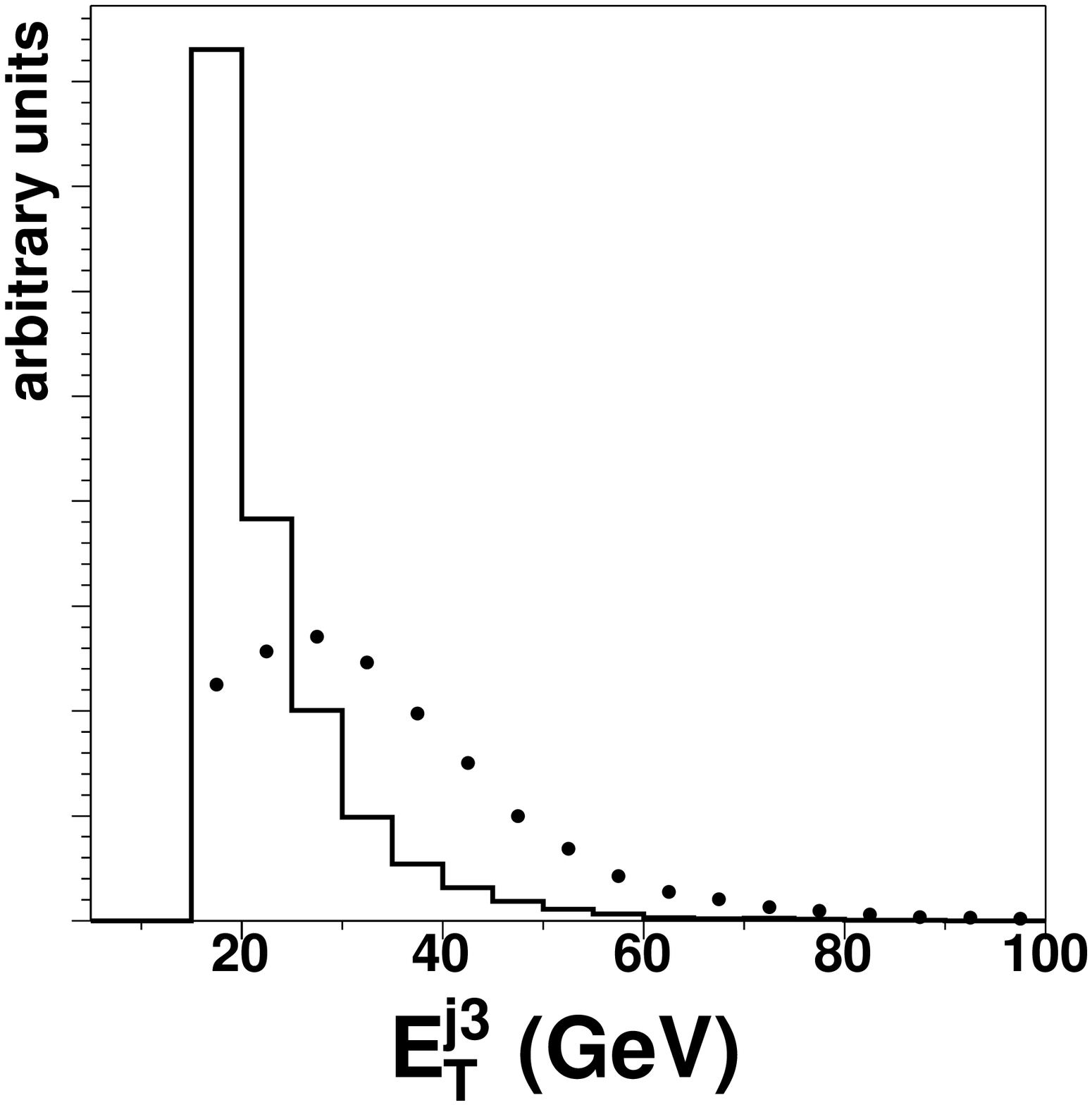}}
\resizebox{1.7in}{!}{ \includegraphics{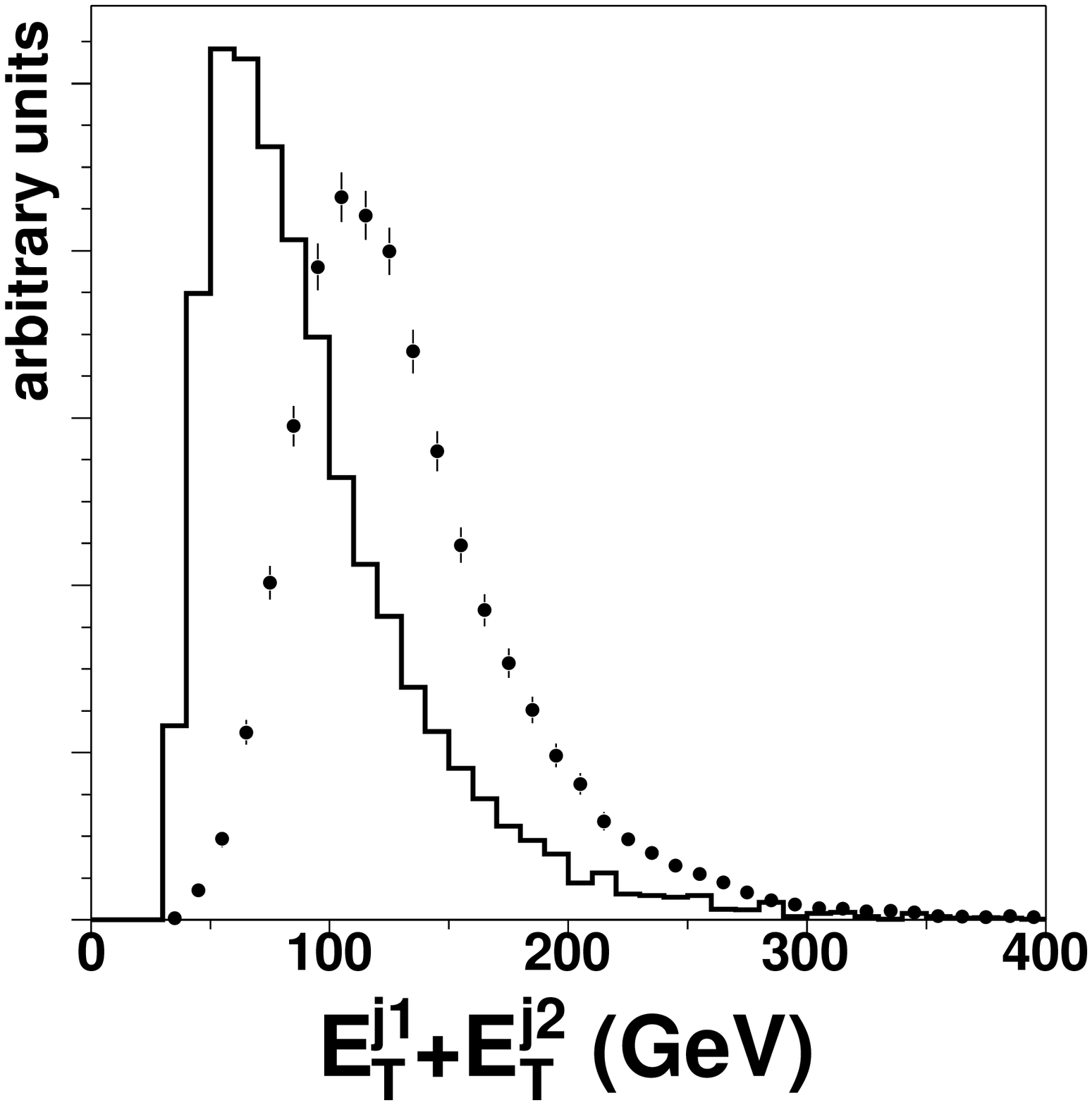}}

  \end{center}
  \caption{\label{fig:top_versus_W3p} Shape comparison of PYTHIA \ttbar\ to ALPGEN +HERWIG $W$+3p\ 
Monte Carlo simulation for the twenty kinematic and topological properties considered for 
 the $W+\geq$3\ jets sample. The distributions are normalized to equal area.
}
\end{figure*}

We consider here a set of twenty properties, defined in Table~\ref{tab:vardef}, that provide good discrimination between
signal and background. Fig.~\ref{fig:top_versus_W3p} compares the distributions 
from PYTHIA \ttbar\ and ALPGEN+HERWIG $W+3$\ parton Monte Carlo for each property. In the calculation of aplanarity 
and sphericity, we calculate the eigenvalues $Q_{i}$\ of the normalized momentum tensor of the event, defined  as 
$\frac{\sum_{i}{p_{i}^{a}p_{i}^{b}}}{{\sum_{i}{p_{i}^{2}}}}$\
where the $a,b$ indices run over the three spatial directions and the summation 
is taken over the five highest \et\ jets, the lepton and the missing transverse energy.  
The variable $M_{W}^{\textrm{rec}}$\ is intended to reconstruct the
invariant mass of the jets from the $W\to jj$\ decay. As we do not correct jets back to parton level, 
our simulation predicts that jets from the $W$\ decay will have an invariant mass close to 66~\gevcii.  
Therefore, we pick the invariant mass of the two jets amongst the three highest \et\ jets that is closest to this value.

\begin{table*}[!btp]
\caption{\label{tab:vardef}
The definition for all the kinematic and topological properties considered in this analysis.
 }
\begin{ruledtabular}
\begin{tabular}{ll}
Property  & Definition \\
\hline
\sht 				& Scalar sum of transverse energies of jets, lepton and \met \\
Aplanarity 			& $3/2 Q_{1}$\\
$\sum p_{z}/ \sum E_{T}$ 	& Ratio of total jet longitudinal momenta to total jet transverse energy\\
min($M_{jj}$)			& Minimum di-jet invariant mass of three highest \et\ jets\\
$\eta_{max}$			& Maximum $\eta$ of three highest \et\ jets \\
$\sum_{i=3}^{n} E_{T,i}$	& Sum \et\ of third highest \et\ jet and any lower \et\ jets\\
min($\Delta R_{jj}$)		& Minimum di-jet separation in $\eta$\ and $\phi$ for three highest \et\ jets\\
$\sum_{i=1}^{n} E_{T,i}$	& Sum \et\ of jets \\
\met				& Missing transverse energy \\
Sphericity 			& $3/2(Q_{1}+Q_{2})$ \\
$M_{event}$			& Invariant mass of jets, lepton and \met \\
$M_{12}+M_{23}+M_{13}$	        & Sum of di-jet invariant masses of three highest \et\ jets\\
$E_{T}^{j1}$			& \et\ of jet with highest \et\\
$E_{T}^{j2}+E_{T}^{j3}$		& Sum of \et\ of jets with second and third highest \et\\
$M_{W}^{rec}$ 			& Di-jet invariant mass closest to 66.0 GeV of three highest \et\ jets\\
$\sum_{i=1}^{3} \eta_{i}^{2}$	& Sum of $\eta^{2}$\ of three highest \et\ jets \\		
$\Delta \Phi_{lm}$ 		& Azimuthal angle between lepton and \met\\
$E_{T}^{j2}$			& \et\ of jet with second highest \et\\
$E_{T}^{j3}$			& \et\ of jet with third highest \et\\
$E_{T}^{j1}+E_{T}^{j2}$		& Sum of \et\ of jets with first and second highest \et\\
\end{tabular}
\end{ruledtabular}
\end{table*}

\begin{figure}[btp]
  \begin{center}
\resizebox{3.63in}{!}{ \includegraphics{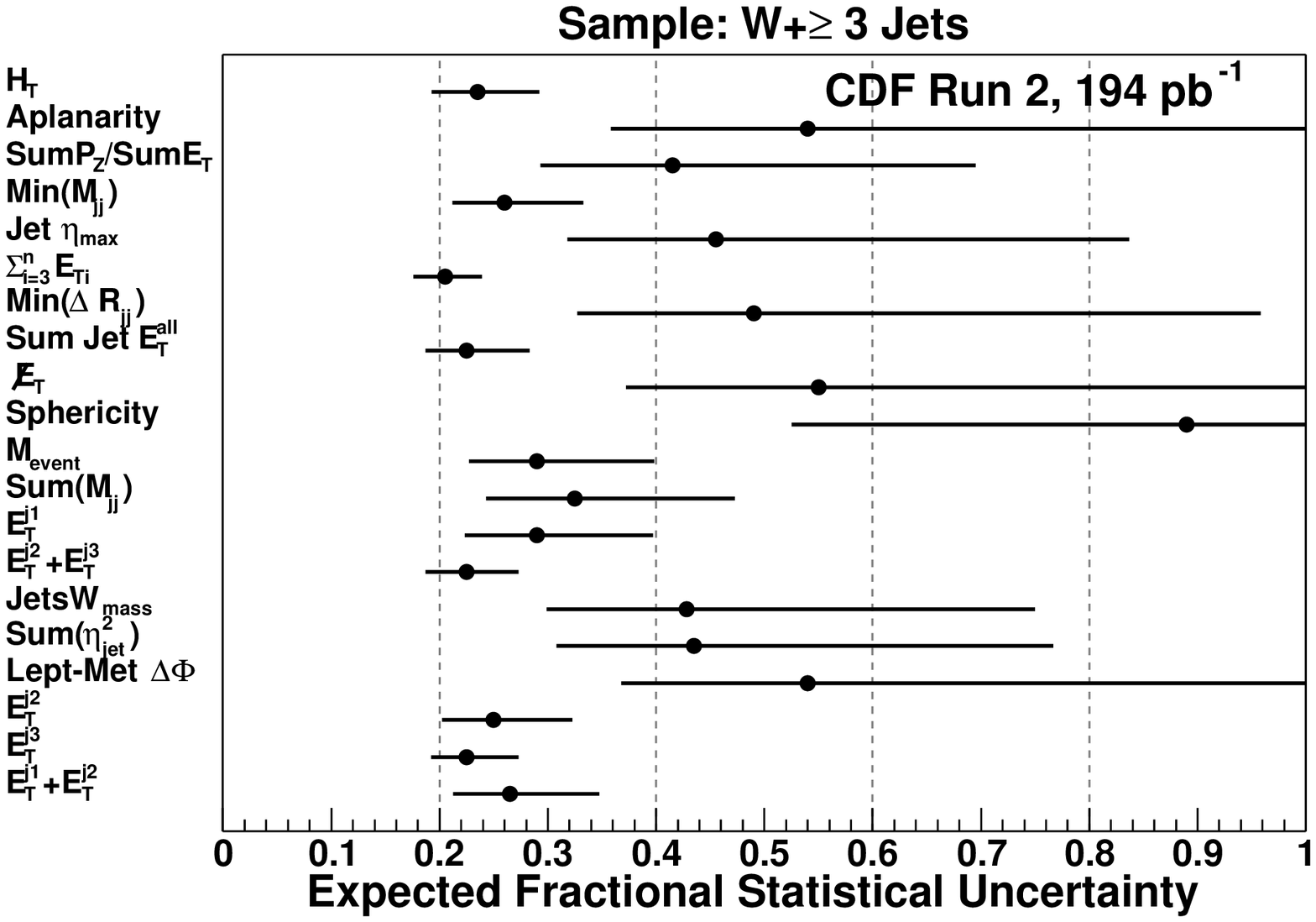}}
  \end{center}
  \caption{\label{fig:exp_error_1d}Expected statistical sensitivity of fits to each of the kinematic 
distributions for the $W+\geq$3\ jets sample. The points mark the
 median of the relative error distribution, the error bars mark the 16-84 percentile interval. }
 
  \end{figure}

\begin{figure}[btp]
  \begin{center}
\resizebox{3.63in}{!}{ \includegraphics{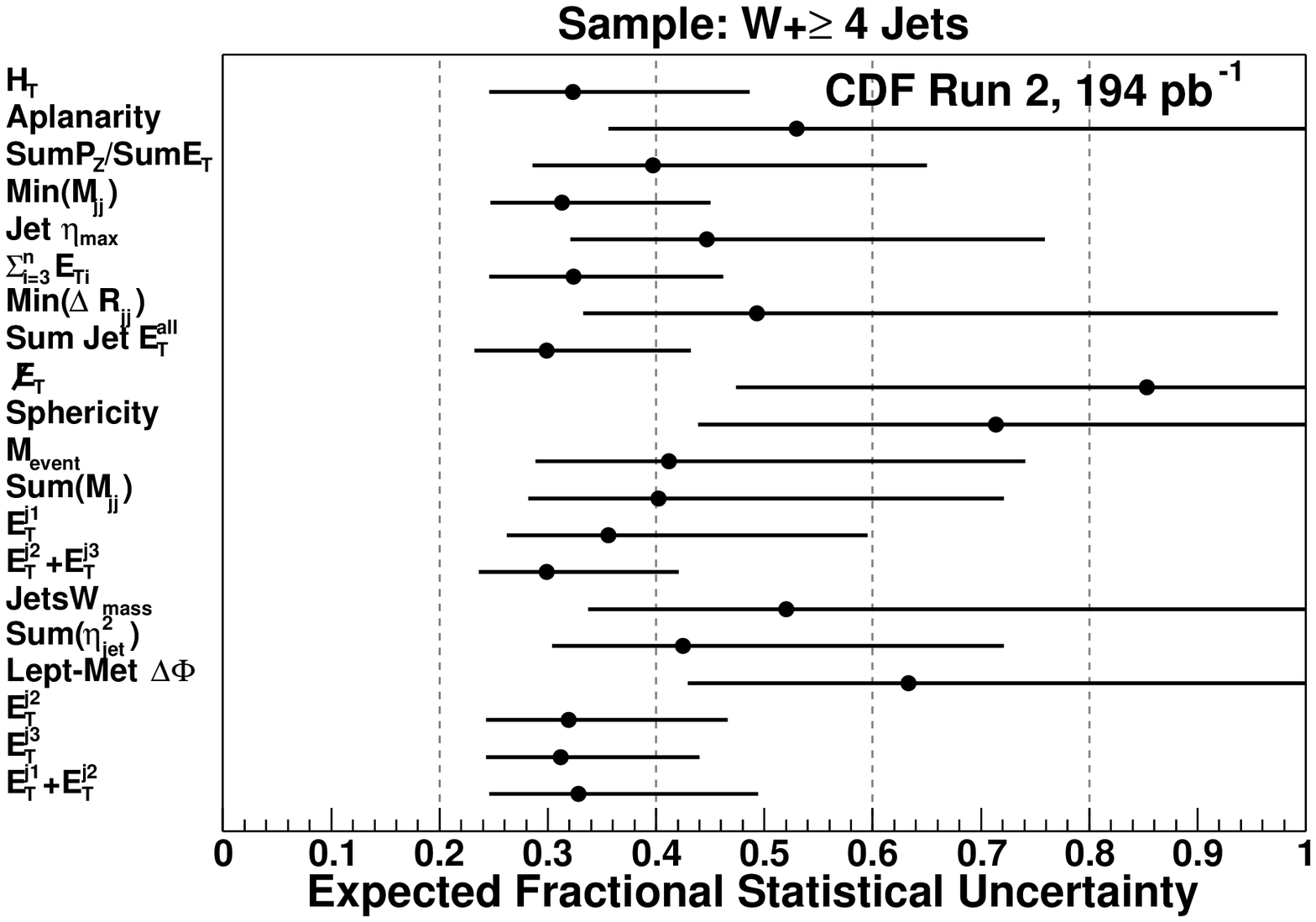}}
  \end{center}
  \caption{\label{fig:exp_error_2d} Expected statistical sensitivity of fits to each of the kinematic  
distributions for the $W+\geq$4\ jets sample.  The points mark the
 median of the relative error distribution, the errors bars mark the 16-84 percentile interval.  } 
  \end{figure}

The expected statistical sensitivity of each single property is estimated {\it a priori} by
constructing simulated experiments of the same size on average as the data sample from Table~\ref{tab:ndata}. 
Each simulated experiment contains $N_{\ttbar}$\ signal \ttbar\ events drawn from a Poisson distribution with
mean given by Table~\ref{tab:ndata}, $N_{q}$\ multi-jet background events drawn from a Poisson distribution with
mean given by Table~\ref{tab:qcdbkg}, and $N_{w}$\ W-like background events drawn from a Poisson distribution
with mean equal to the remainder.
In every simulated experiment, we perform a separate binned maximum likelihood fit for each of the twenty single properties.
The expected statistical uncertainty on the number of \ttbar\ events is shown in Fig.~\ref{fig:exp_error_1d} for all
twenty single properties in the $W+\geq$3 jets sample.  A similar sensitivity plot for the $W+\geq$4 jets sample is shown in 
Fig.~\ref{fig:exp_error_2d}. 

We choose to use the total transverse energy in the event, \sht, since it is both one of the observables
that provides good discrimination between events containing top decays and events from background processes,
and since it has been commonly used in other analyses for this purpose~\cite{ttbardilepton,svxruniipaper}.
We note that the sum of the jet transverse energies or the transverse energy of the third most energetic jet 
have similar statistical power.  From a fit to the \sht\ distribution in the $W+\geq$3 jets sample, we expect to obtain a statistical uncertainty
in the range 19-29\% for 68\% of data-sized experiments, with a median at 23.5\%. 

Although the $W+\geq$4 jets sample has an improved signal to background ratio, we find a 
larger expected statistical uncertainty in the range 25-48\% for 68\% of data-sized experiments,
with a median of 32\%.  The lower sensitivity is due to both lower statistics - 45\% of 
the \ttbar\ events fail the 4th jet requirement - and reduced discriminating power- the increased jet activity means that 
$W+\geq$4 jet events have larger \sht\ and are therefore more similar to top pair production.
Finally, we note that the systematic uncertainty, discussed in Section~\ref{sec:sys}, is also about 20\% larger, in part due to the increased sensitivity of the 
selection to the jet energy scale.


\subsection{\label{subsec:nnvar} Artificial Neural Network}

The ANN that we develop is a feed-forward network \cite{perceptron} with one intermediate (hidden) layer and one output node.
Training of the network is performed with 4000 PYTHIA \ttbar\ and 4000 $W$+3 parton ALPGEN+HERWIG  Monte Carlo events
that pass the selection requirements. During the iterative training, the weights of the network are adjusted in order to 
minimize a mean squared error function~\cite{cata15}: 
  
 \begin{displaymath}
 E=\frac{1}{N}\sum_{i=1}^{N}{(O_{i}-t_{i})^{2}}
 \end{displaymath}  
 
\noindent where $N$ is the number of events in the training sample, $O_{i}$ is the output of the
network and $t_{i}$ is the desired target value for the $i$-th event. We choose a target value
of 1.0 for signal events and 0.0 for background events.  
We use the back-propagation training method from the JETNET~\cite{jetnet} software package, with 
a pruning option turned on that has the effect of adding a regularization term to 
the error function in order to discourage unnecessary weights. 
The iterative training is halted at the point where the error function has the lowest value on an independent
sample of Monte Carlo simulated events.  This protects the ANN from effects due to statistical fluctuations in the
training sample.
%After we complete training, another statistically independent set of simulated events is used to check the quality of the training.

\begin{figure}[htbp]
  \begin{center}
 \resizebox{3.5in}{!}{ \includegraphics{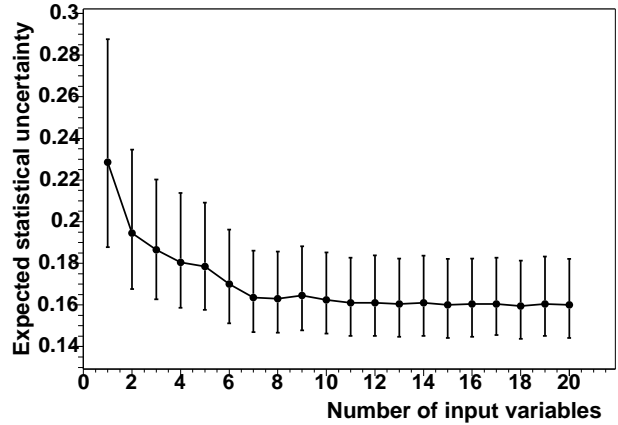}}
 \end{center}
\caption{\label{fig:stat_error_ann} Expected statistical sensitivity for ANNs with 
the number of inputs ranging from 1-20 for the $W+\geq$3\ jets sample.  
The points represent the median in the relative error distribution,
 error bars mark the 16-84 percentile interval.
  } 
  \end{figure}

\begin{figure}[htbp]
  \begin{center} 
\resizebox{3.5in}{!}{\includegraphics{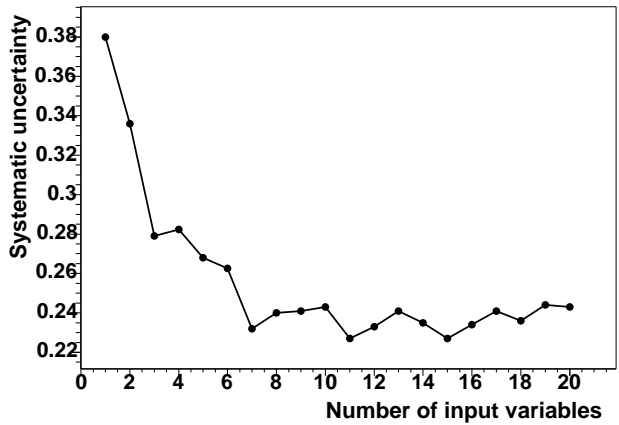}}
  \end{center}
  \caption{\label{fig:sys_error_ann} Average systematic uncertainties for
 ANNs with the number of inputs ranging from 1-20 for the $W+\geq$3\ jets sample.  
  } 
  \end{figure}

For inputs to the ANN, we consider many different combinations of twenty kinematic and topological properties 
described in Table~\ref{tab:vardef}.  
The performance of each artificial neural network is tested {\it a priori} by constructing simulated
experiments as before, where now we simply treat the output of the ANN as a single discriminant.  
We show that the addition of more inputs to the ANN reduces the expected statistical uncertainty in Fig.~\ref{fig:stat_error_ann}
and the average systematic uncertainty, described in Section~\ref{sec:sys}, in Fig.~\ref{fig:sys_error_ann}.  
In either case, there is little gain beyond seven inputs.  
For each increment in the number of inputs, one extra property is added in the order given in Table~\ref{tab:vardef}. 
The network with one input uses the kinematic property $H_{T}$.
We note that this order is somewhat arbitrary, as there are other combinations that would give similar performance at each stage.

Although simplicity may not be a stringent requirement~\cite{lessons}, 
we choose a seven input network as the minimal configuration yielding good performance. 
The properties chosen are the first seven listed in  Table~\ref{tab:vardef}: (1) the total transverse energy in
the event \sht, (2) the event aplanarity, (3) the ratio between total jet longitudinal 
momenta and the total jet transverse energy, (4) the minimum di-jet invariant mass of the three highest \et\ jets, 
(5) the maximum jet rapidity of the three highest \et\ jets, 
(6) the sum of transverse energy of the third highest \et\ jet and any other lower \et\ jets, and (7) the minimum di-jet separation. 
For these seven input properties, we compare the average statistical and systematic uncertainties for ANNs with 1 to 10 nodes 
in the hidden layer.  We choose a 7-7-1 ANN configuration, which consists of seven input properties, 
seven hidden nodes and one output unit.  We expect to obtain a statistical uncertainty
in the range 15-19\% for 68\% of data-sized experiments, with a median at 16.5\%. 
This is a relative improvement of 30\% with respect to the \sht\ distribution alone. 

For the $W+\geq$4 jet sample, which has higher signal to background ratio but lower signal acceptance, 
we train a second 7-7-1 ANN with the same seven input properties.  We use $W$+4p ALPGEN+HERWIG Monte Carlo 
to model the kinematics of the background. We find a 
larger expected statistical uncertainty, in the range 19-28\% for 68\% of data-sized experiments,
with a median of 23\%.  The lower sensitivity is due to both lower statistics - 45\% of 
the \ttbar\ events fail the 4th jet requirement - and reduced discriminating power- the increased jet activity means that 
$W+\geq$4 jet events are topologically and kinematically more similar to top pair production. Even so,
we note that the sensitivity here is comparable to that from the single \sht\ distribution in the $W+\geq$3 jet sample.

Finally, in order to check that our fit procedure is unbiased, we constructed simulated experiments 
with input \ttbar\ signal cross sections ranging from 1~pb to 12~pb.  In all cases, we find 
that the average measured \ttbar\ cross section using the 7-7-1 ANN is consistent 
with the input \ttbar\ cross section.

%\begin{figure}[!htbp]
%  \begin{center} 
%\resizebox{3.2in}{!}{ \includegraphics{fig5.eps}}
%  \end{center}
%  \caption{\label{fig:exp_lin_curve} Measured \ttbar\ cross section using the 7-7-1 ANN versus input \ttbar\ cross section
%for simulated experiments.  Points mark the average cross section, the error bars mark the 16-84 percentile interval, and
%the line indicates a gradient of unity.
%  } 
%  \end{figure}

%------------------------------------------------------------------
\section{\label{sec:mcmodel} Check of Monte Carlo modeling}

The method described in the previous section relies on the accurate
modeling of kinematic and topological quantities by Monte Carlo generators and on
the accurate description of the detector response by the simulation of the CDF detector.
We compare kinematic and topological properties of
the mutually exclusive $W$+1 jet, $W$+2 jet, and $W$+3 jet samples with our model. 
A Kolmogorov-Smirnov (KS) statistic is used to quantify the 
quality of the agreement.

\begin{figure*}[btp]
  \begin{center}
\resizebox{2.32in}{!}{ \includegraphics{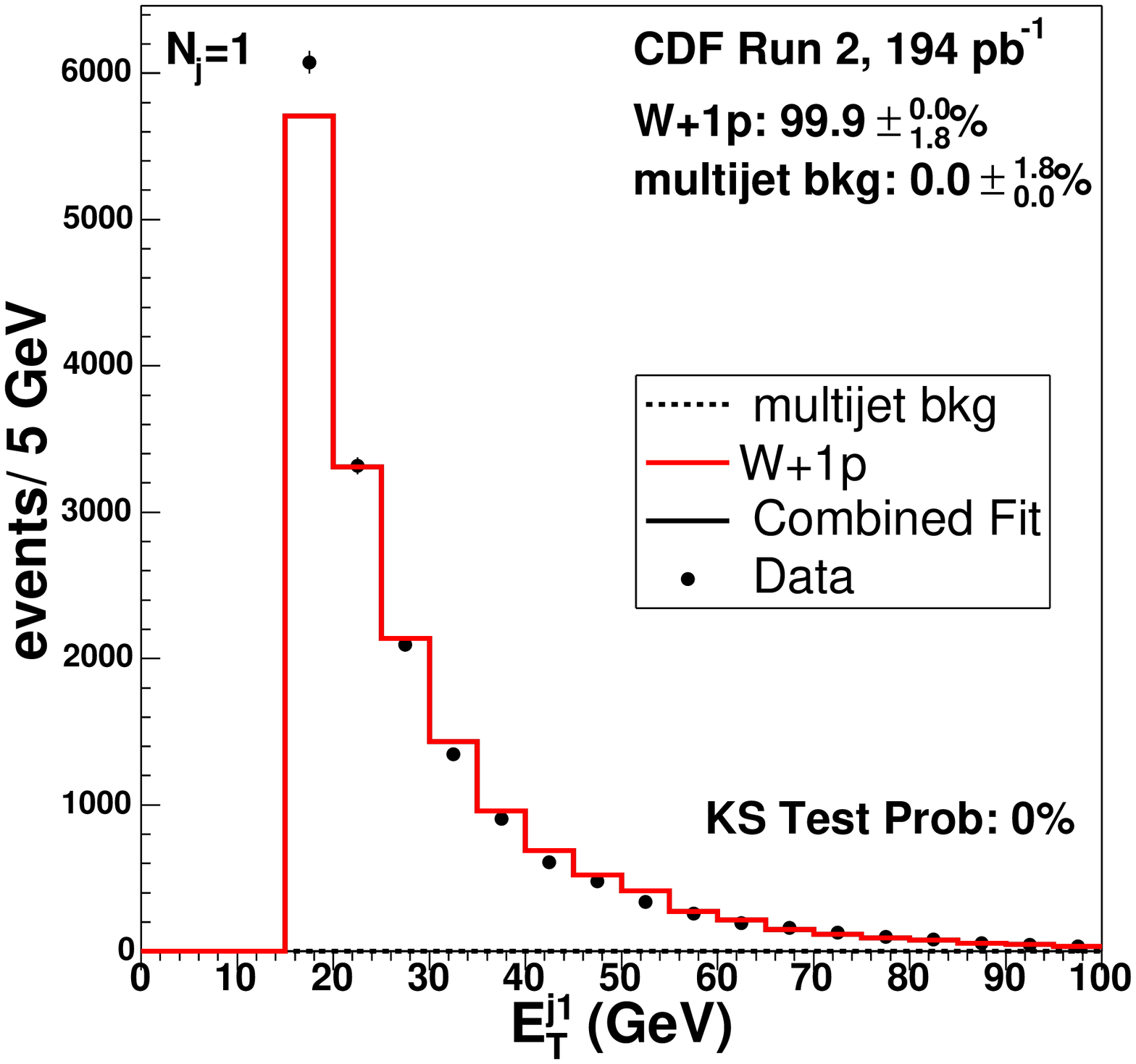}}
\resizebox{2.32in}{!}{ \includegraphics{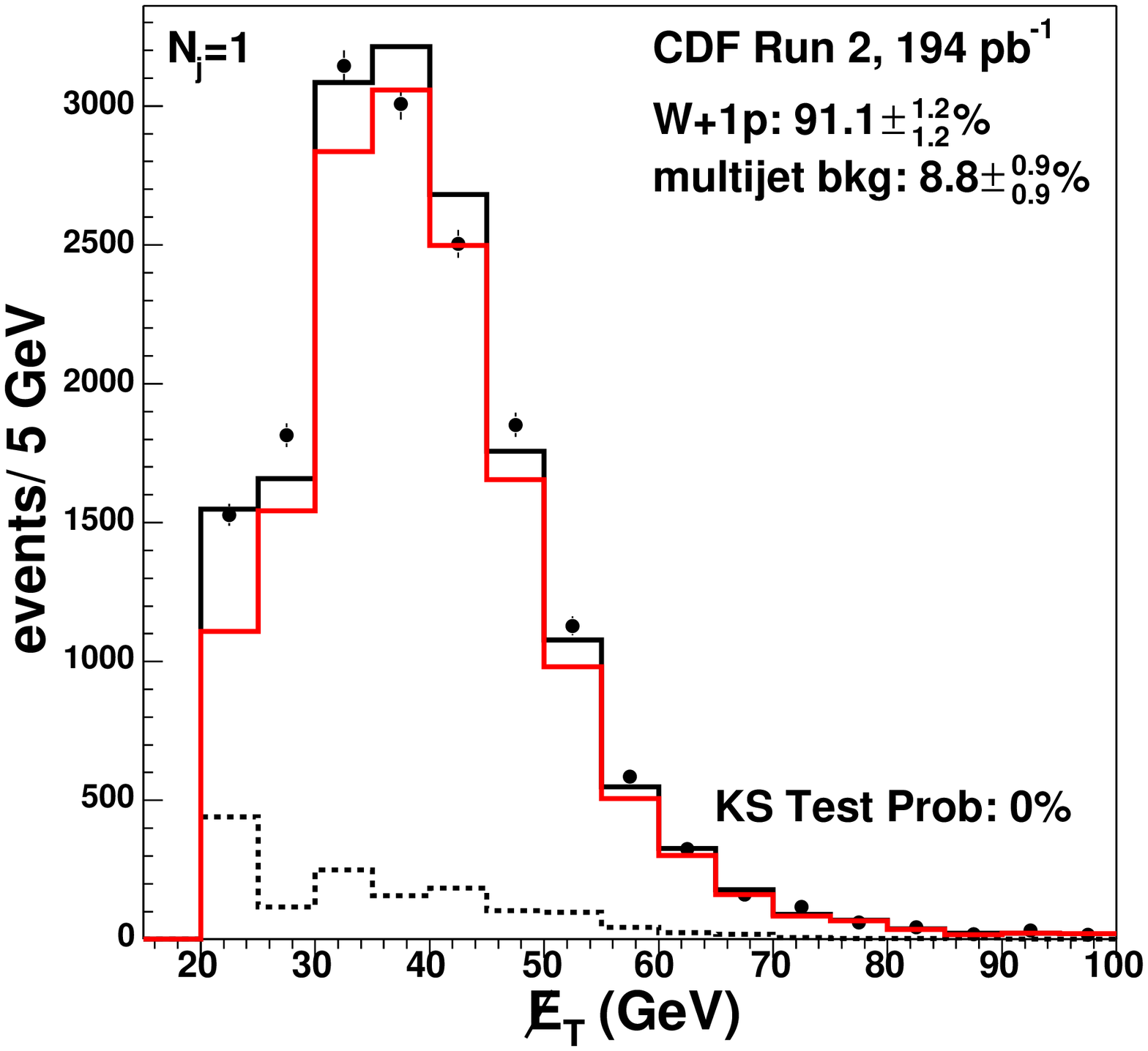}}
\resizebox{2.32in}{!}{ \includegraphics{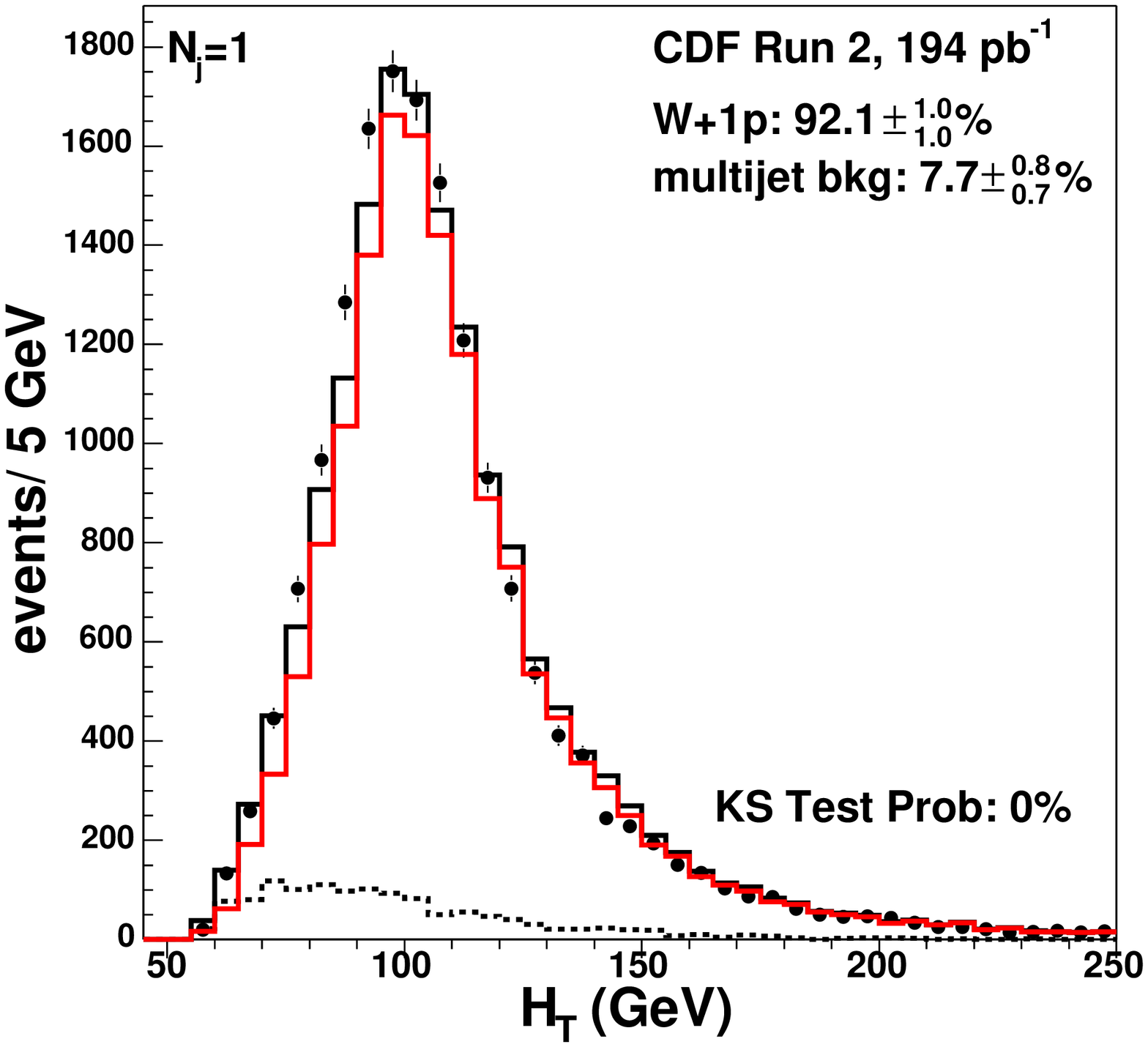}}
\resizebox{2.32in}{!}{ \includegraphics{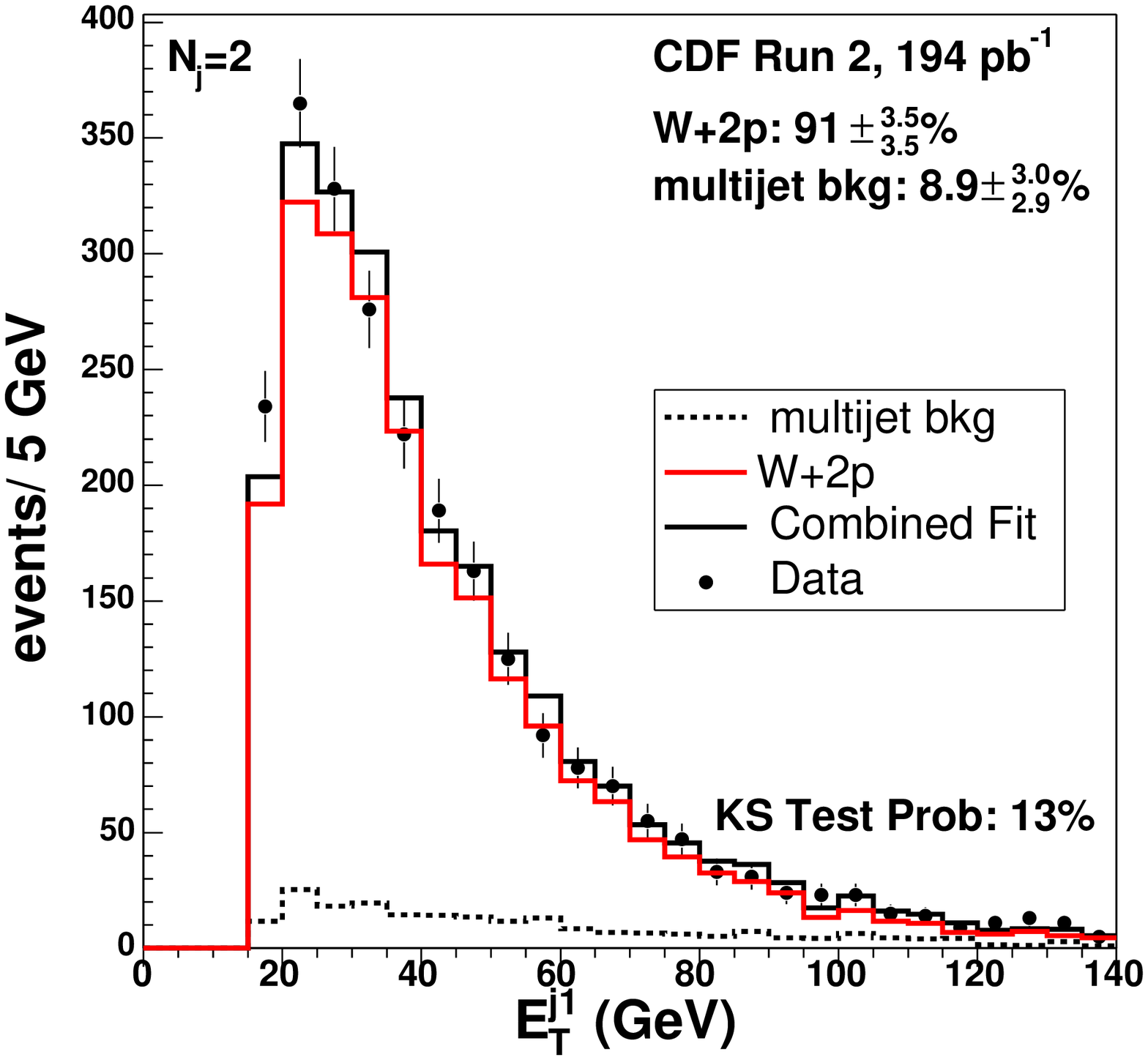}}
\resizebox{2.32in}{!}{ \includegraphics{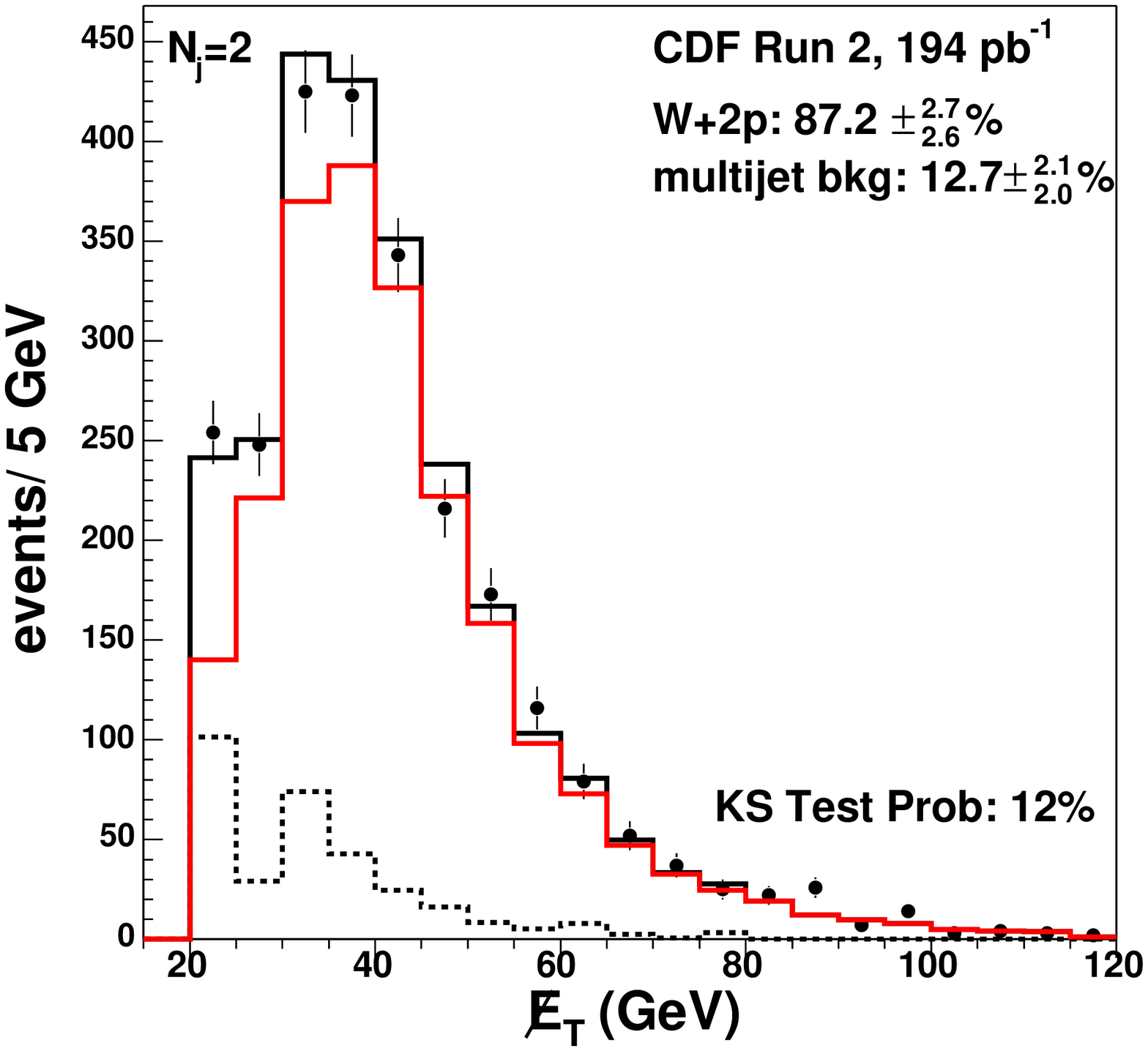}}
\resizebox{2.32in}{!}{ \includegraphics{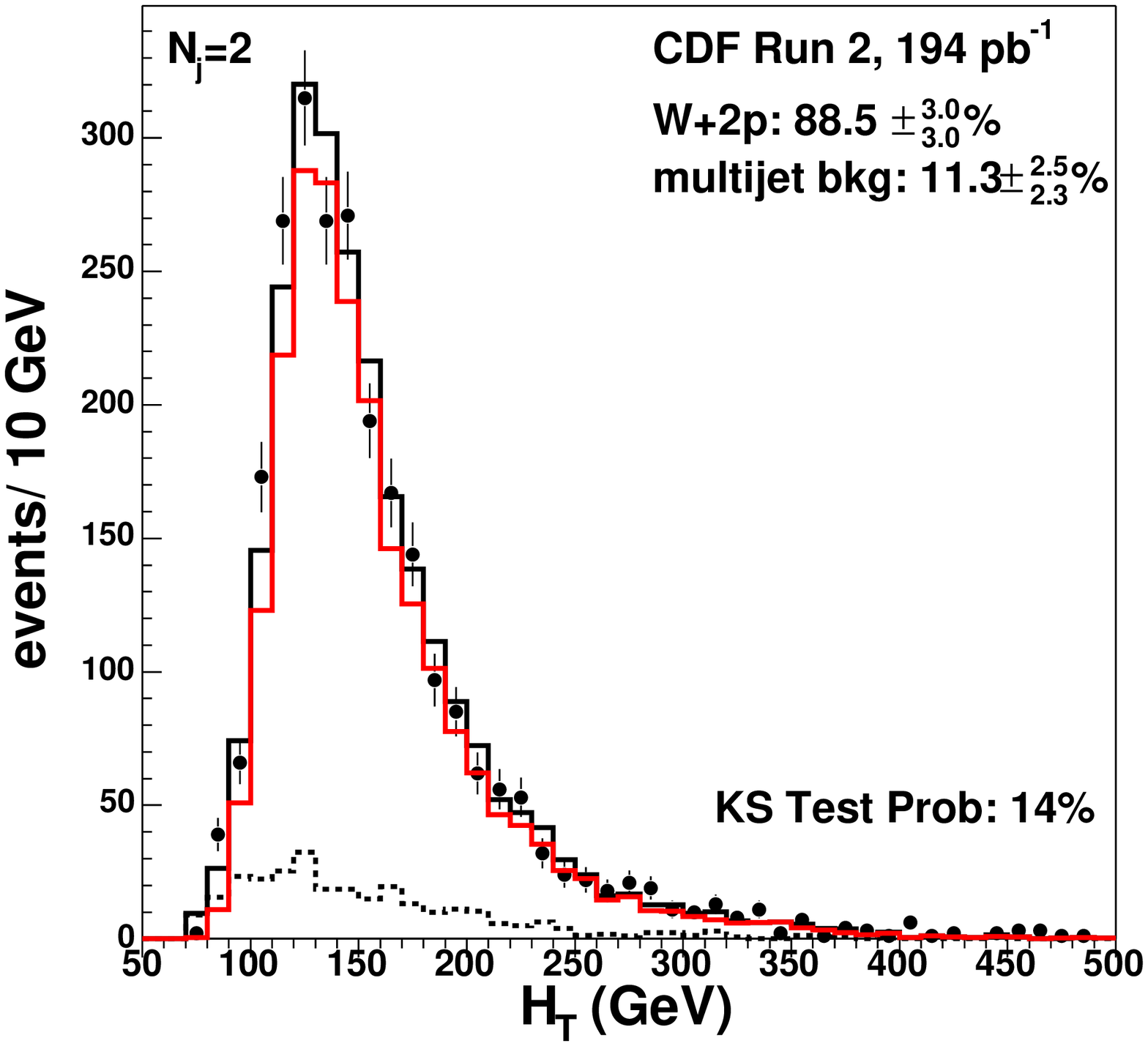}}
  \end{center}
  \caption{\label{fig:mc_validation_123} The leading jet \et\, \met, and \sht\ distributions for
$W+1$ and $W+2$ jet events compared to the prediction from ALPGEN+HERWIG $W+1p$ and
$W+2p$ Monte Carlo and multi-jet background distributions. A binned maximum likelihood fit is performed 
to the data, allowing the number of QCD background events to fluctuate. Note that the \met\ distribution
is sculpted between 20 and 30~GeV by the multi-jet background rejection of Section~\ref{subsec:furtherbkg}. 
  } 
  \end{figure*}

In the $W+1$\ and $W+2$\ jet samples, we neglect the \ttbar\ contribution 
as this is expected to be negligible, as shown in Table~\ref{tab:ndata}. Fig.~\ref{fig:mc_validation_123} shows the 
leading jet \et, \met\ and \sht\ distributions for $W+1$\ jet\ and $W+2$\ jet data events compared to the 
prediction from ALPGEN+HERWIG $W+1p$\ and $W+2p$\ Monte Carlo respectively, and our model of the QCD multi-jet background
from non-isolated lepton data.   We observe better agreement between data and our model 
in the $W$+2 jet sample than in the $W$+1 jet sample.  As we noted in Section~\ref{sec:bkg}, 
our $W+n$ parton model of the $W+\geq n$\ jets background approximates the contributions from higher-order matrix elements with a parton shower, which does not alter the kinematics of the $W$\ boson.   This effect is most pronounced in the $W$+1 jet region, which has the largest relative change in the shape of the $W$ boson \pt\ between $W$+n parton and $W$+(n+1) parton. With the larger statistics available in the 1 and 2-jet bins, 
the QCD multi-jet background is allowed to float and a two-component binned maximum likelihood fit 
is performed to the data.

\begin{figure*}[btp]
  \begin{center}
  
\resizebox{2.32in}{!}{ \includegraphics{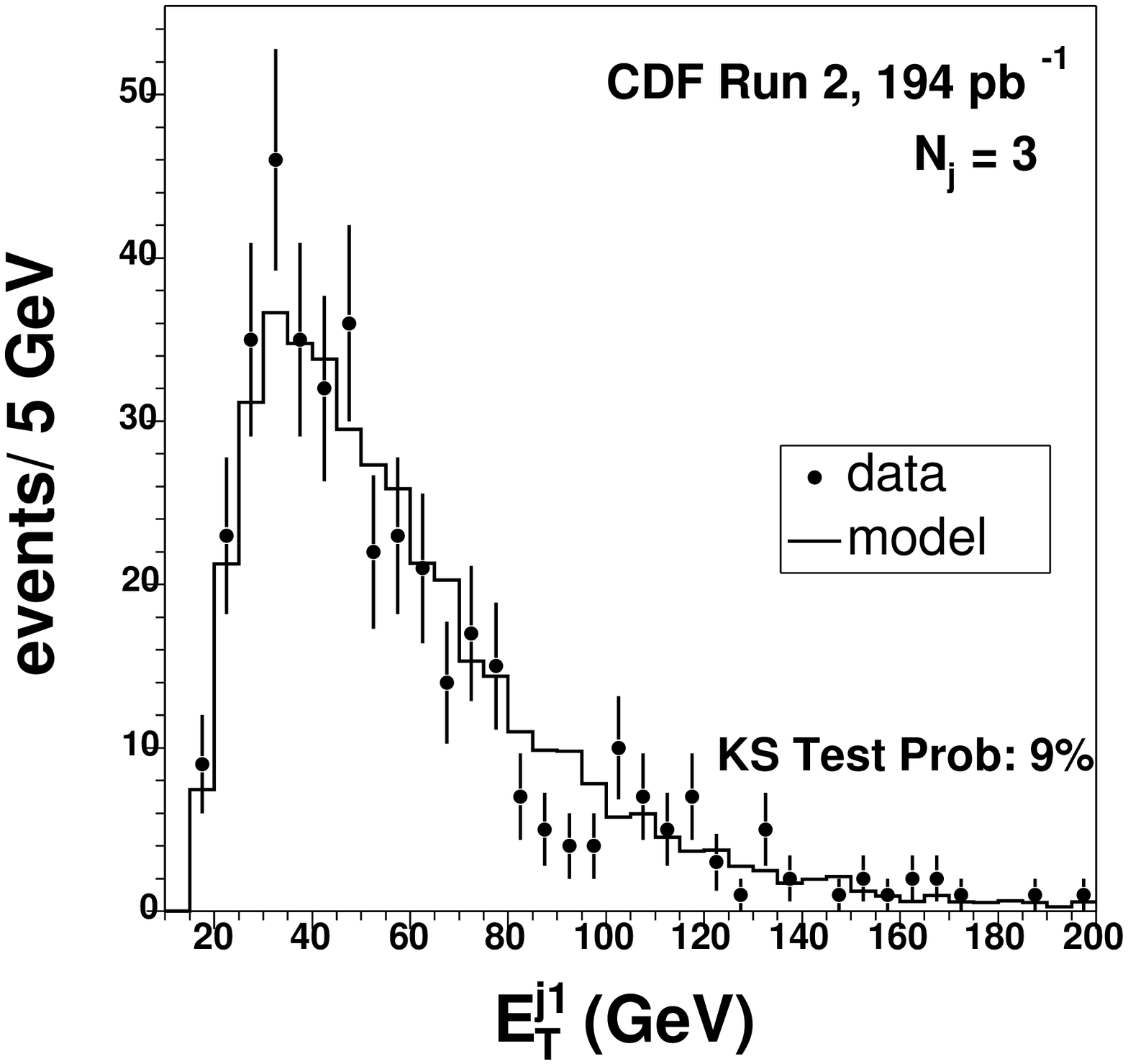}}
\resizebox{2.32in}{!}{ \includegraphics{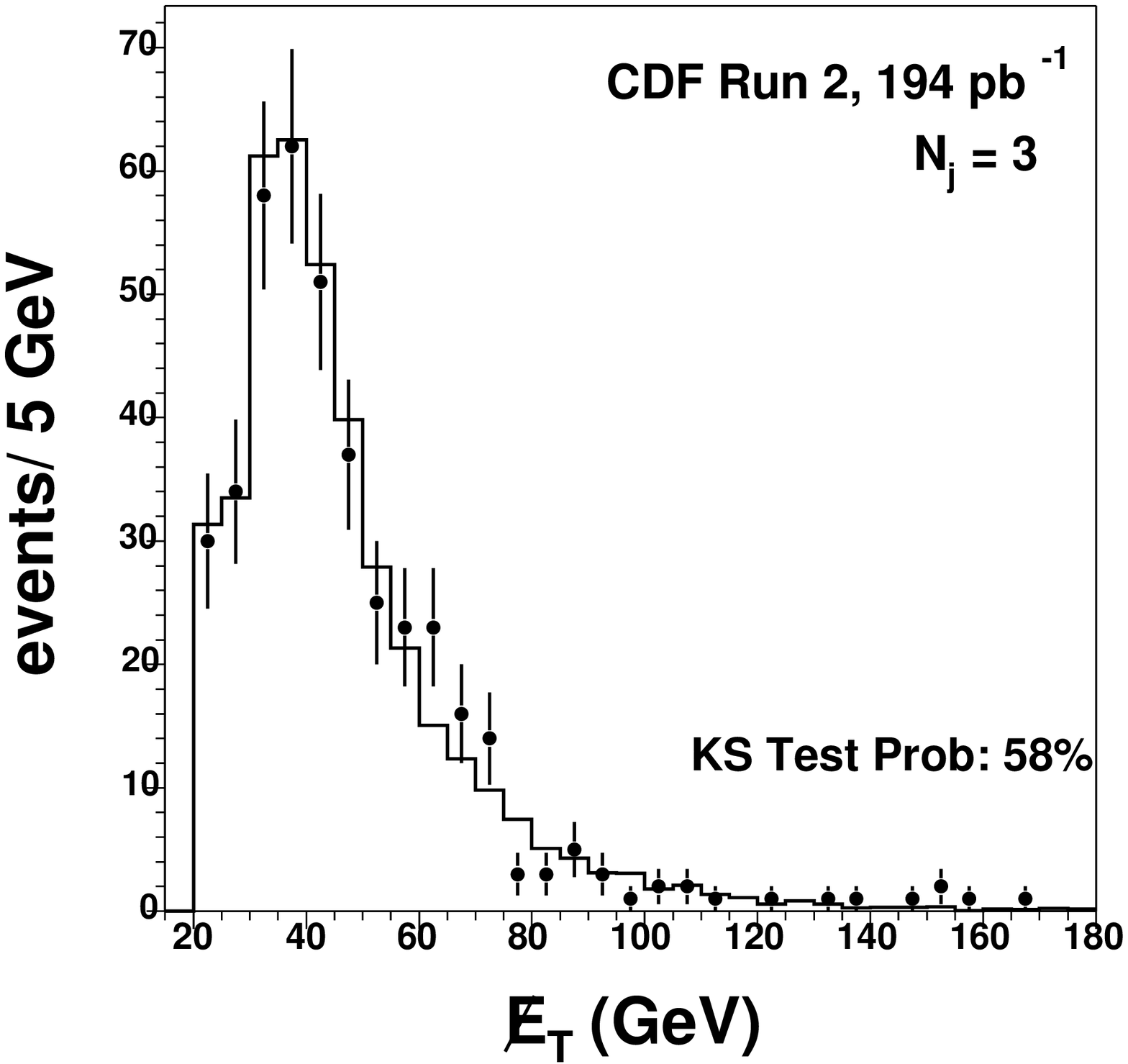}}
\resizebox{2.32in}{!}{ \includegraphics{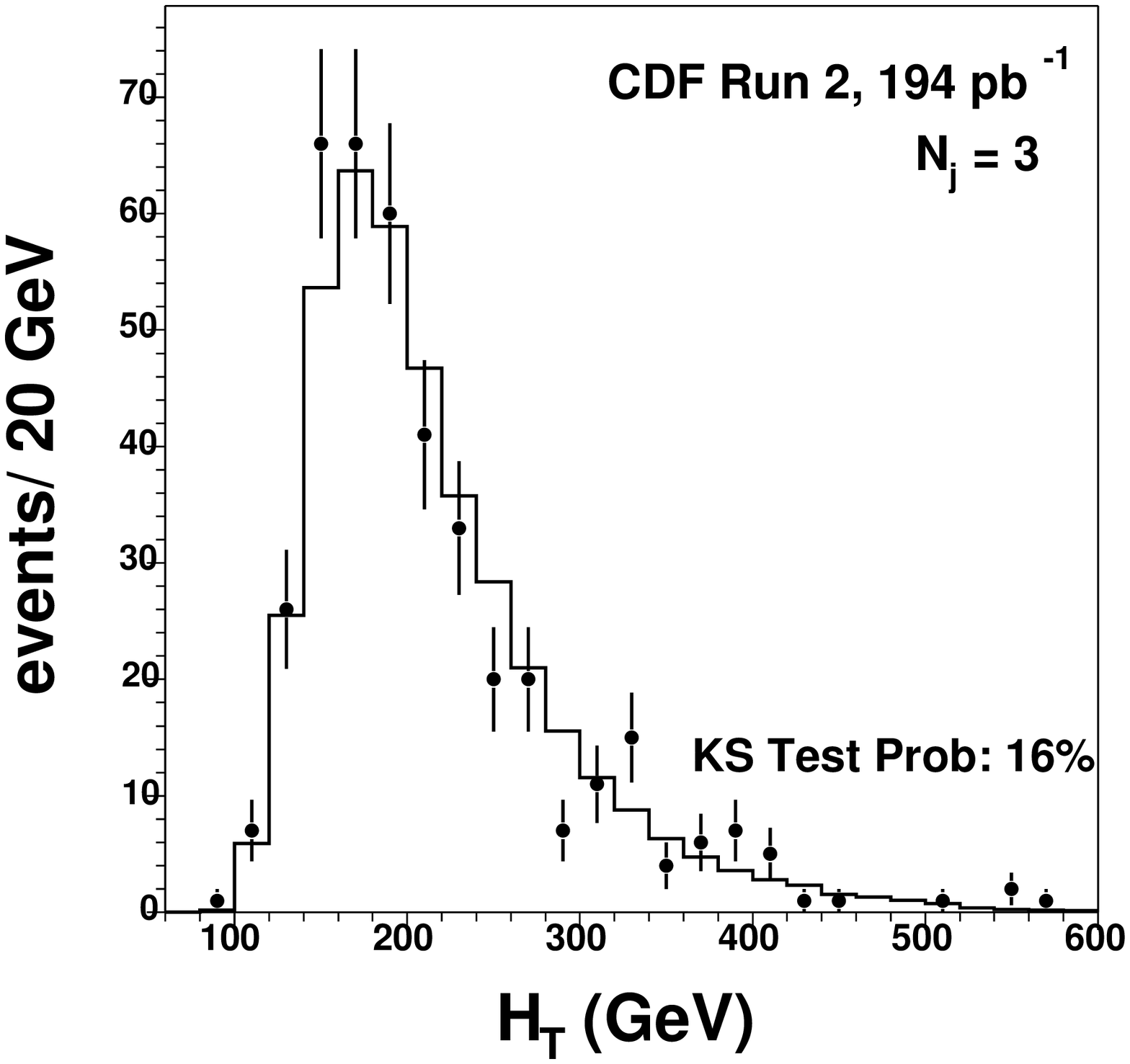}} 
\resizebox{2.32in}{!}{ \includegraphics{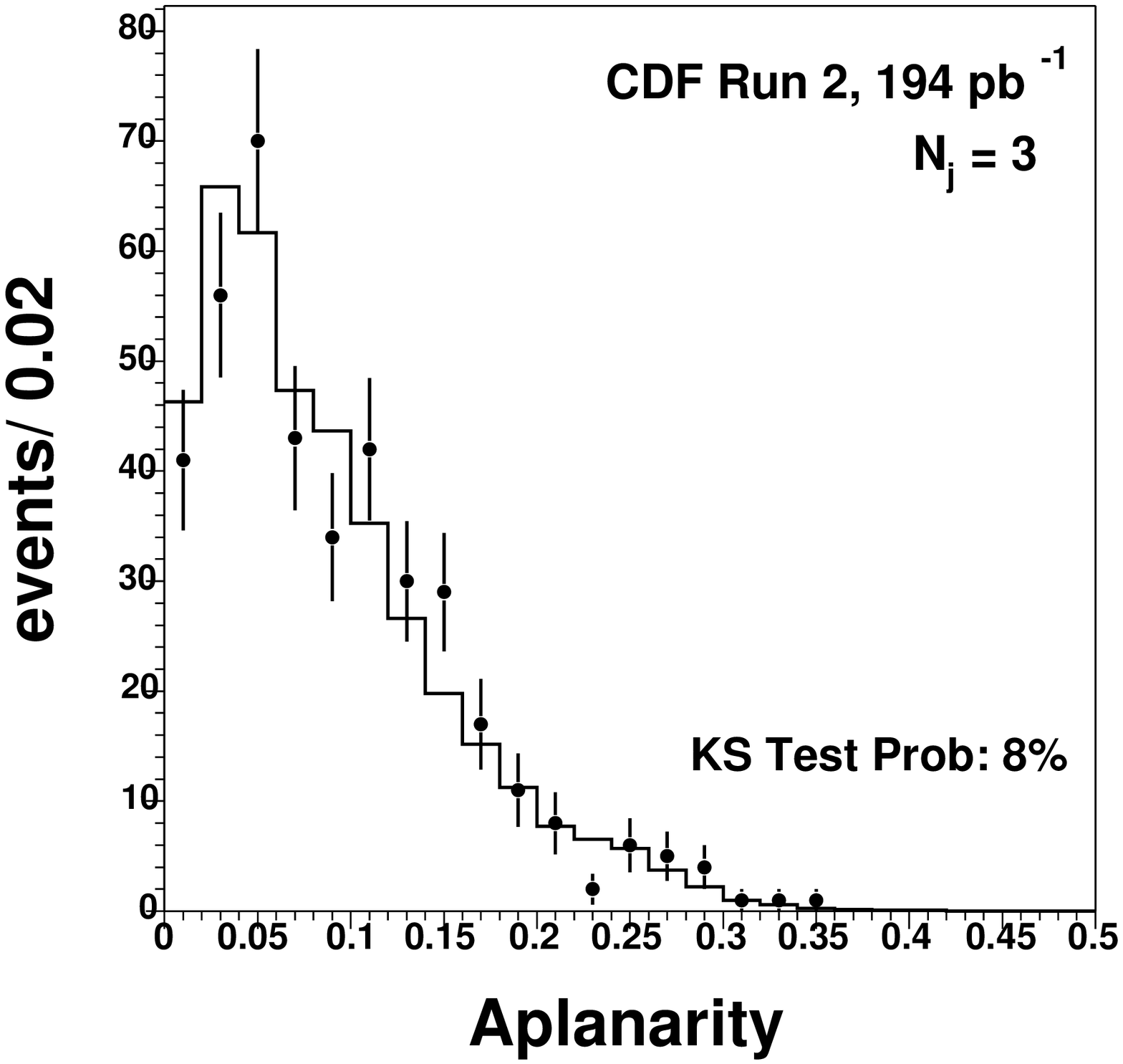}}
\resizebox{2.32in}{!}{ \includegraphics{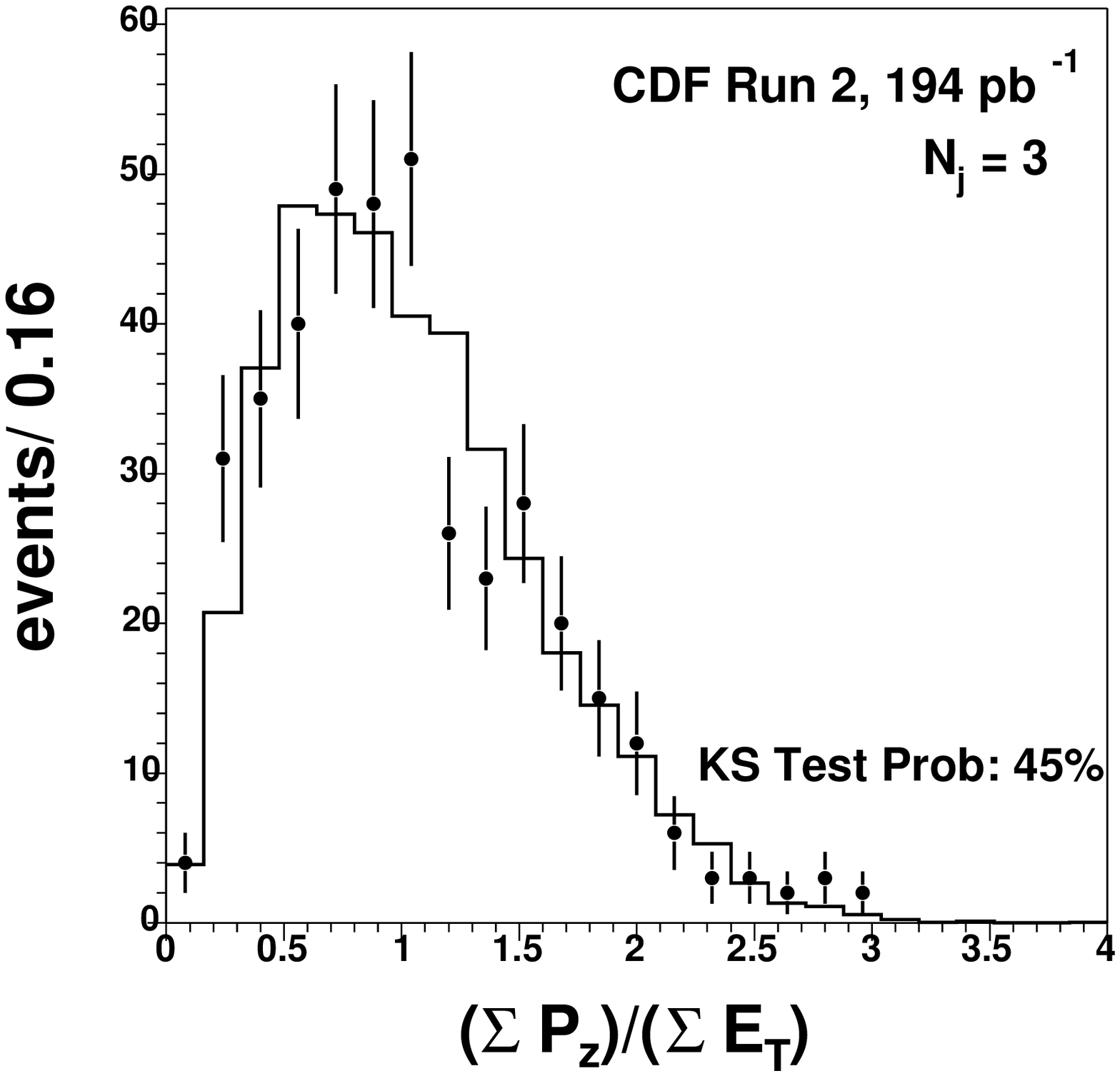}}
\resizebox{2.32in}{!}{ \includegraphics{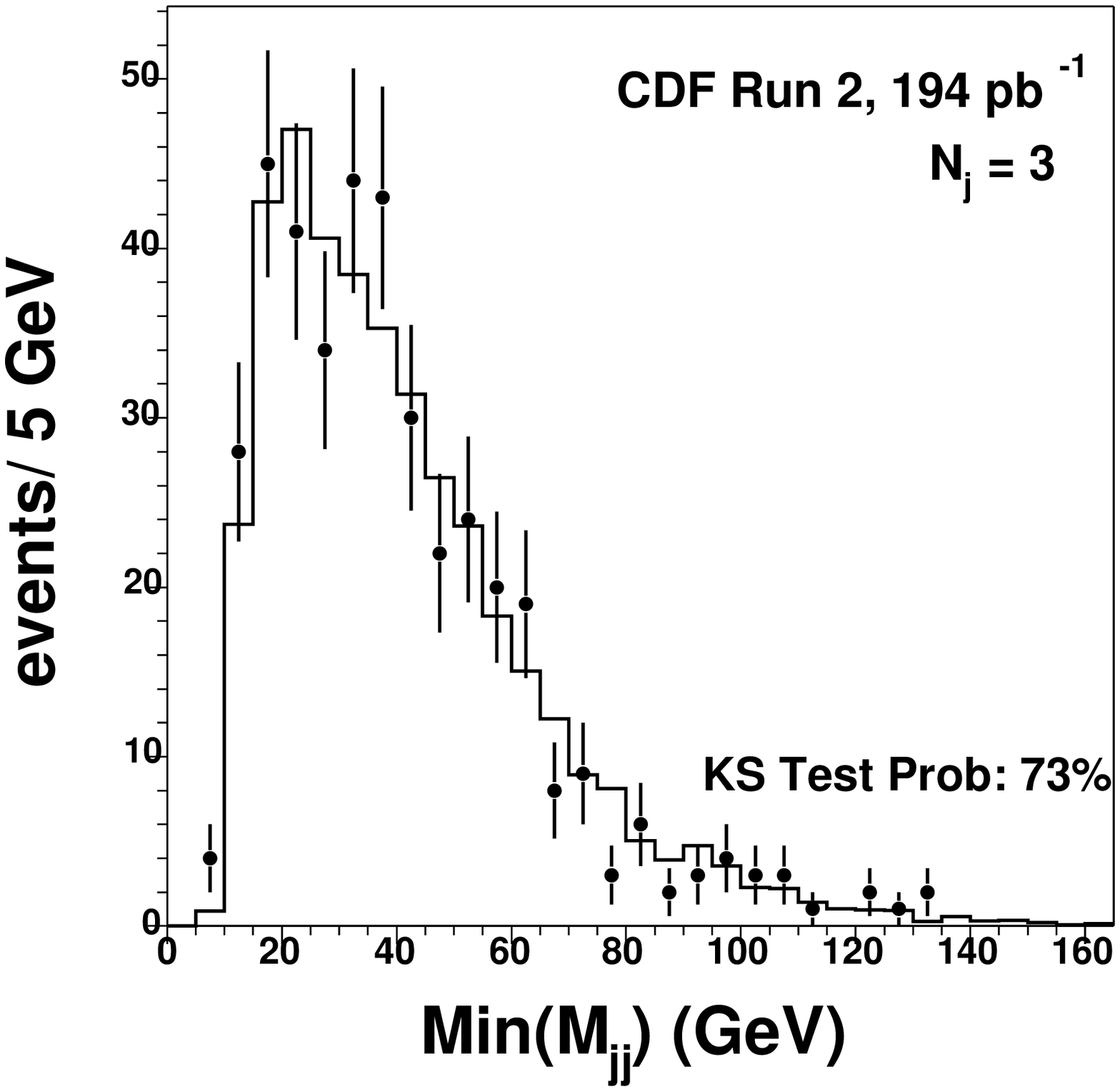}}
\resizebox{2.32in}{!}{ \includegraphics{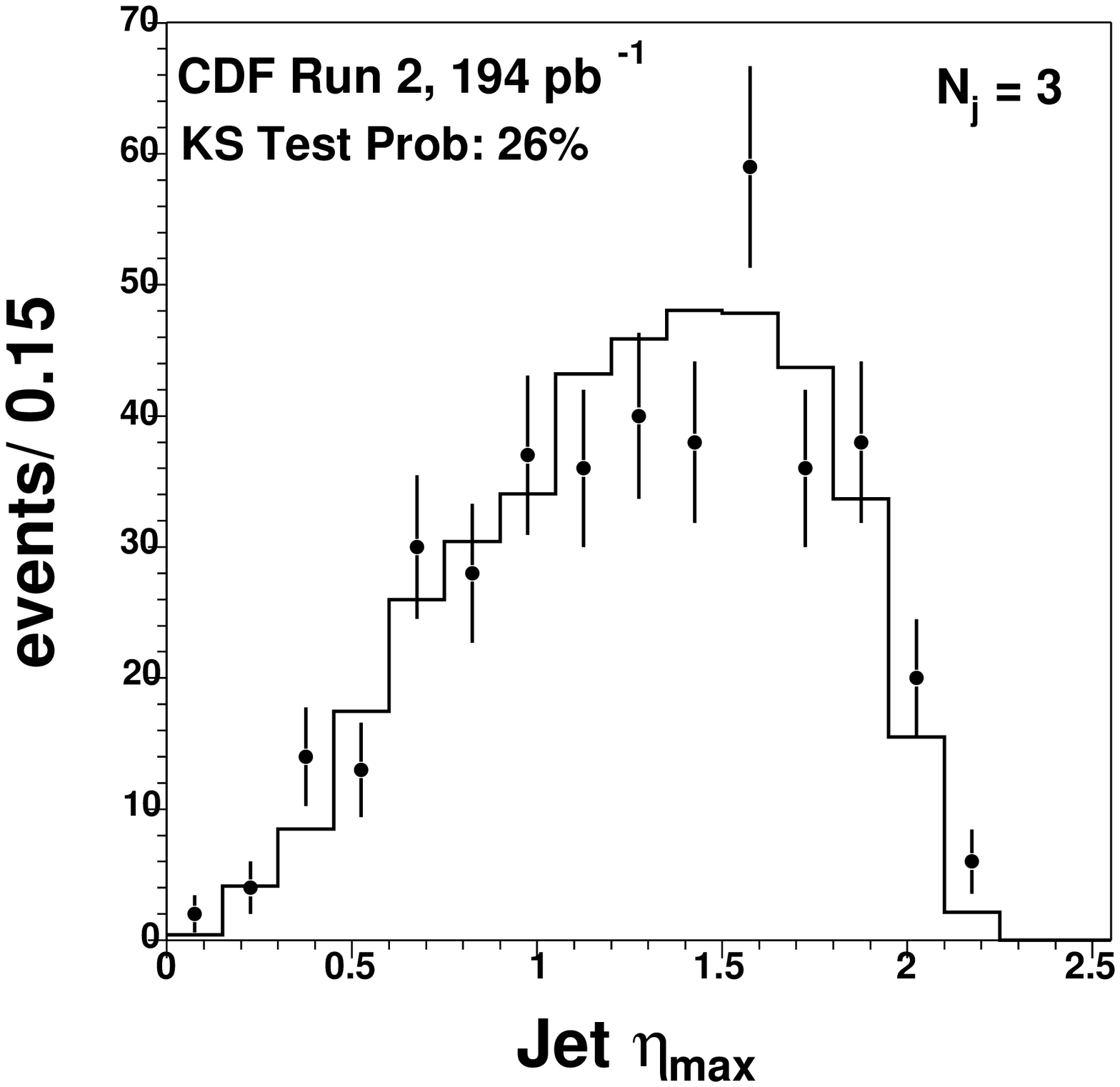}}
\resizebox{2.32in}{!}{ \includegraphics{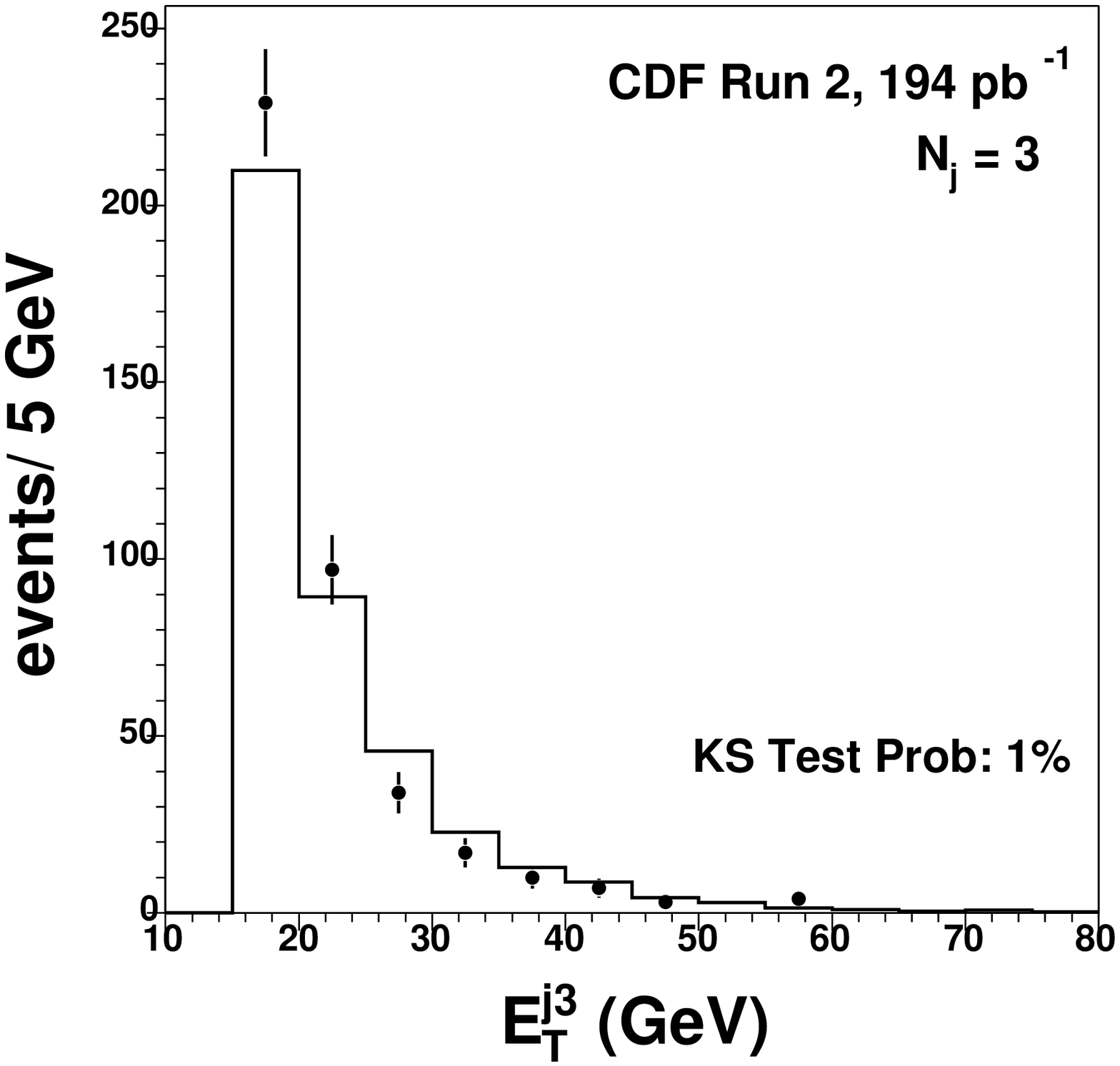}}
\resizebox{2.32in}{!}{ \includegraphics{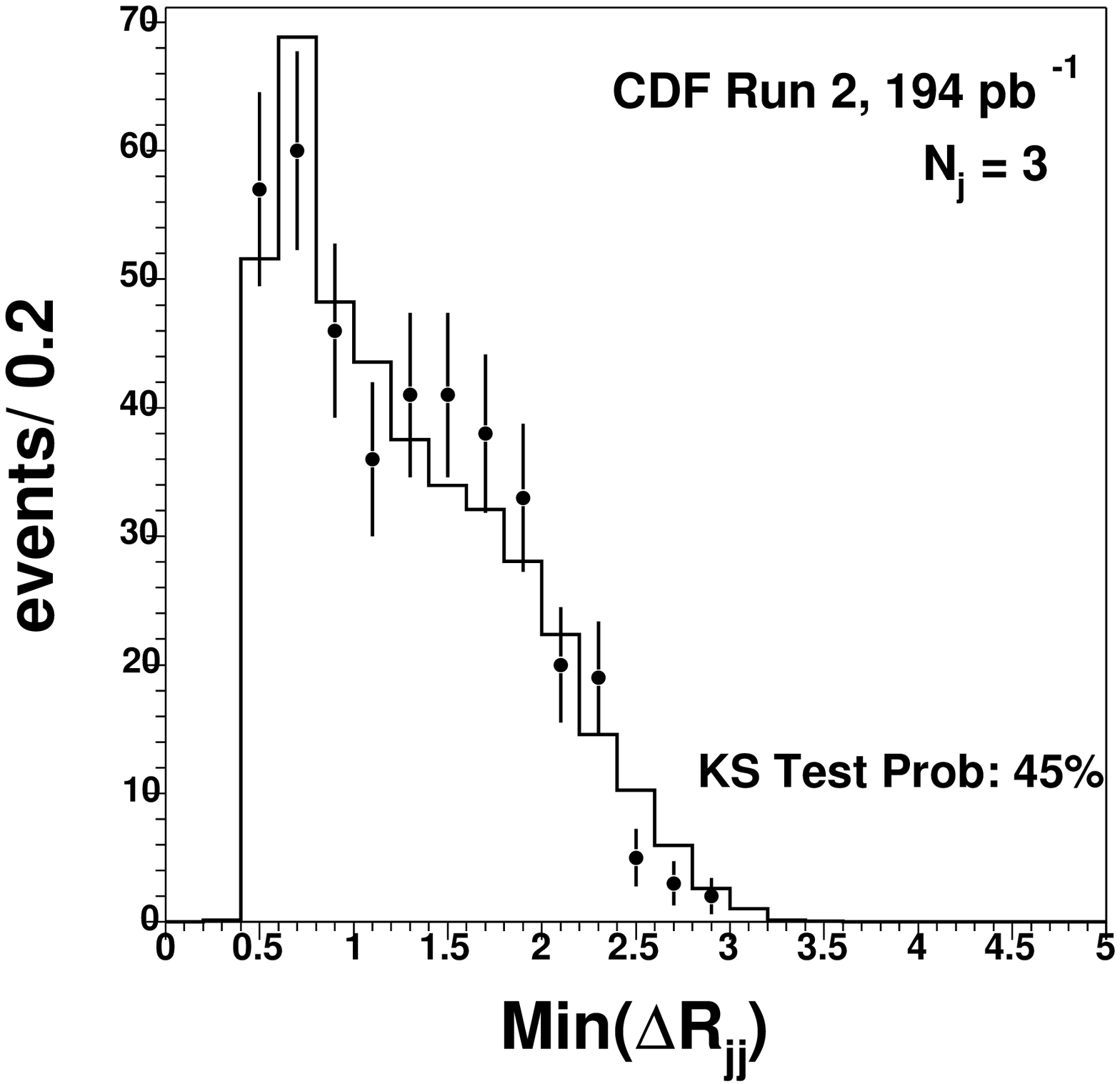}}
\end{center}
  \caption{\label{fig:mc_validation_ann} The \met, leading jet \et, and the seven ANN input
  distributions in $W+3$ exclusive jet events compared to the 
 predictions from ALPGEN+HERWIG $W$+3p Monte Carlo, multi-jet background,and PYTHIA \ttbar\ Monte Carlo.
The model here is a combination of 84\% $W$+jets simulated events with 6\% multi-jet background, and
10\% \ttbar\ simulated events.
}   
\end{figure*}

\begin{figure}[btp]
\begin{center}
\resizebox{2.2in}{!}{ \includegraphics{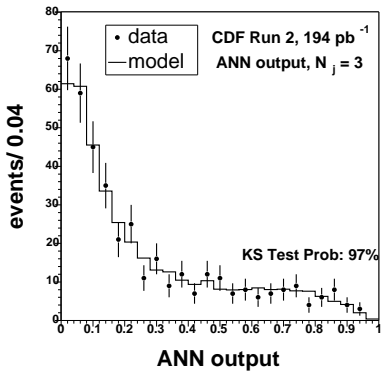}}
\end{center}
\caption{\label{fig:ann} The distribution for the ANN output for
$W+3$ exclusive jet events compared to the prediction from ALPGEN+HERWIG $W$+3p Monte Carlo, multi-jet background 
 and PYTHIA \ttbar\ Monte Carlo. The model here is a combination of 84\% $W$+jets simulated events 
with 6\% multi-jet background, and 10\% \ttbar\ simulated events.
}   
\end{figure}

As discussed in the previous section, we use events with three or more jets for our 
\ttbar\ cross section measurement. In the $W+3$\ jet sample, we expect a contribution of only 
about 10\% from \ttbar, as shown in Table~\ref{tab:ndata}.  This latter region is top-depleted but 
otherwise kinematically and topologically identical to the majority of the background in the signal sample. 
Therefore we use events with exactly three jets to make a complete comparison of
all the discriminating properties and the correlations between them.  
Fig.~\ref{fig:mc_validation_ann} shows the distributions for the leading jet \et, 
\met, \sht, as well as other ANN input properties for $W+3$ exclusive jet events compared to the 
prediction from the ALPGEN+HERWIG $W$+3p Monte Carlo, multi-jet background and PYTHIA \ttbar\ Monte Carlo.
The model here is not the result of a binned maximum likelihood fit but rather has the \ttbar\ fraction 
fixed to 10\% as expected  for a top mass of 175 GeV/$c^{2}$  in Table~\ref{tab:ndata}, the multi-jet 
background to the 6\% estimate from Table~\ref{tab:qcdbkg}, and the W+jets background as the 
remaining 84\%.  A similar comparison for the output of the ANN in the  $W+3$\ jet exclusive sample 
is shown in Fig.~\ref{fig:ann}. Overall, the KS test values indicate good agreement between data and the Monte Carlo simulation. 
 
\begin{figure*}[btp]
  \begin{center}
\resizebox{2.2in}{!}{ \includegraphics{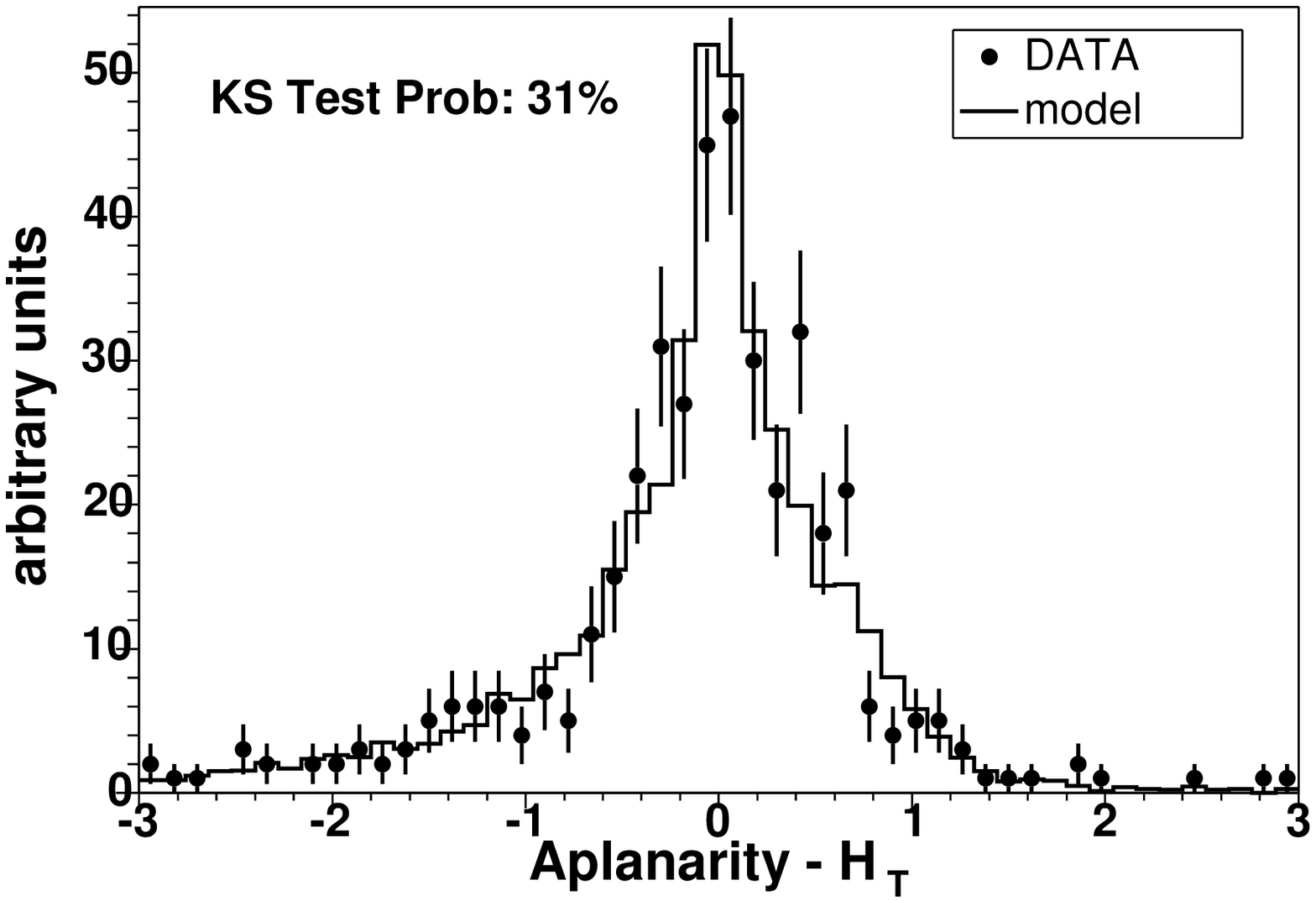}}
\resizebox{2.2in}{!}{ \includegraphics{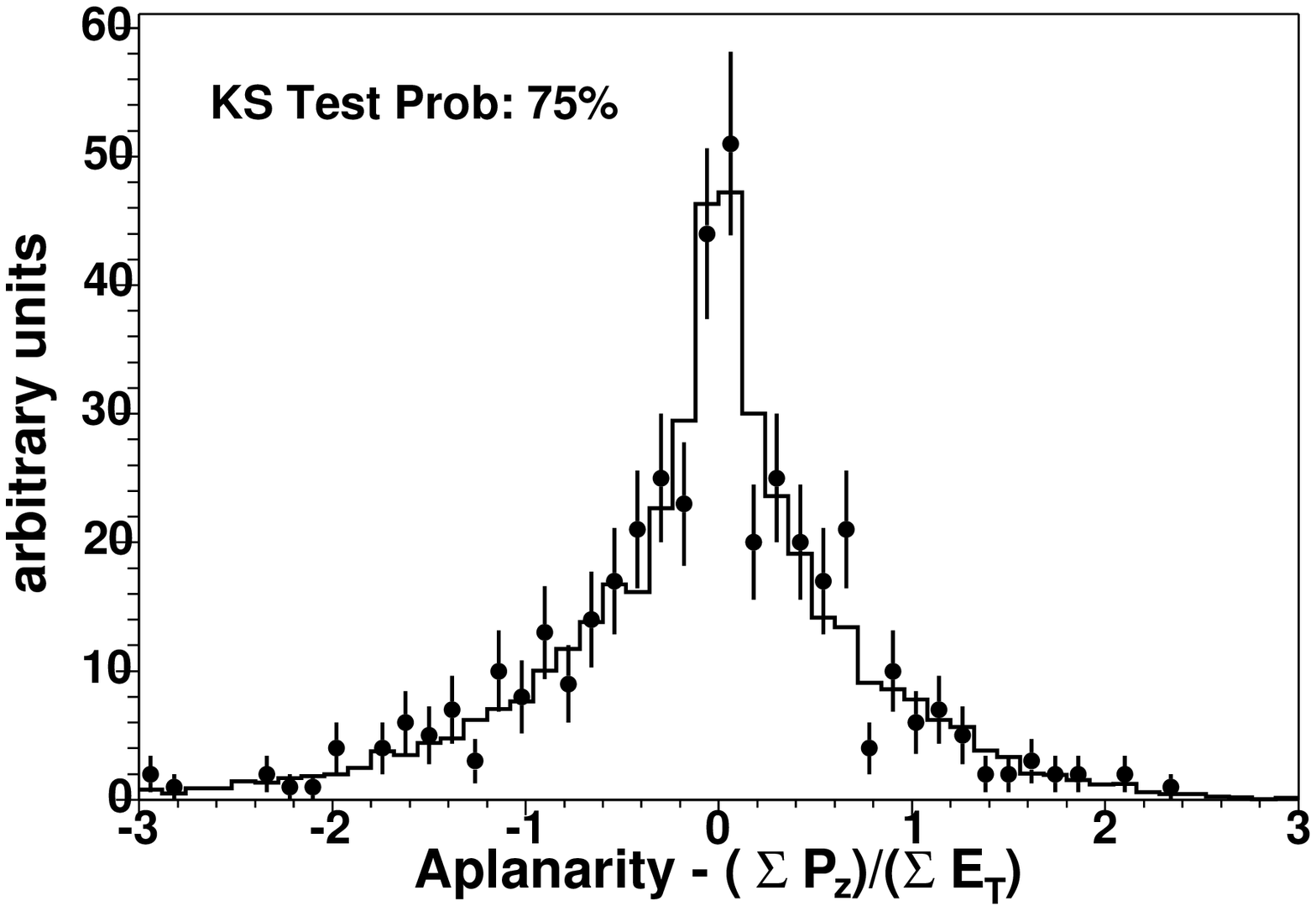}}
\resizebox{2.2in}{!}{ \includegraphics{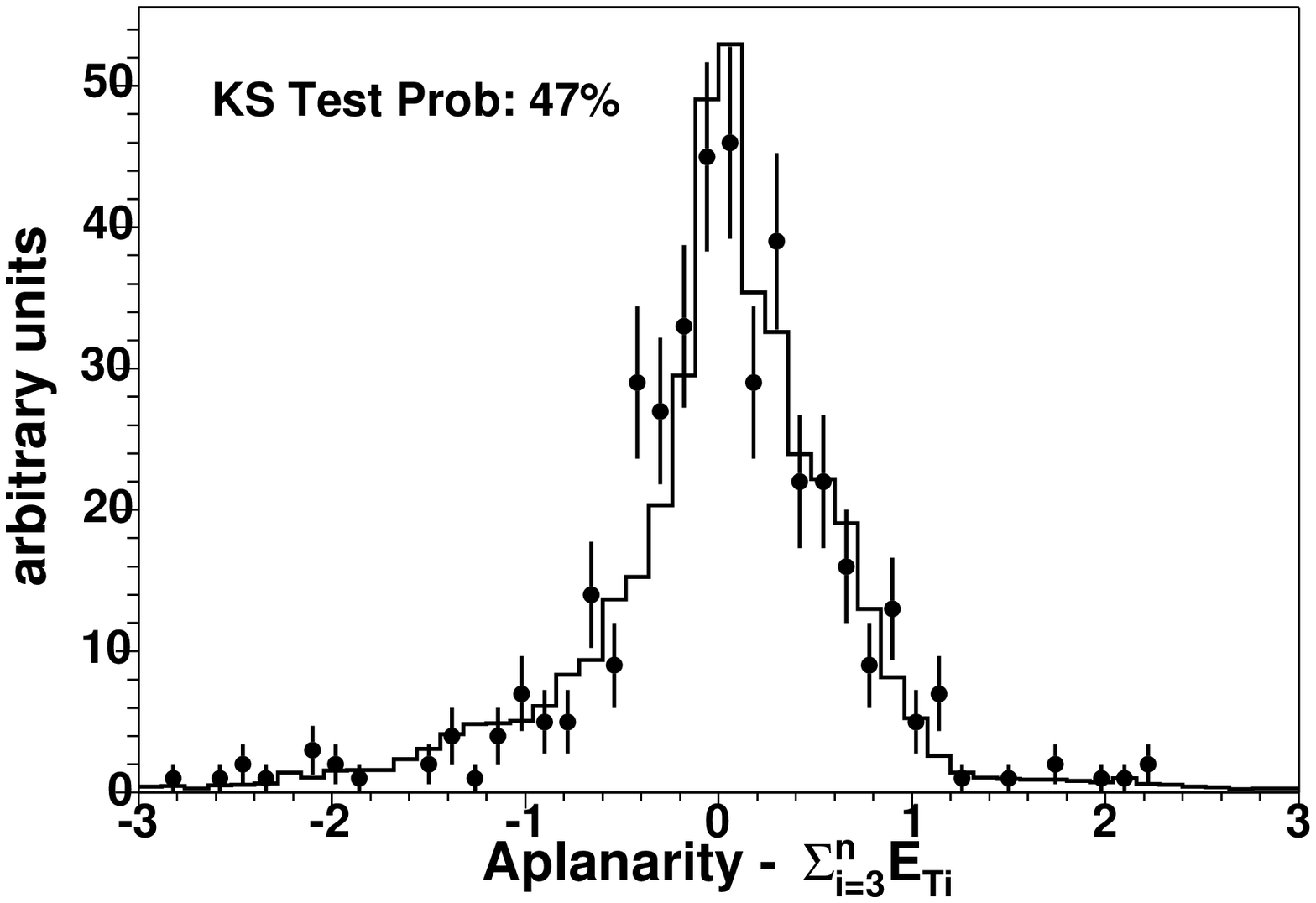}}
\resizebox{2.2in}{!}{ \includegraphics{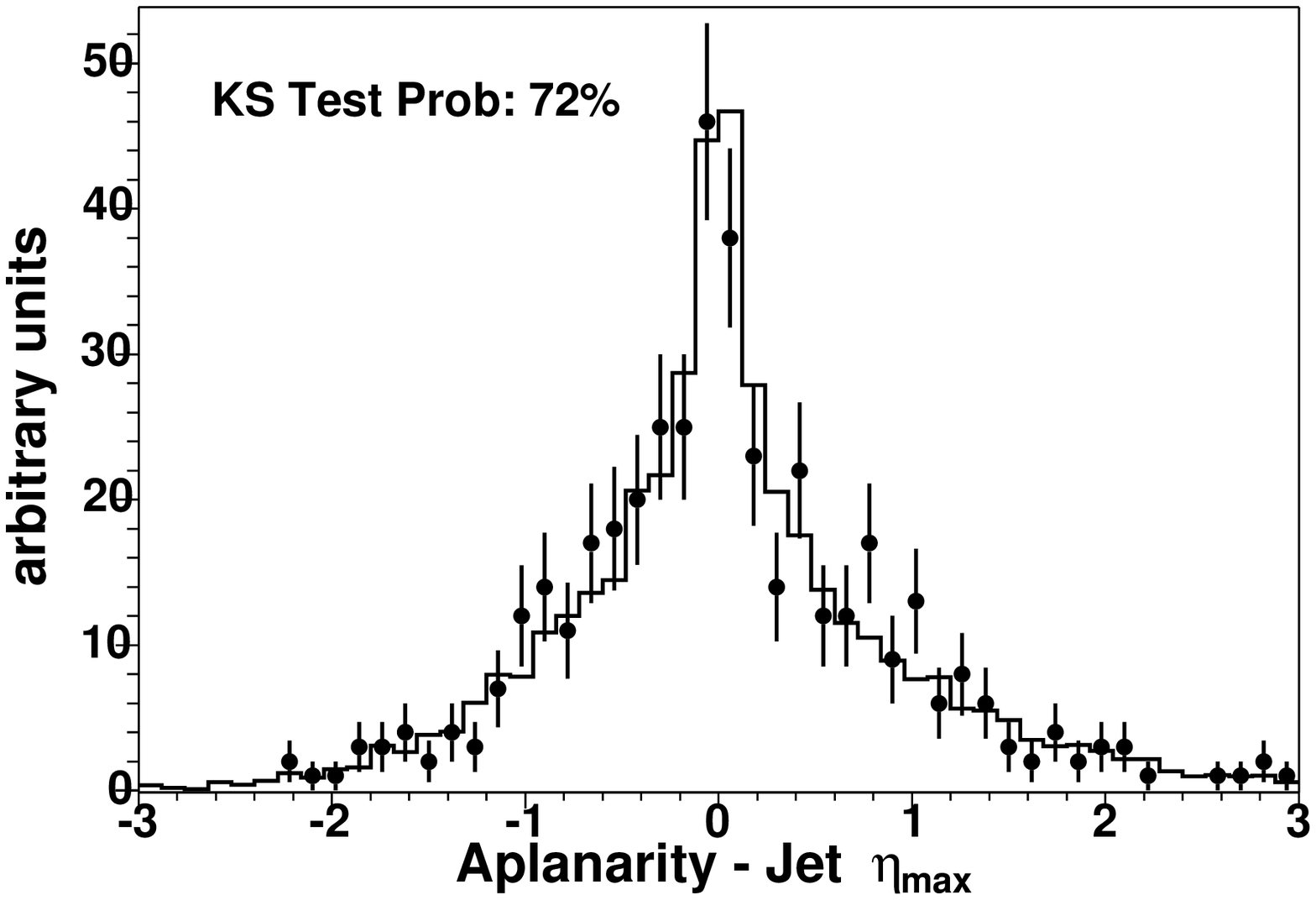}}
\resizebox{2.2in}{!}{ \includegraphics{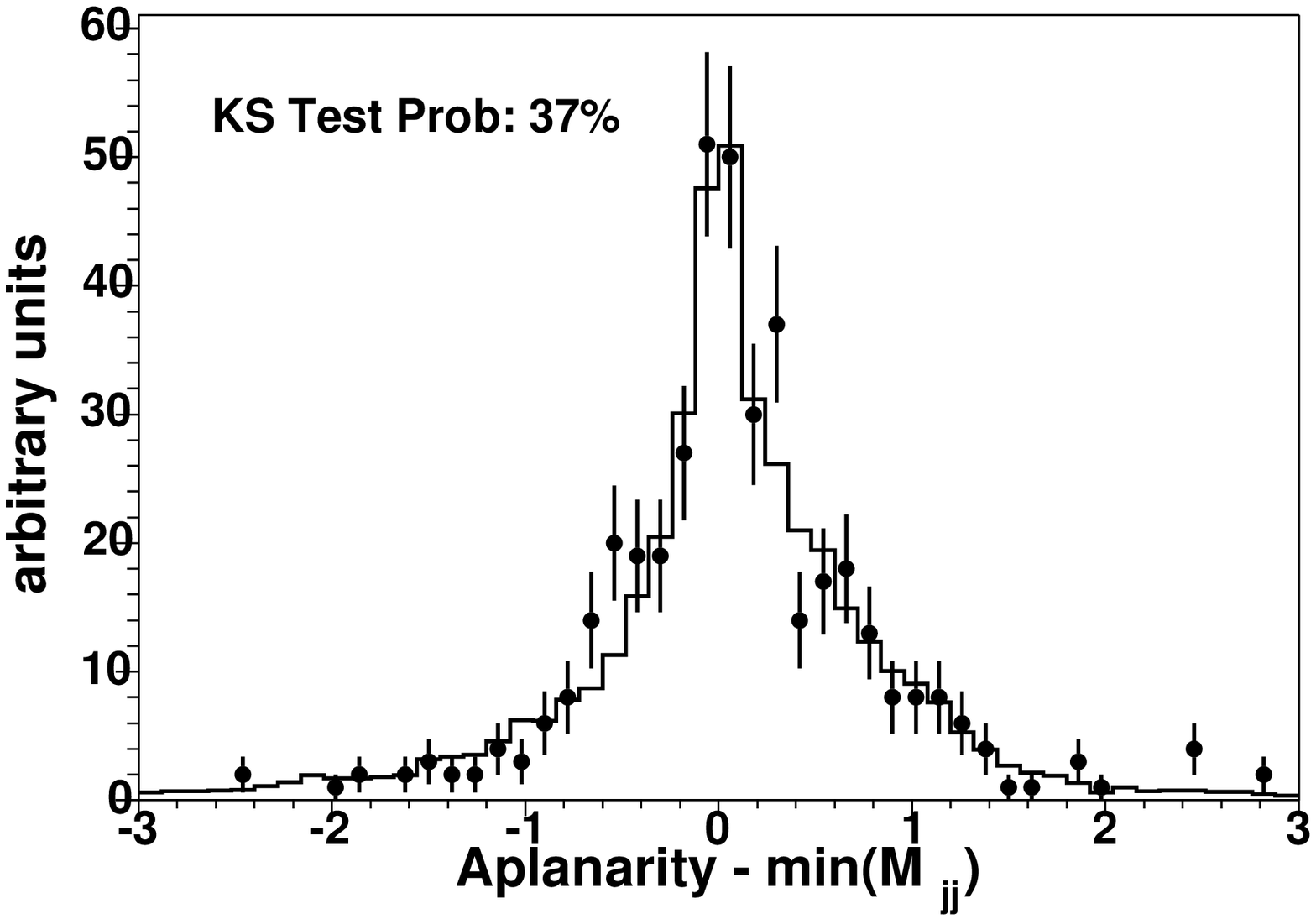}}
\resizebox{2.2in}{!}{ \includegraphics{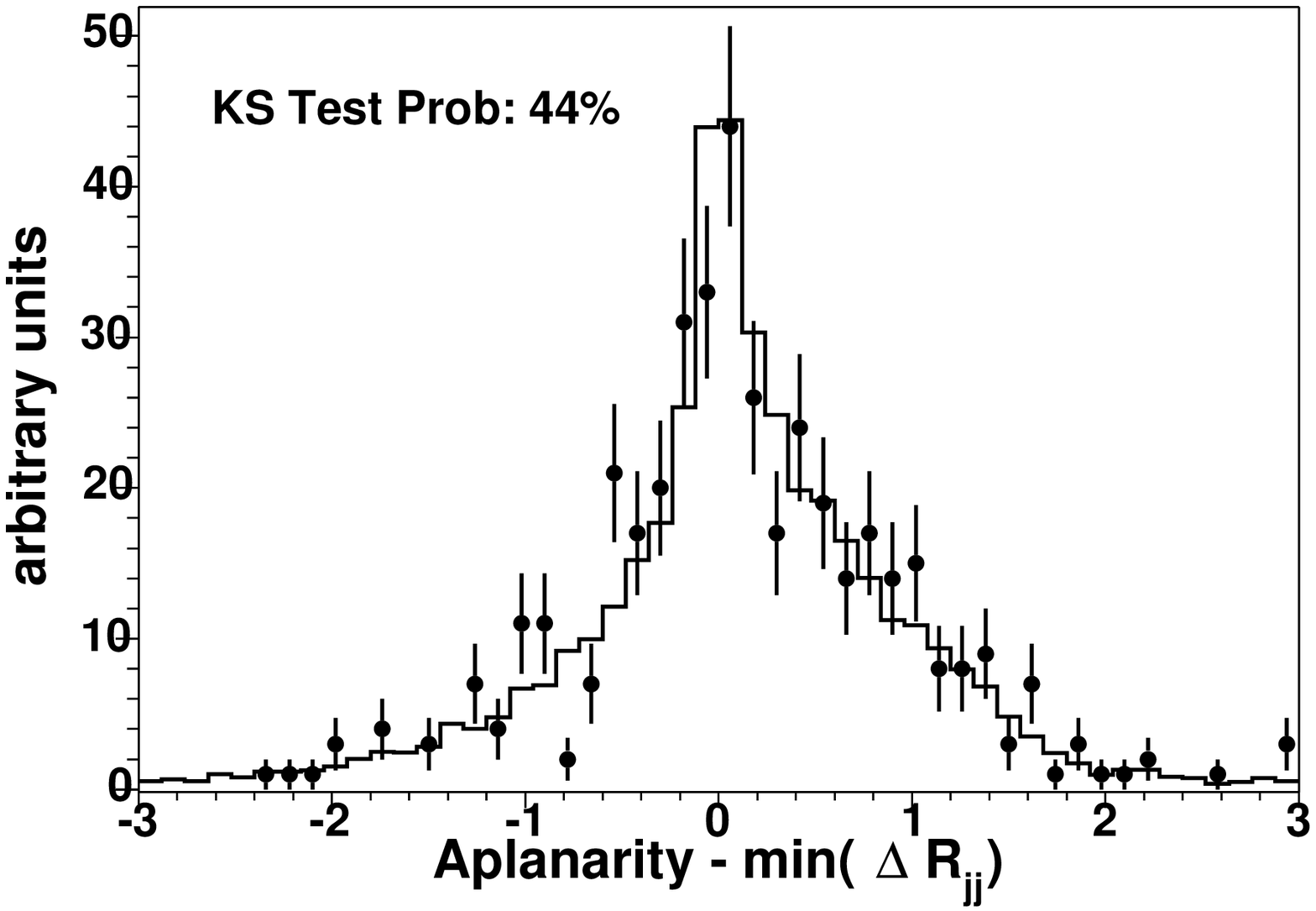}}
\resizebox{2.2in}{!}{ \includegraphics{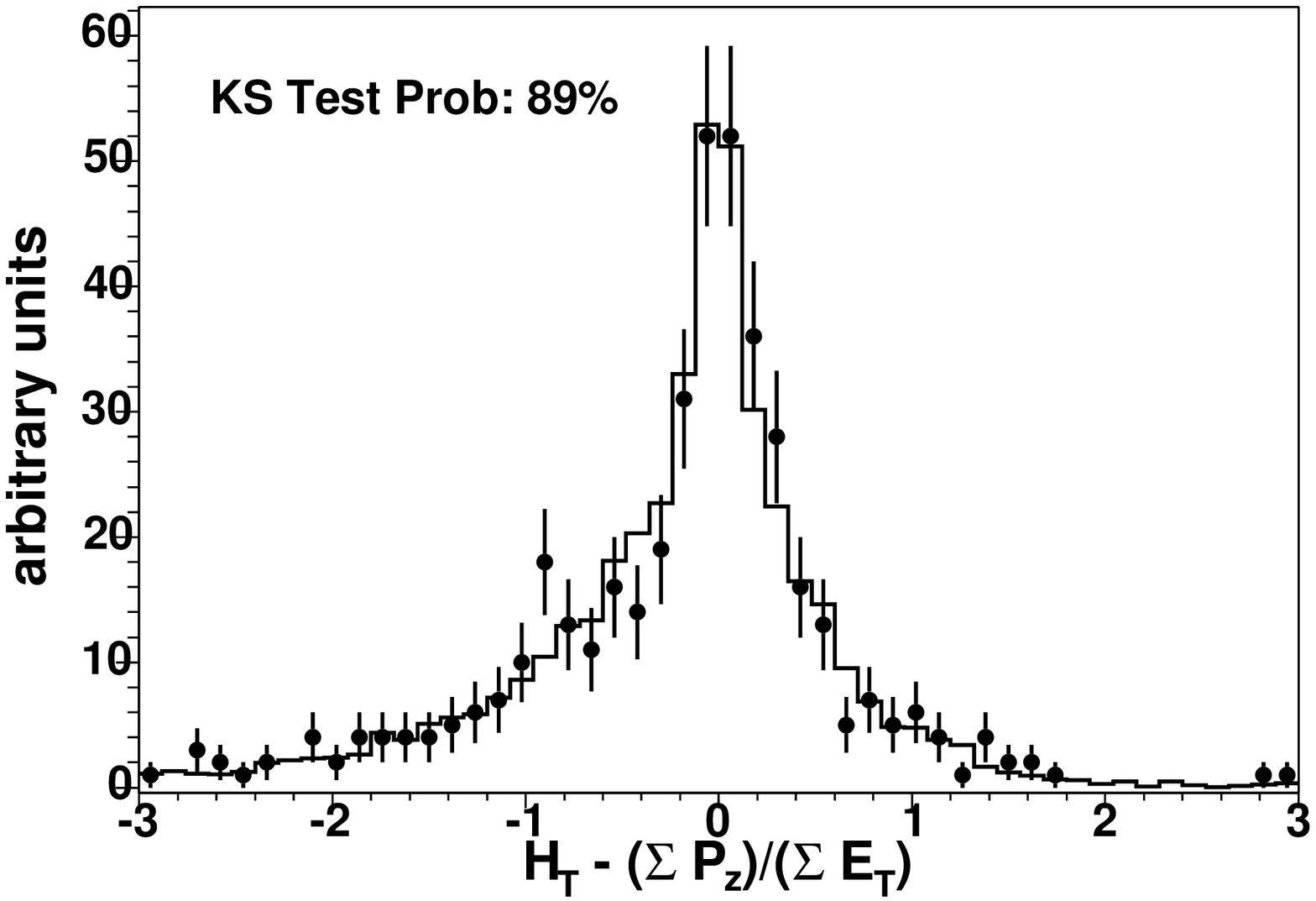}}
\resizebox{2.2in}{!}{ \includegraphics{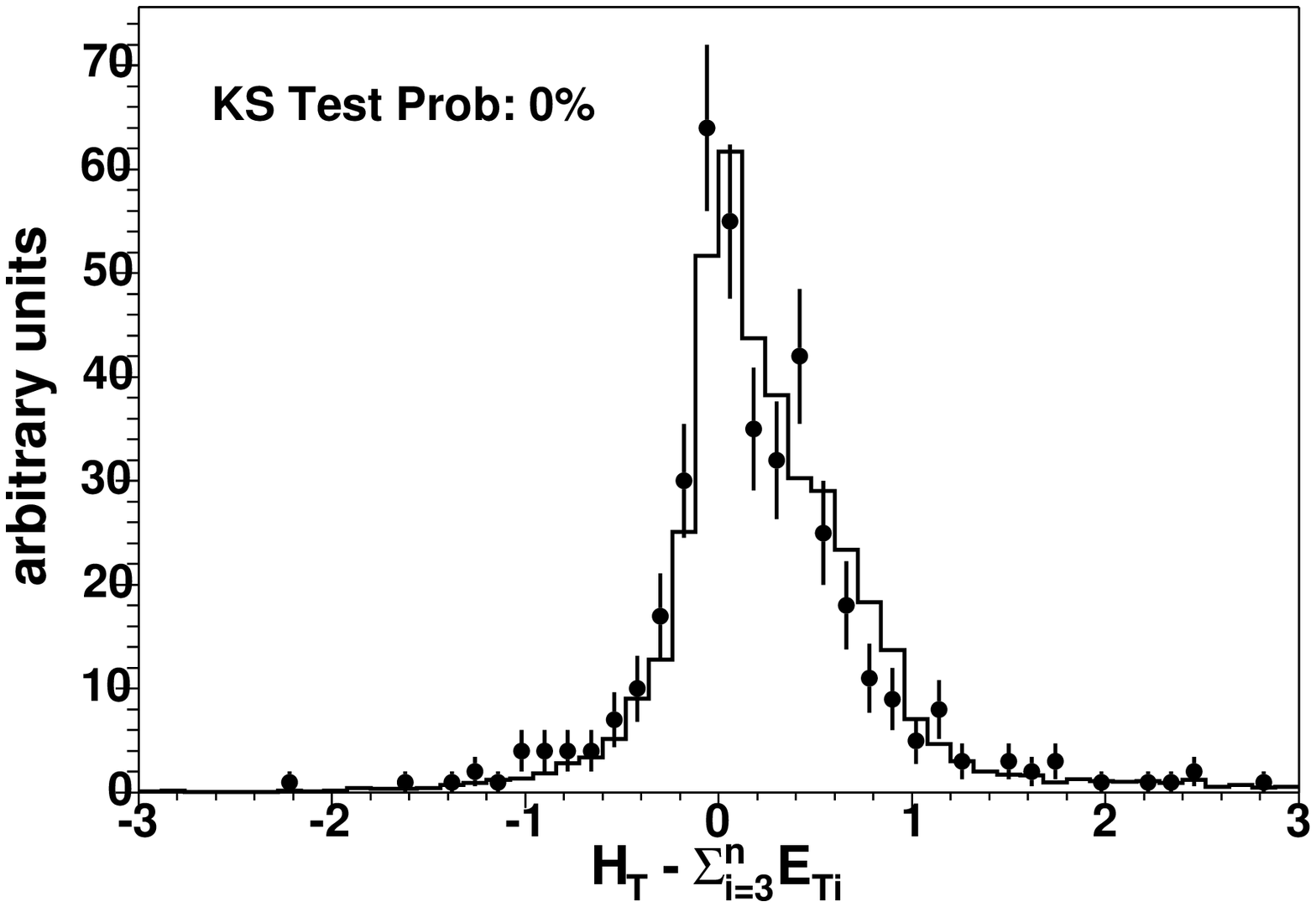}}
\resizebox{2.2in}{!}{ \includegraphics{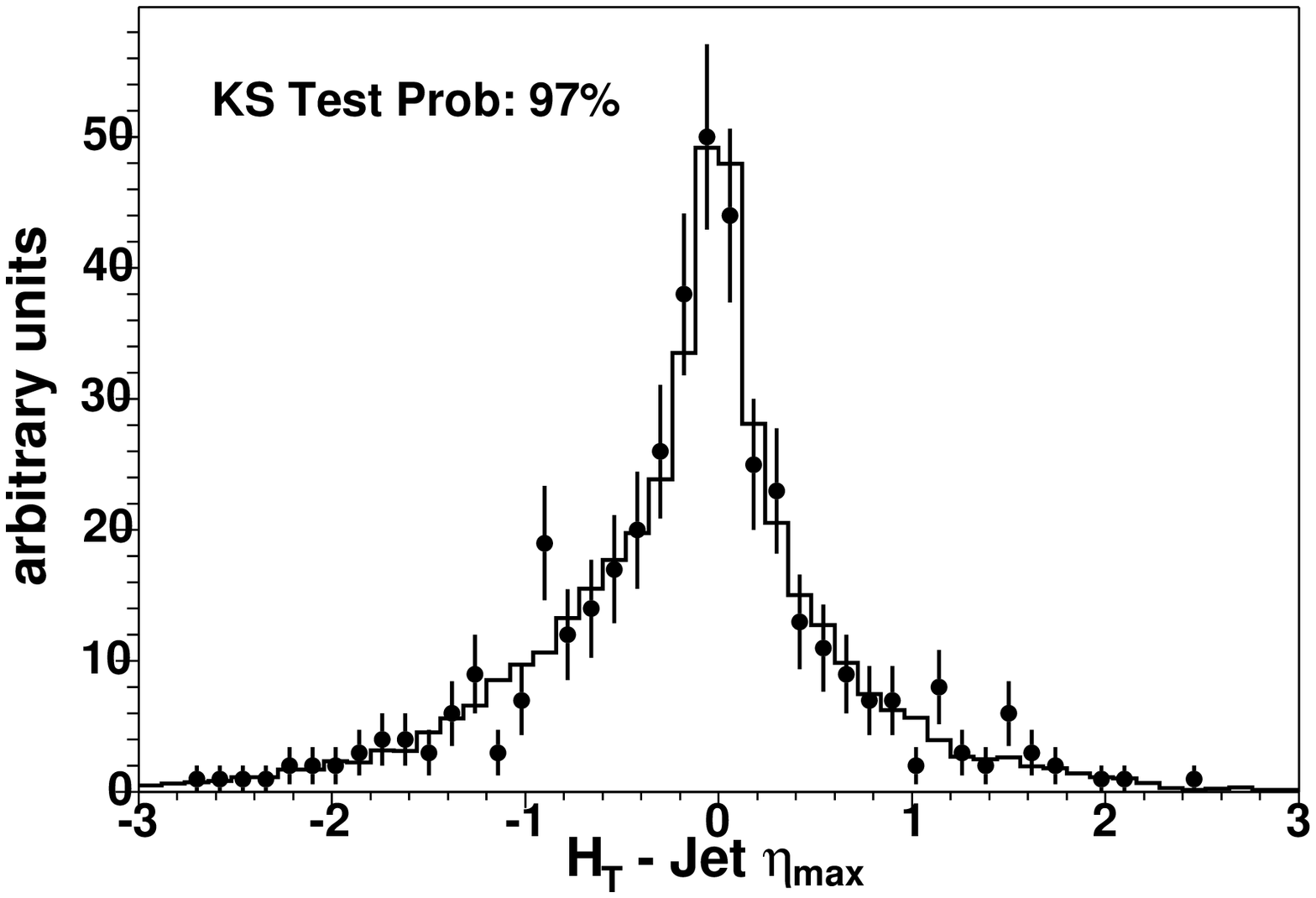}}
\resizebox{2.2in}{!}{ \includegraphics{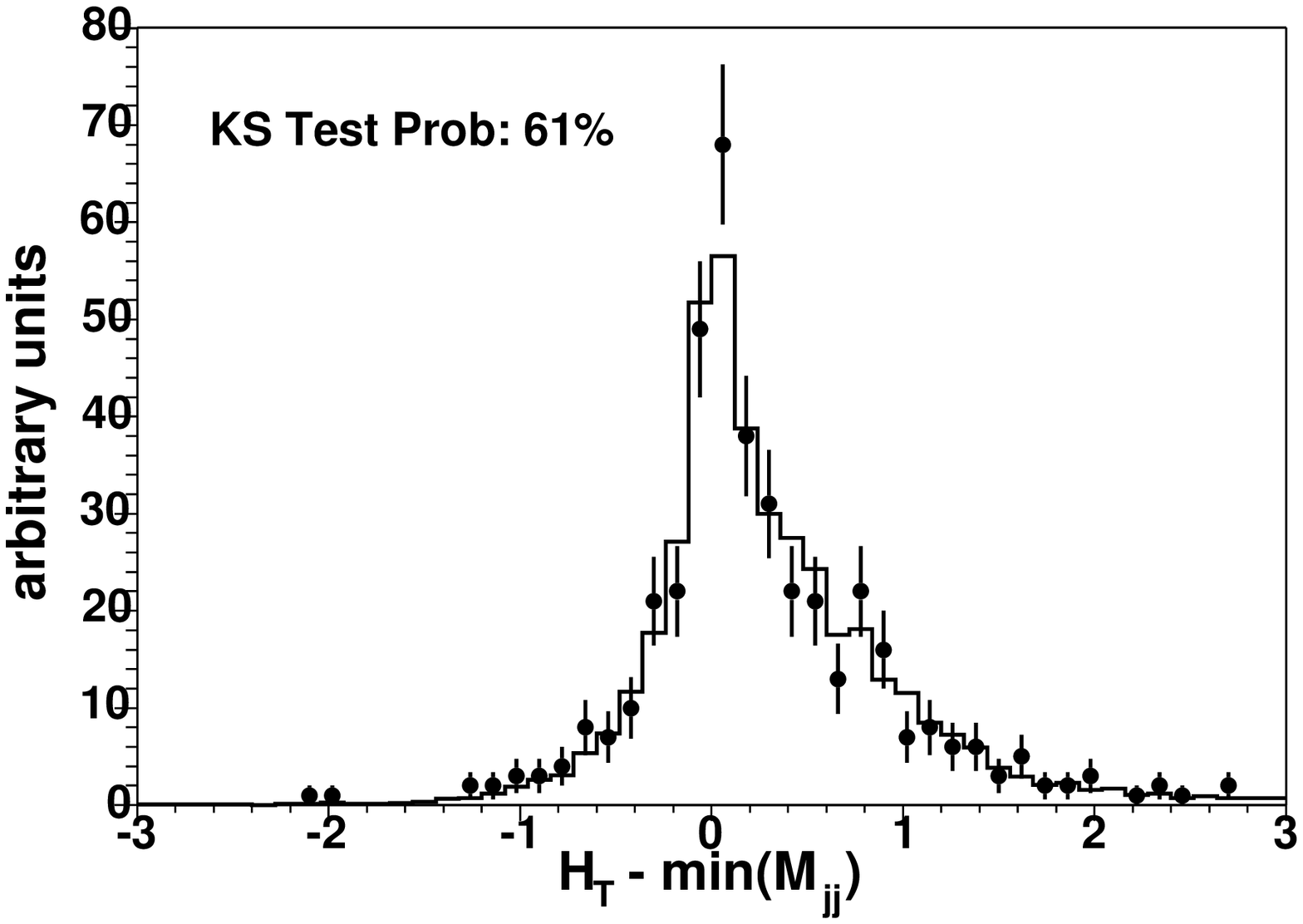}}
\resizebox{2.2in}{!}{ \includegraphics{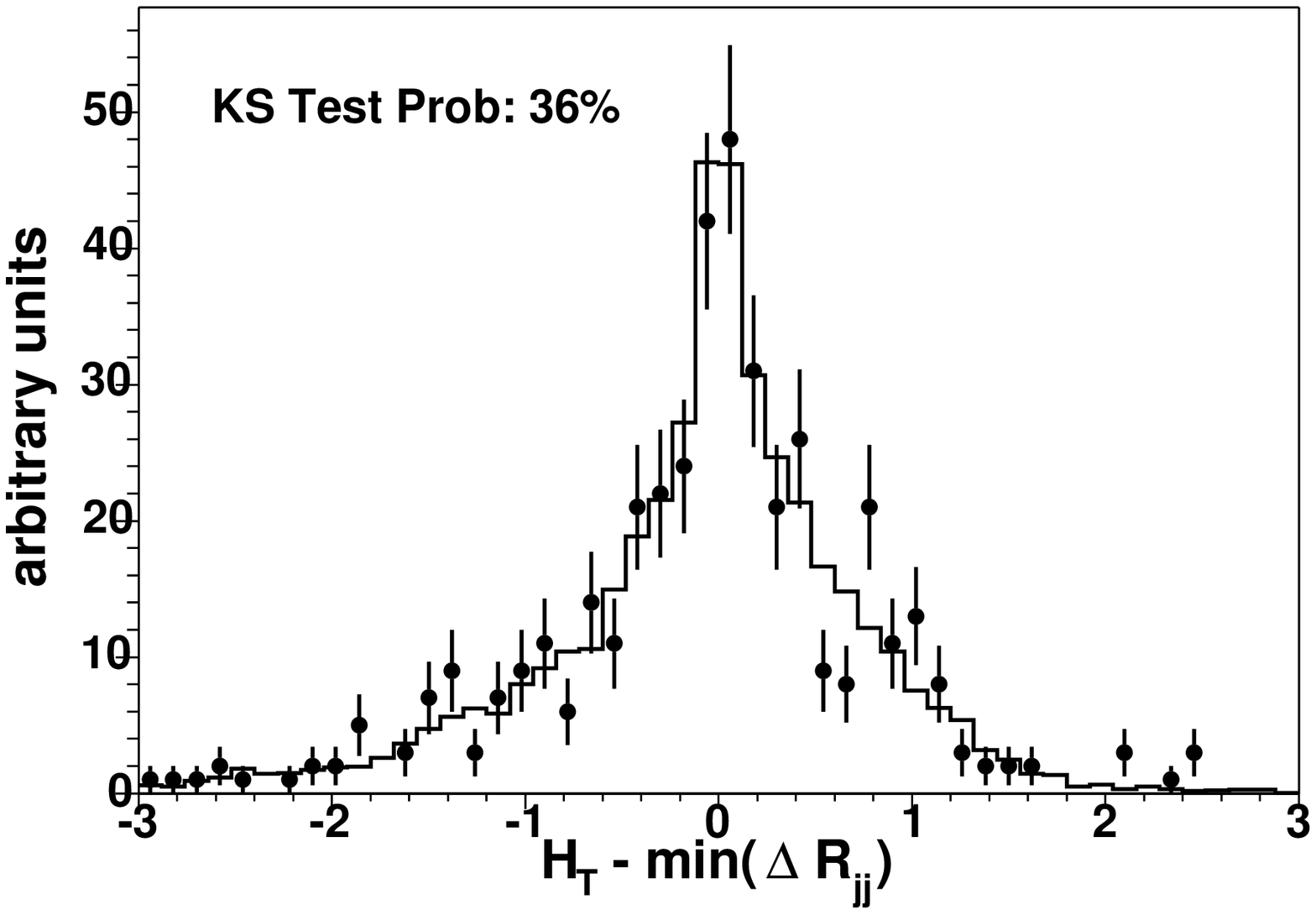}}
\resizebox{2.2in}{!}{ \includegraphics{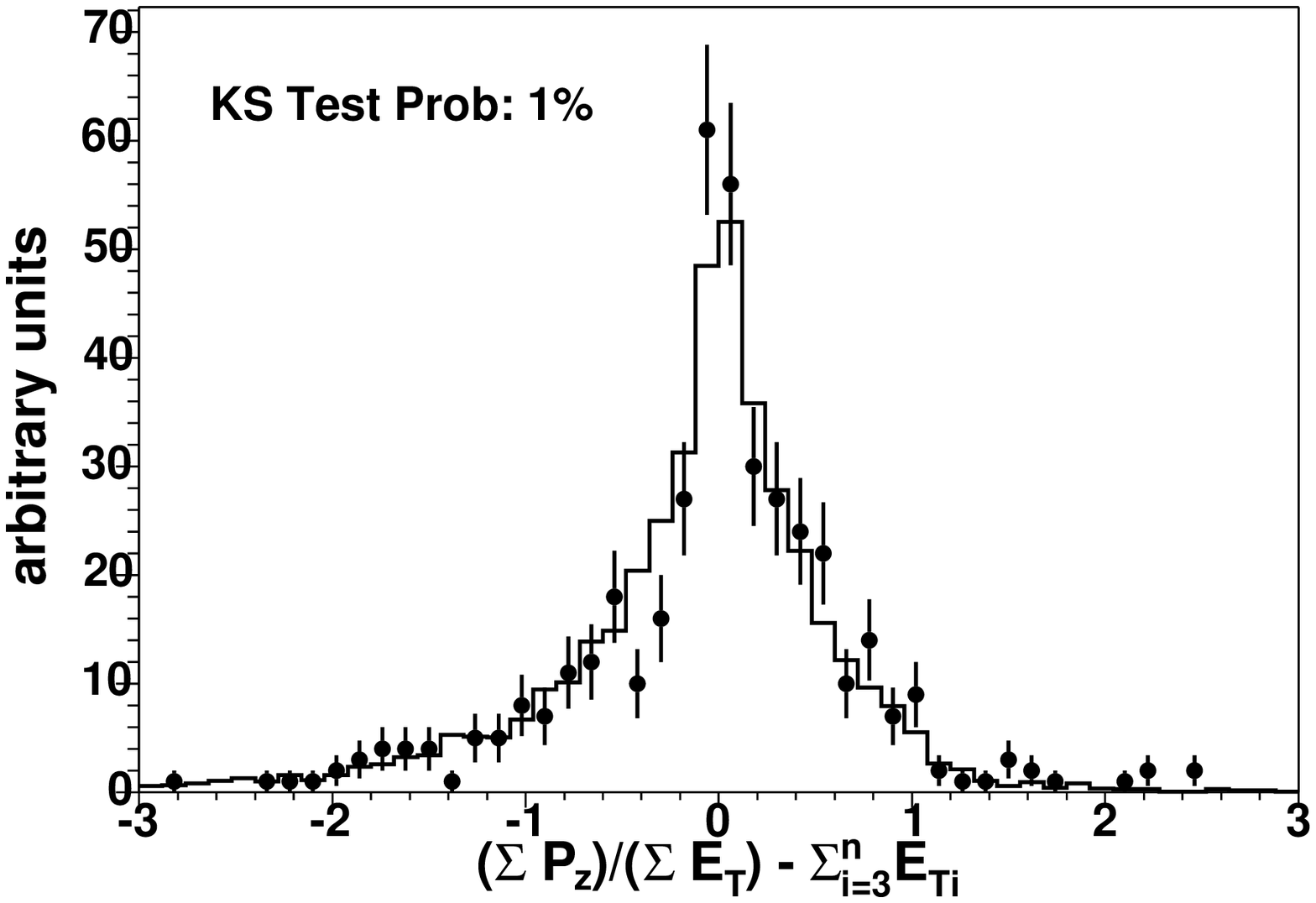}}
  \end{center}
  \caption{\label{fig:mc_validation_corr1} The distributions for the event-by-event correlations 
 between some of the seven ANN input properties for the $W+3$ exclusive jet events compared to the 
 predictions from ALPGEN+HERWIG $W$+3p Monte Carlo and PYTHIA \ttbar\ Monte Carlo. The model here is
  is a combination of 10\% \ttbar\ and 90\% $W$+jets simulated events.
 } 
  \end{figure*}

\begin{figure*}[btp]
  \begin{center}
\resizebox{2.2in}{!}{ \includegraphics{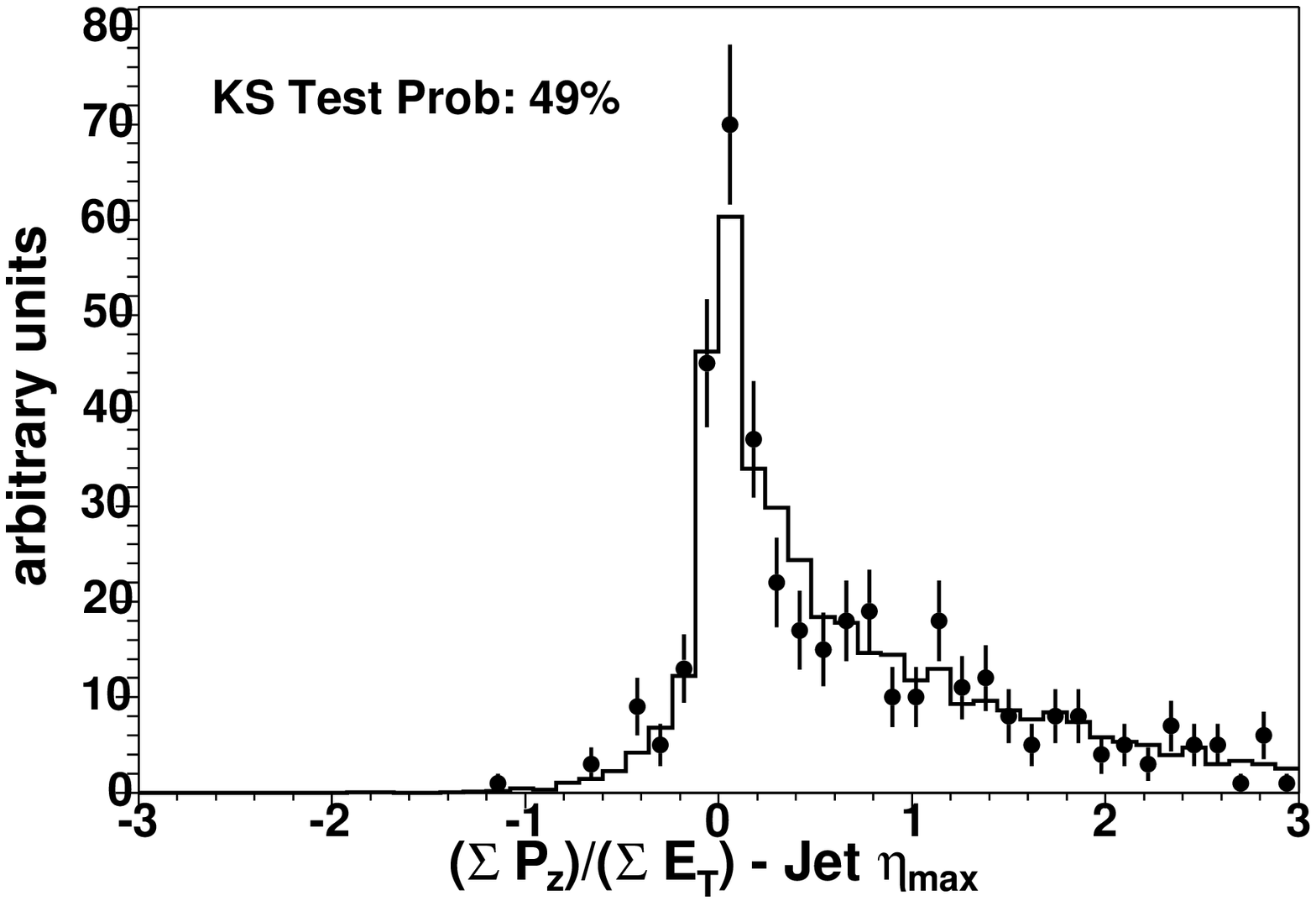}}
\resizebox{2.2in}{!}{ \includegraphics{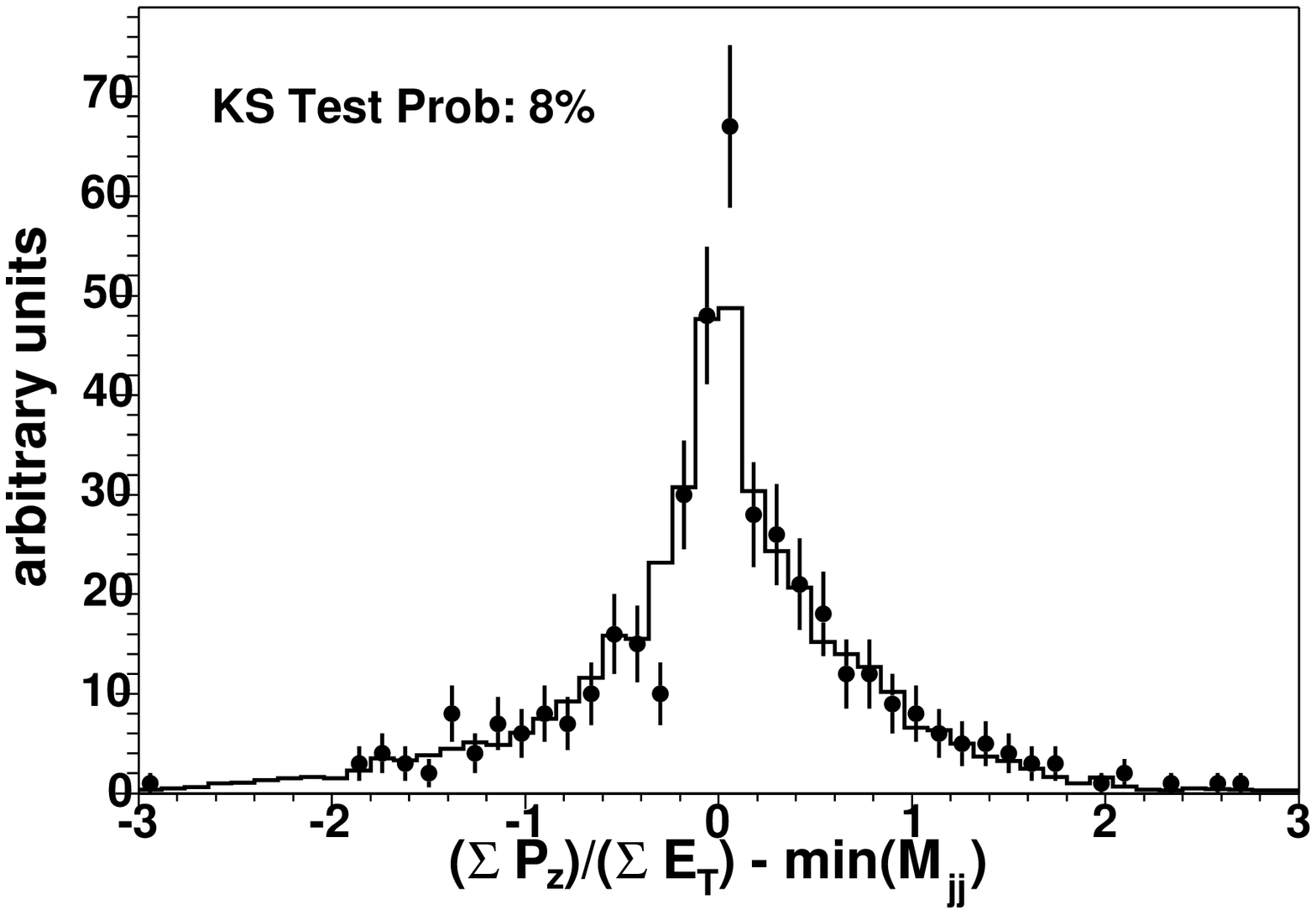}}
\resizebox{2.2in}{!}{ \includegraphics{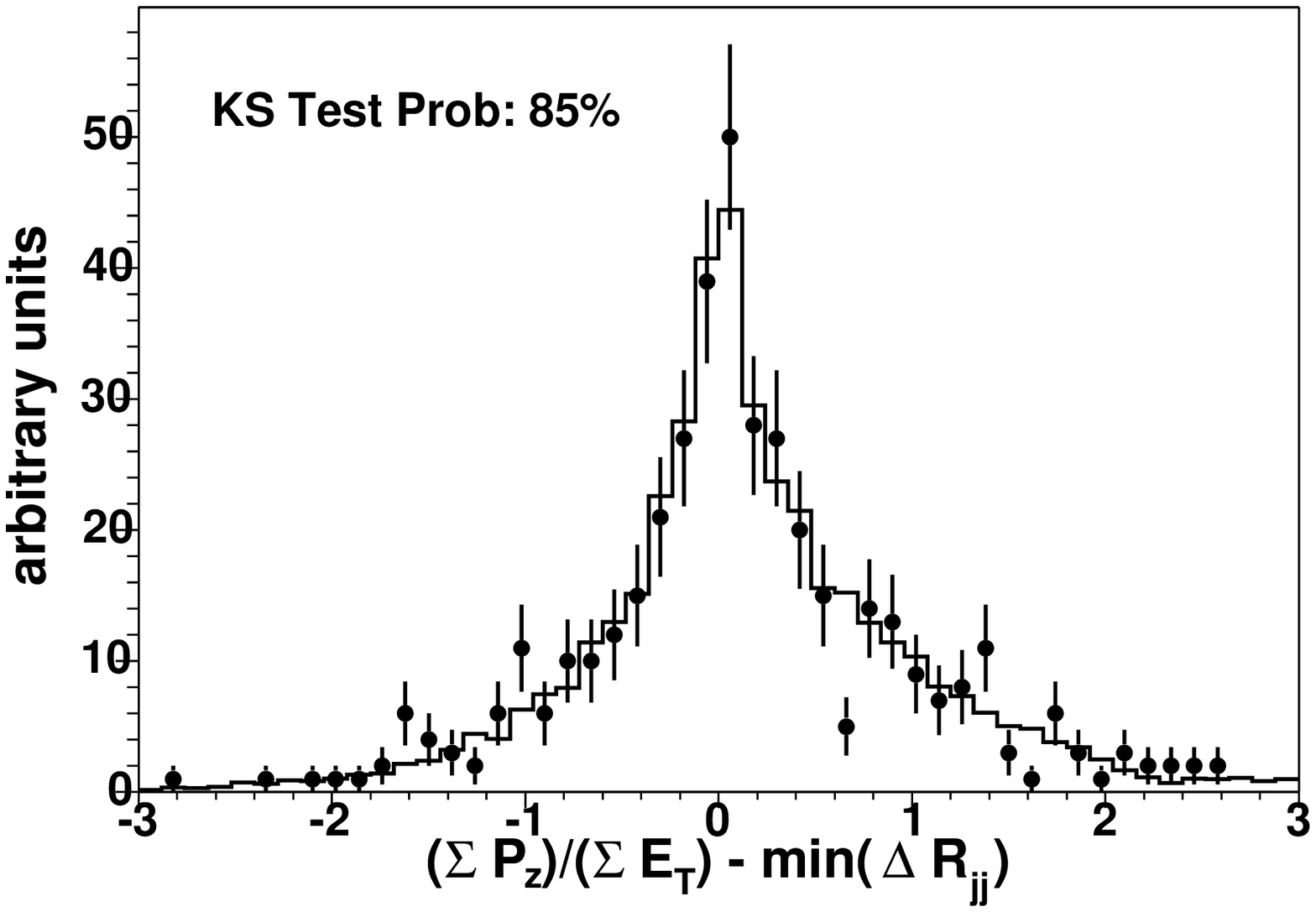}}
\resizebox{2.2in}{!}{ \includegraphics{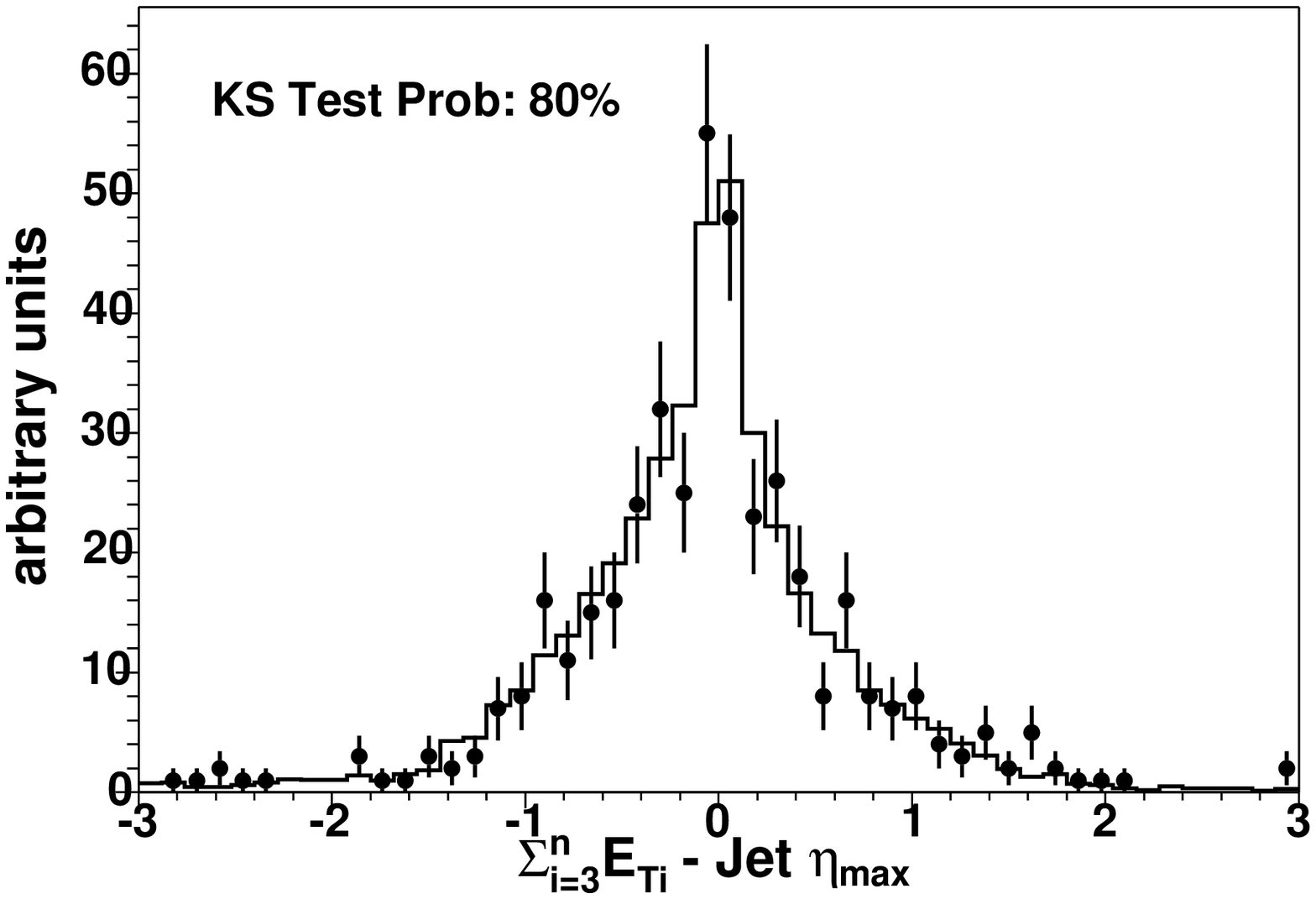}}
\resizebox{2.2in}{!}{ \includegraphics{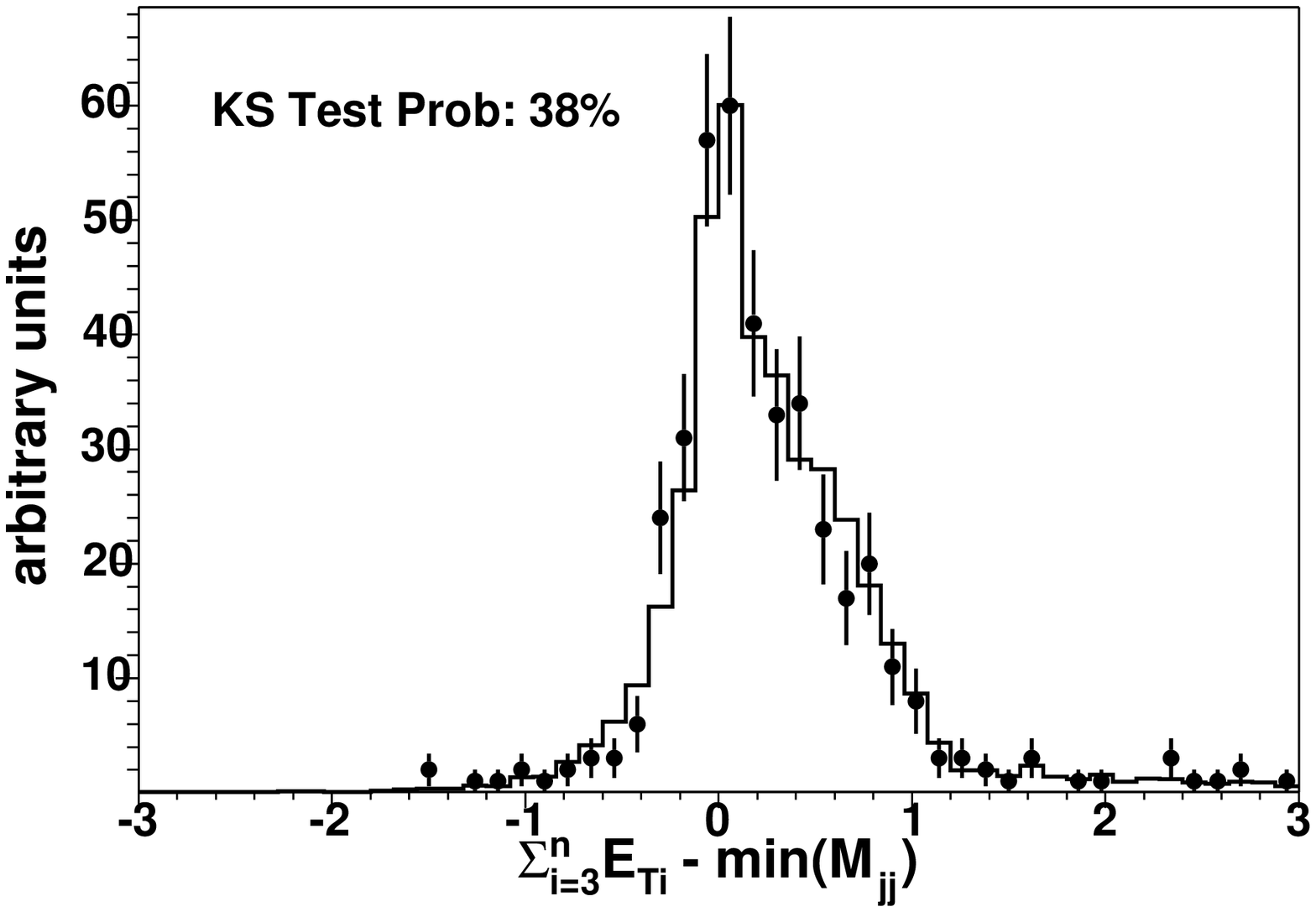}}
\resizebox{2.2in}{!}{ \includegraphics{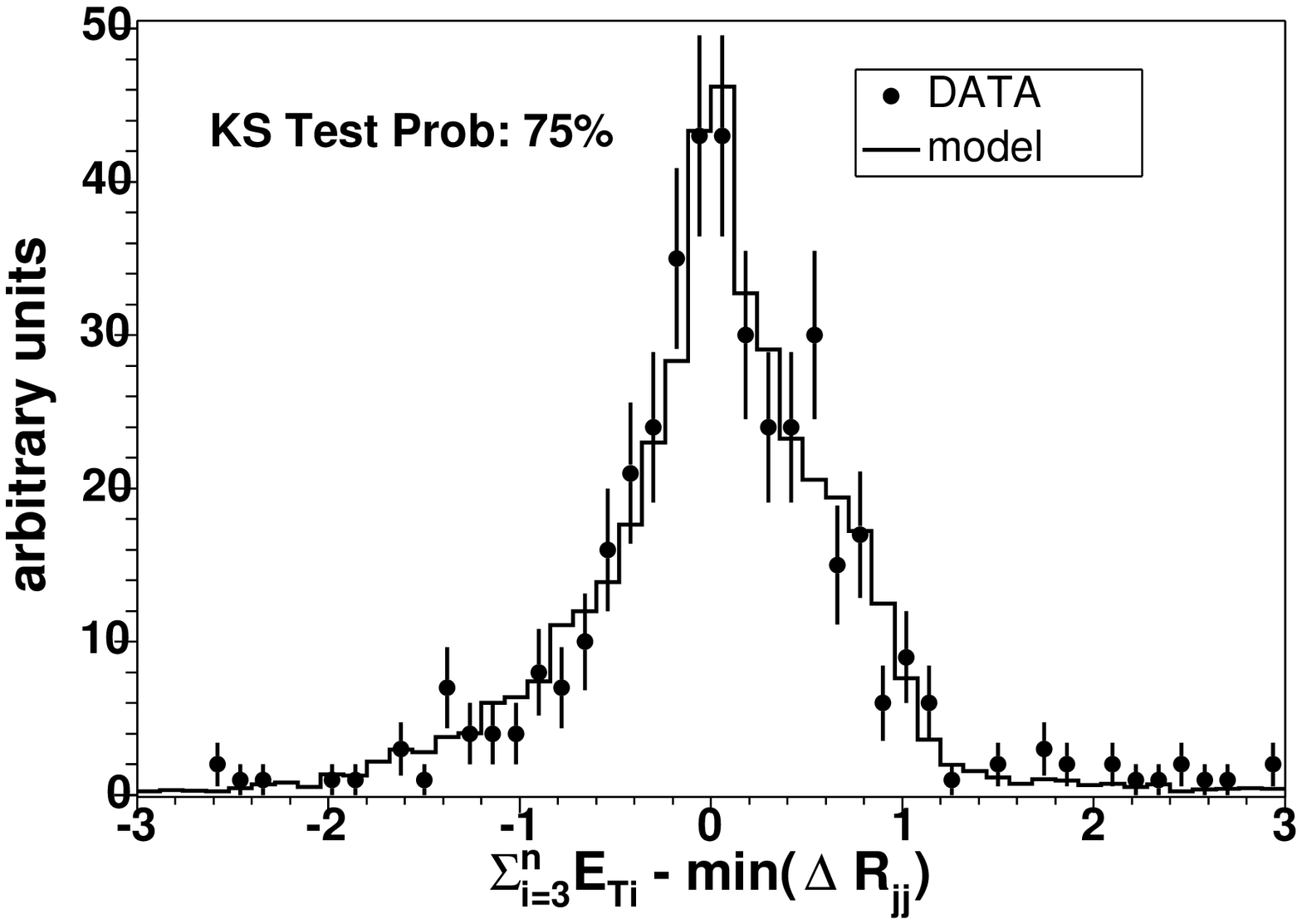}}
\resizebox{2.2in}{!}{ \includegraphics{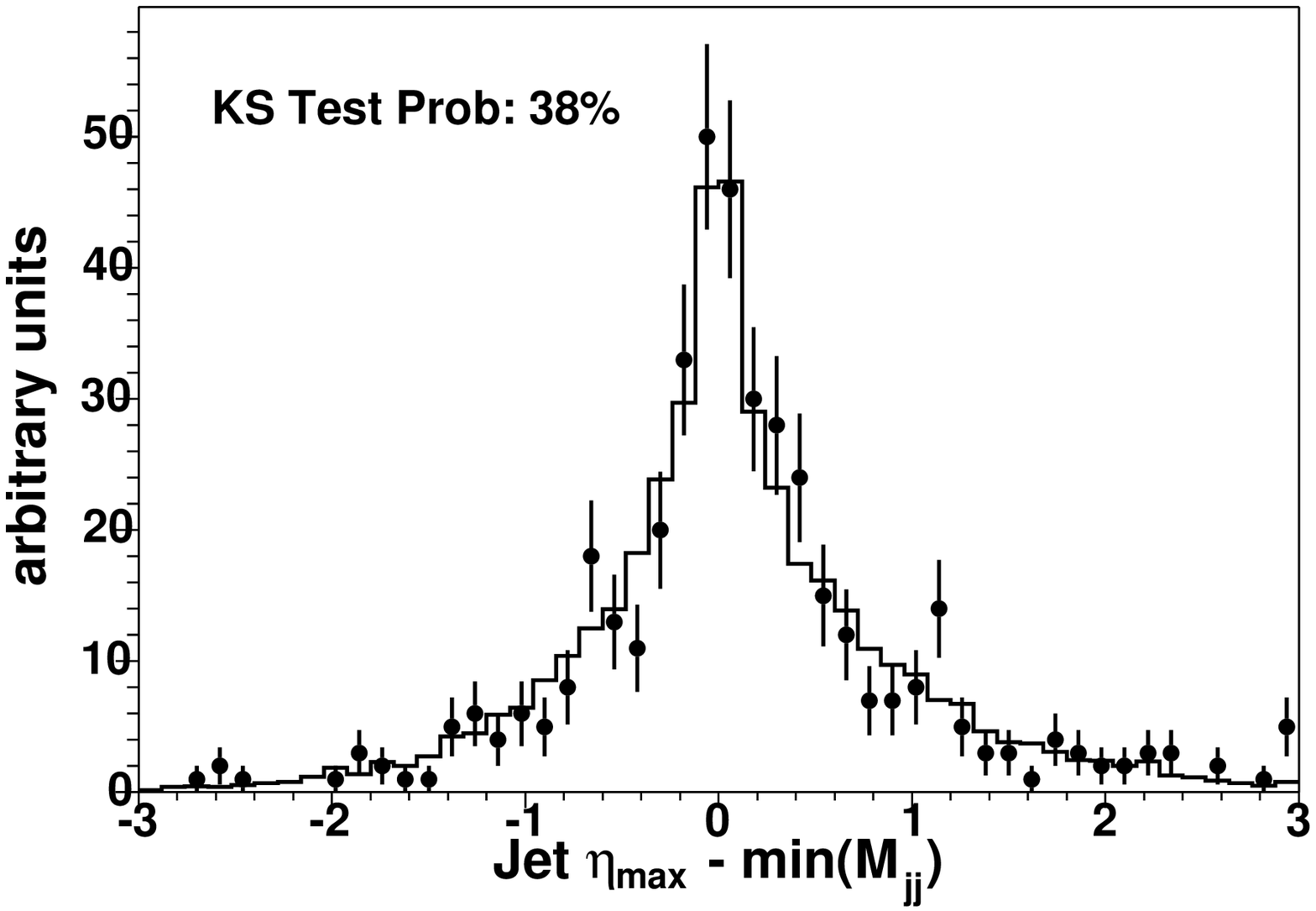}}
\resizebox{2.2in}{!}{ \includegraphics{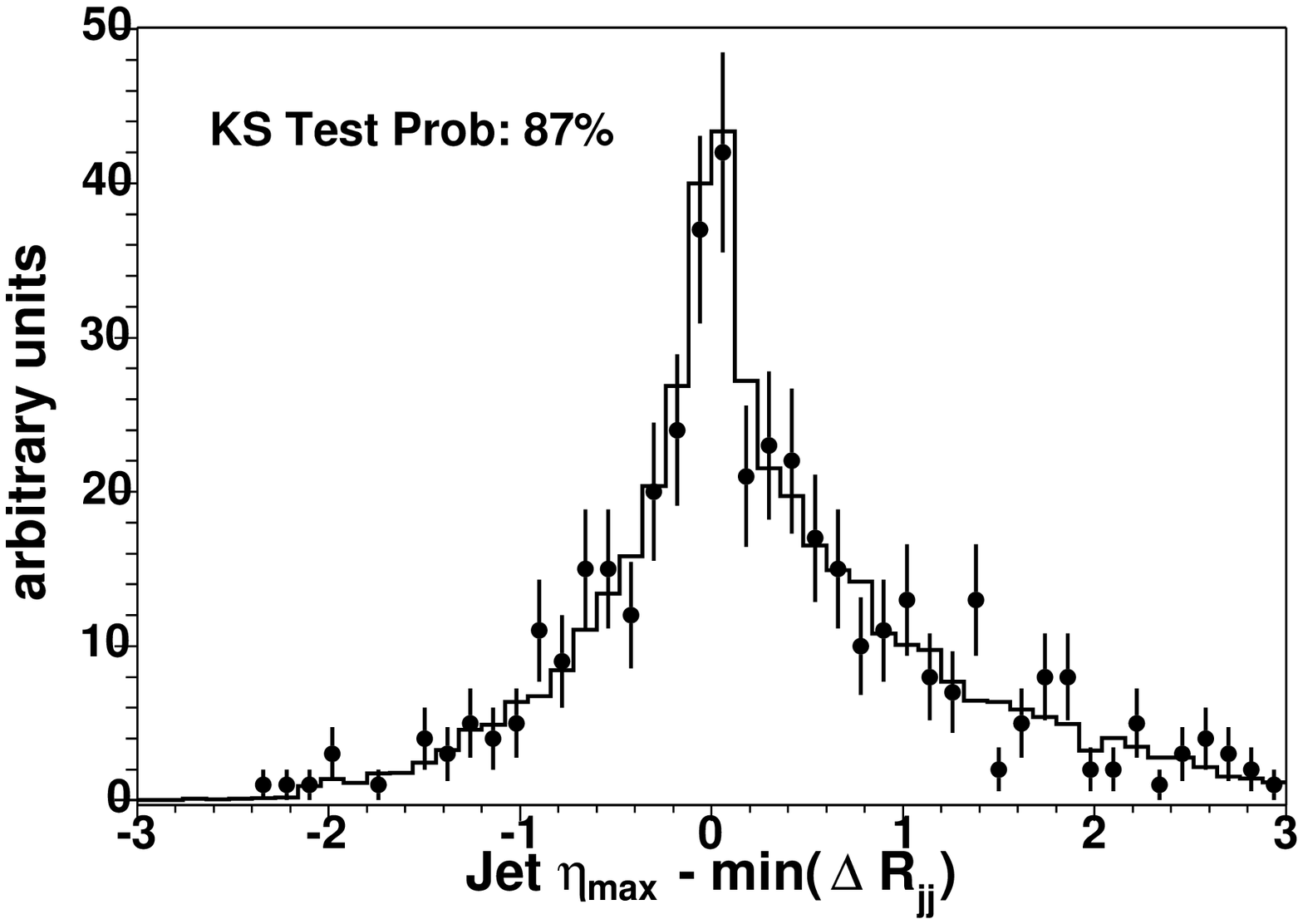}}
\resizebox{2.2in}{!}{ \includegraphics{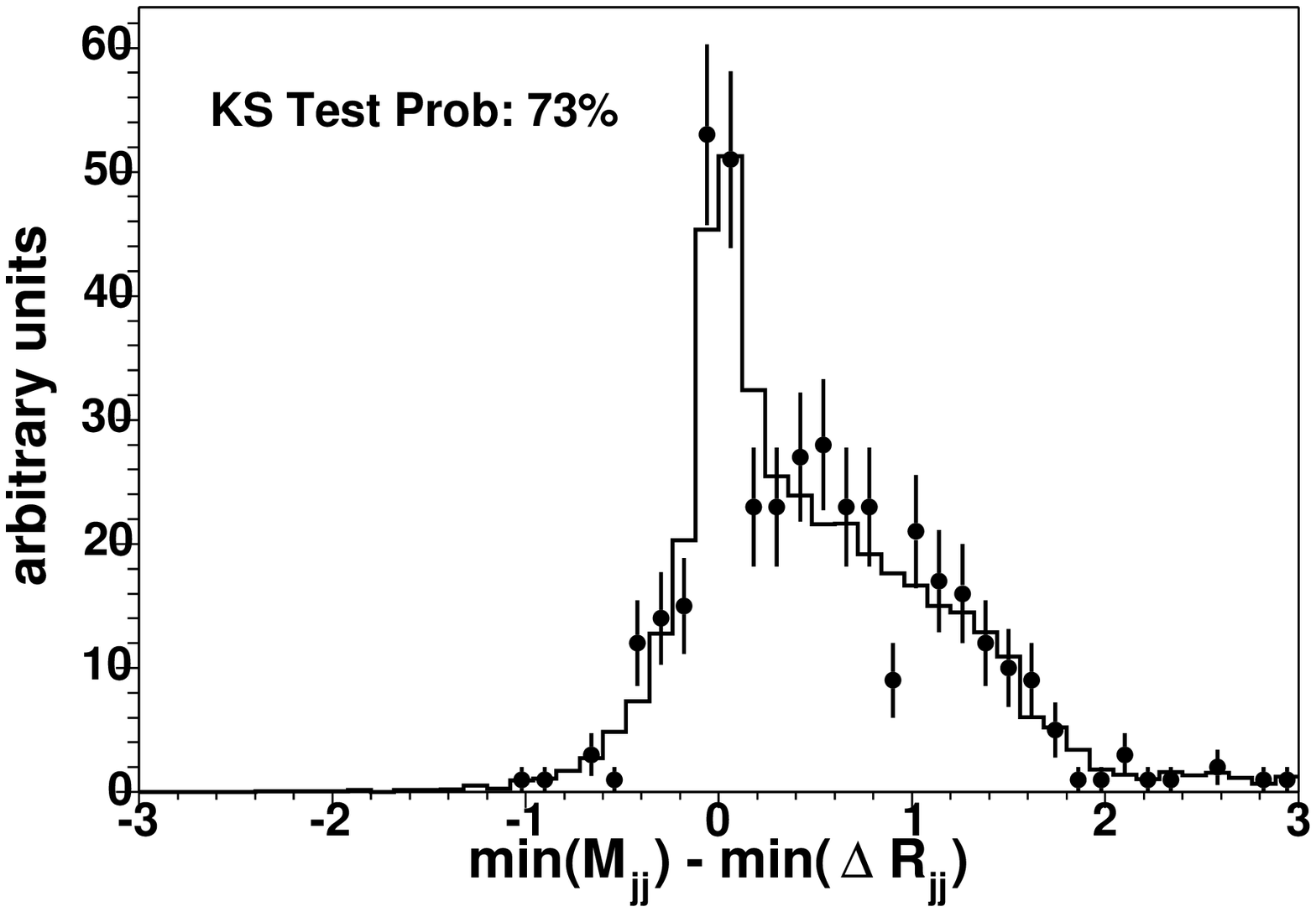}}
  \end{center}
  \caption{\label{fig:mc_validation_corr2} The distributions for the event-by-event correlations 
 between some of the seven ANN input properties for the $W+3$ exclusive jet events compared to the 
 predictions from ALPGEN+HERWIG $W$+3p Monte Carlo and PYTHIA \ttbar\ Monte Carlo. The model here is
  is a combination of 10\% \ttbar\ and 90\% $W$+jets simulated events.
 } 
  \end{figure*}

The correlations between the various kinematic and topological properties also provide information that we use 
in our multivariate approach. We have 
looked at the pair correlations for the 7 input properties used in the ANN. For two generic 
variables $x$ and $y$, a correlation variable $corr(x,y)$ is defined on event by event basis:
 \begin{equation}
 corr(x,y)= 
   \frac{(x-\overline{x})\cdot(y-\overline{y})}{(\Delta x\cdot\Delta y)^{1/2}},
 \end{equation}  
 \noindent where $\overline{x}$ is the average value in the $x$ variable and 
 $\Delta x = \overline{(x-\overline{x})^{2}}$.  Fig.~\ref{fig:mc_validation_corr1} and 
 Fig.~\ref{fig:mc_validation_corr2} show the distributions for the event-by-event correlations 
 between the 7 ANN input properties for $W+3$ exclusive jet events, compared to the 
 predictions from the ALPGEN+HERWIG $W$+3p Monte Carlo and PYTHIA \ttbar\ Monte Carlo. 
 The model here is a combination of 10\% \ttbar\ and 90\% W+jets simulated events.
Overall, the KS test values indicate agreement between data and the Monte Carlo simulation.

%------------------------------------------------------------------
\section{\label{sec:sys} Systematic Uncertainties}

Our measurement of the top pair production cross section is sensitive 
to systematic effects having an impact on the signal acceptance, on the 
shape of various kinematic distributions, and the luminosity. 
This last uncertainty is 5.9\%, 
where 4.4\% comes from the acceptance  and operation of the luminosity monitor and 4.0\% from the calculation of the total \ppbar\ 
cross section~\cite{cdflumi}.

Acceptance systematics fall into sub-categories of those that affect the efficiency of the trigger
 and lepton  identification, and those that affect the efficiency for passing the \met\ and 
jet ${E_T}$ cuts. We quote such systematics in percent (\%) as the relative change 
in the \ttbar\ acceptance:

\begin{itemize}

\item {\bf Lepton Identification Efficiency}.  
For electrons, we consider the uncertainties on the electron energy scale and resolution,
the electron momentum scale and resolution, the amount of material in the detector, and the conversion removal efficiency.
For muons, we consider the uncertainties on the muon momentum scale and resolution, 
the modeling of geometrical coverage of the muon detectors, and the cosmic ray removal efficiency. 
We estimate an uncertainty of 2\% from these effects.

\item {\bf Lepton Isolation}.
$Z\rightarrow\ell\ell$ candidates provide a clean sample of high \pt\ leptons that can be used to
estimate a correction factor for the difference in lepton identification efficiency between data 
and simulation. However, the leptons from \ttbar\ decays tend to be less isolated than the leptons 
from $Z$\ decays. To account for this different environment, we calculate the correction factor 
as a function of lepton isolation for $Z$\ events and then use the lepton isolation distribution 
in \ttbar\ PYTHIA Monte Carlo to obtain an appropriately weighted correction factor. We estimate an 
uncertainty of 5\%, which is dominated at the present time by the small statistics in the $Z$\ data sample.

\item {\bf Jet Energy Scale.} 
We estimate an uncertainty of 4.7\%, the average of the changes in acceptance
from shifting the jet energy scale by the uncertainty discussed in Section~\ref{subsec:jet}. 

\item {\bf ISR/FSR.} Jets due to initial state gluon radiation (ISR) may be produced in addition 
to the jets from the top decay products. We estimate the uncertainty associated with the modeling 
of ISR by taking half the change in acceptance for two Monte Carlo samples. These have different 
$\Lambda_{\textrm QCD}$ values and K-factors for the transverse momentum scale of the ISR evolution.
The range of variation\footnote{We vary PYTHIA parameters PARP(61) from 0.100 to 0.384~MeV 
(default 0.192~MeV) and PARP(64) from 0.5 to 2.0 (default 1.0) respectively.} 
was determined by taking the extremes of a range determined by a study of Drell-Yan 
$Z \rightarrow \ell\ell$\ events in data and Monte Carlo. The uncertainty from the modeling of 
final state gluon radiation (FSR) is estimated by applying these same variations to the FSR evolution. 
In addition to the hard scattering process and initial and final state radiation, 
remnants of the proton and anti-proton interaction affect event kinematics.  
We use a PYTHIA \ttbar\ Monte Carlo sample, where the parameters used to 
describe the charged particle multiplicity in di-jet data have been re-tuned assuming less ISR~\cite{RickFieldref}, to estimate 
the uncertainty from the modeling of the underlying event. We estimate a total uncertainty of 3\% from these effects.

\item {\bf Parton Distribution Functions.}  The uncertainty in the distribution of the proton(anti-proton) momentum
amongst its constituent partons affects the relative contributions of the $q\bar{q}$\ and $gg$\ 
processes to \ttbar\ production as well as the momentum of the \ttbar\ system. 
In the CTEQ parametrization the parton distribution functions (PDFs) are described by
20 independent eigenvectors. In a next to the leading order (NLO) version of PDFs, CTEQ6M, 
a 90\% confidence interval is provided for each eigenvector.  
For the maximum and minimum value of each eigenvector, 
we compute a new acceptance by re-weighting our default CTEQ5L PYTHIA \ttbar\ sample.
We add in quadrature the difference between the weighted acceptance for the twenty eigenvectors with respect to 
the weighted acceptance from the central CTEQ6M and find an uncertainty of 0.5\%.  
The dominant contribution is from the eigenvector most closely associated with
the gluon distribution function at large-$x$, which changes the contribution of the $gg$\ process from 11\% to 21\%.  
We also take the difference of 1.0\% between the acceptance for the leading order CTEQ5L, with a 5\% contribution from the $gg$\ process,
and the central value from next to leading order CTEQ6M, with a 15\% contribution from the $gg$\ process.
We find a consistent value for the acceptance from CTEQ5L and the alternative 
MRST set~\cite{MRST}.  The uncertainty from $\alpha_{s}$\ is estimated by comparing the weighted
 acceptance for MRST with $\alpha_{s}=0.1125$\ and $\alpha_{s}=0.1175$, which is 1.0\%. 
Adding these three contributions in quadrature, we obtain a total uncertainty of 1.5\%.

\item{\bf Generator.} We compare PYTHIA to HERWIG, after correcting for the lack 
of QED FSR from leptons in HERWIG and for the default HERWIG $W \rightarrow \ell \nu$\ branching 
ratio of 11.1\%. We find an acceptance uncertainty due the choice of \ttbar\ event generator 
of 1.4\%.

\end{itemize}

To evaluate the effect of systematic changes in the shapes of kinematic distributions,
we use simulated experiments, as described in Section~\ref{subsec:1var}.
In this case, we fit the simulated ``data" distribution to signal and background distributions 
from our default model, and also to signal and background distributions from a model with a particular 
systematic effect applied. For example, an alternative shape for the ANN output distribution is 
obtained by processing a set of Monte Carlo simulated events modified according to a particular 
systematic effect with the network trained using the nominal Monte Carlo samples. 
The average difference in the fitted number of signal events,
relative to the expected number listed in Table~\ref{tab:ndata}, is quoted in percent (\%) as a systematic uncertainty.  
\begin{itemize}

\item {\bf Jet Energy Scale.} A change in the jet energy scale affects the total transverse energy and simultaneously 
five of the seven kinematic properties used in the ANN.  Fig.~\ref{fig:jes_htann} demonstrates that for an increase in the jet energy scale, the \sht\ distribution for the \ttbar\ signal shifts upward significantly, while the 
the distribution for the $W$+jets background remains almost unchanged. This is due to the large number of $W$+jets background events 
adjacent to the event selection threshold.  For instance, a systematic increase in the jet energy scale means
that many $W$+jets background events with a third jet that previously just failed the kinematic requirement 
will now pass the event selection.  These new events tend to have low values of \sht\ and so compensate 
for the increased \sht\ of the original $W$+jets background events.  
Fig.~\ref{fig:jes_htann} also shows that the better separation afforded by 
the ANN means that the ANN technique is less sensitive to this effect.
We estimate an uncertainty of 26\% for $H_{T}$\ and 17\% for the ANN.

\begin{figure}[htbp]
  \begin{center}
\resizebox{3.2in}{!}{ \includegraphics{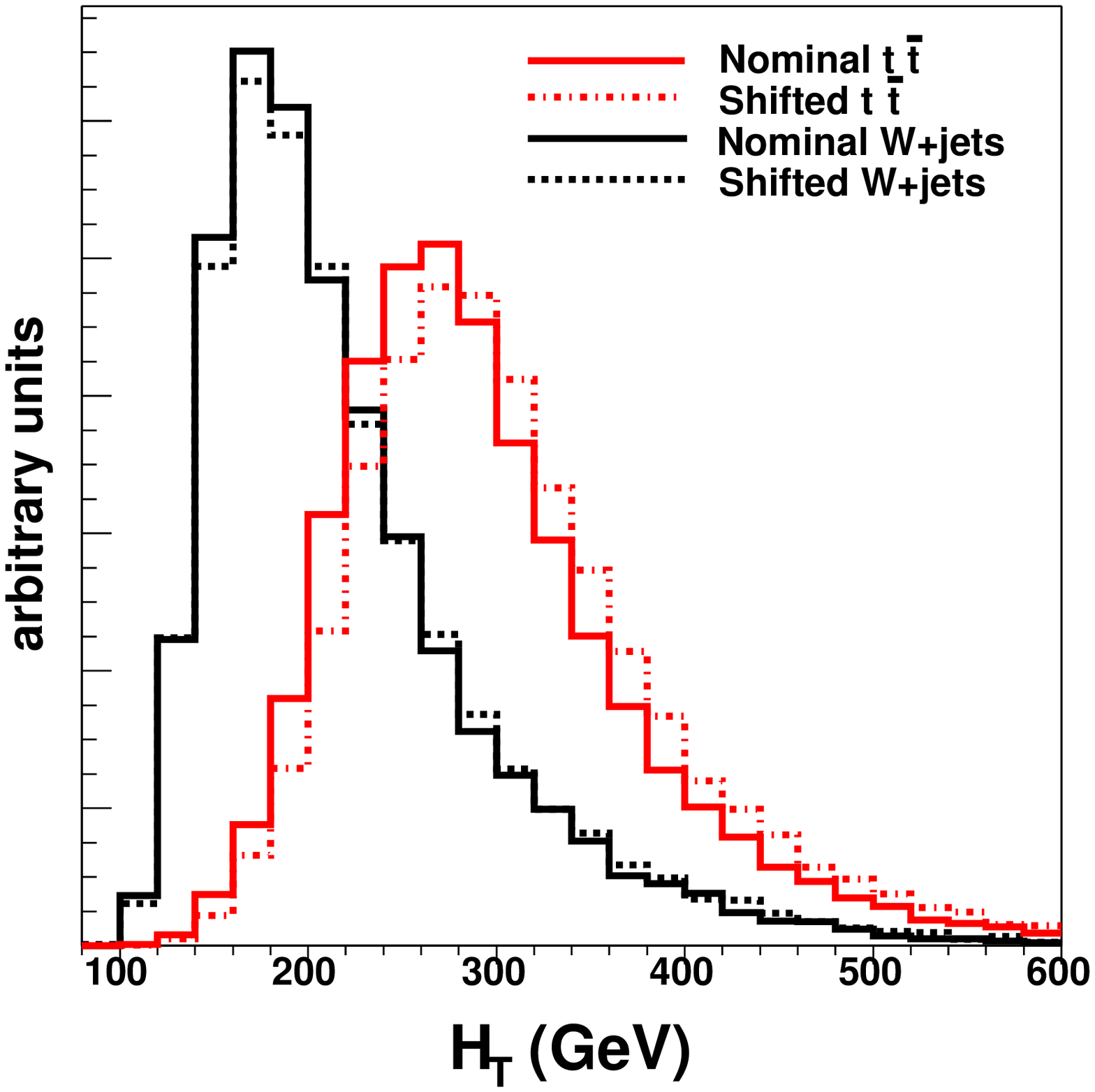}}
\resizebox{3.2in}{!}{ \includegraphics{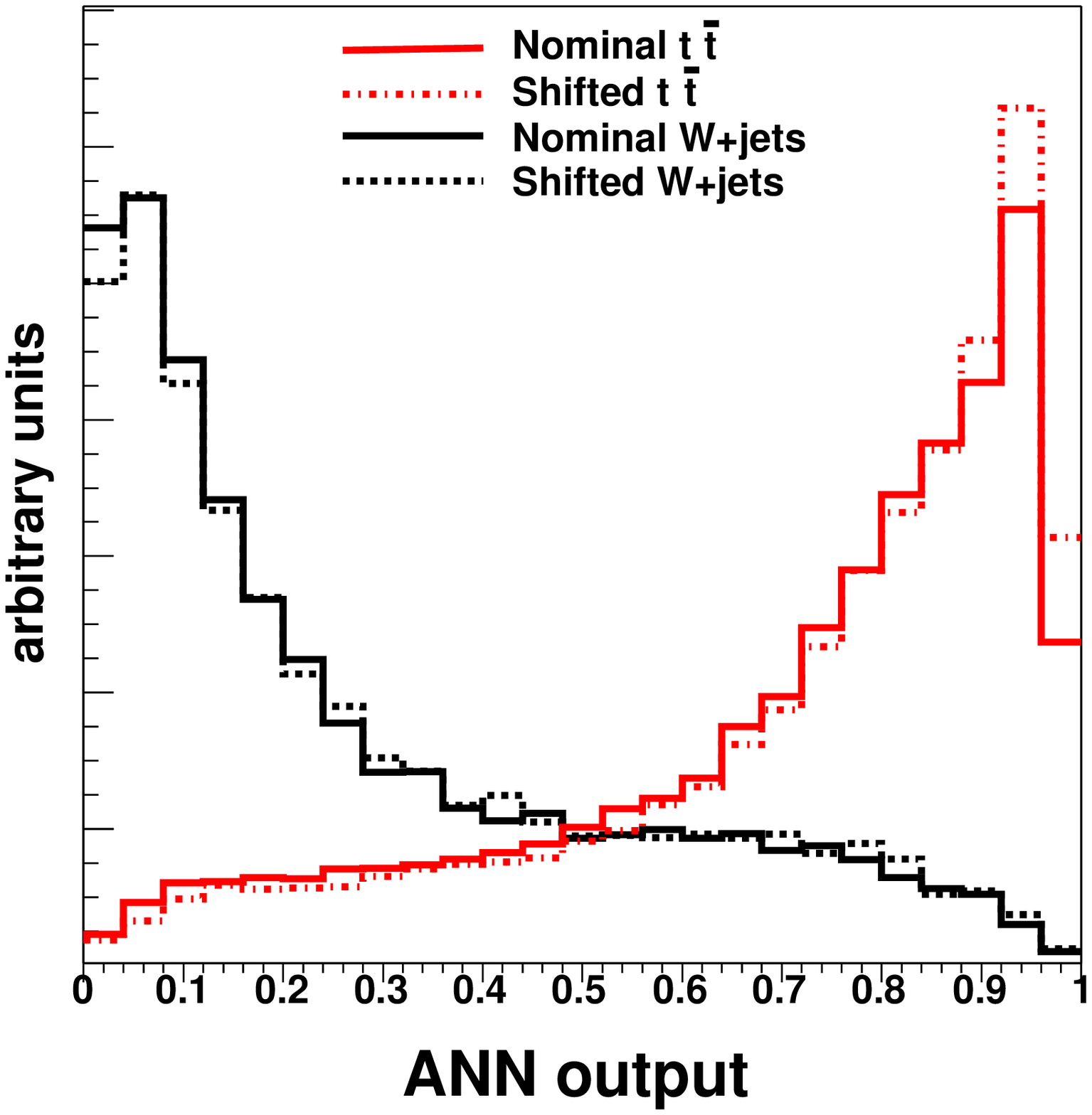}}
  \end{center}
  \caption{\label{fig:jes_htann} The \sht\ and ANN distributions for
the default jet energy scale and a positive shift corresponding to the uncertainty on the jet energy scale. 
The distributions are normalized to equal area.
}
\end{figure}

\item $\mathbf{W}${\bf +jets background.}
The uncertainty on the $W$+jets background shape is calculated from ALPGEN+HERWIG samples having 
different values for the scale of momentum transfer, $Q^{2}$,  in the hard scattering process. 
This affects the initial parton distribution functions and the relative weight of diagrams in 
the leading order matrix element.  We find that the largest change arises between using our 
default $Q^{2}= M_{W}^{2}+\sum_{i} p_{T,i}^{2}$, 
which changes on  an event-by-event basis, and setting $Q^{2}=4M_{W}^{2}$ which 
is the same for every event. We estimate an uncertainty of 24.6\% for $H_{T}$\ and 10.2\% for the ANN.  

\item {\bf QCD multi-jet background.} We first recall that we expect electrons 
from unidentified photon conversions to form a large fraction of this background in the electron 
channel, as we discussed in Section~\ref{sec:bkg}.  Therefore, we use the identified conversions in data to provide a model alternative 
to our default electron and muon non-isolated data samples. 
For the uncertainty on the multi-jet background 
normalization, we vary the contribution by $^{+100}_{-50}$\% around the central value 
listed in Table~\ref{tab:qcdbkg}.  We assign this level of uncertainty from the difference between 
our estimates listed in Table~\ref{tab:qcdbkg} and the amounts of multi-jet
background preferred by a fit to the higher statistics $W$+1 jet and $W$+2 jet regions in Section~\ref{sec:mcmodel}.

\item {\bf Other electroweak backgrounds.} We estimate this systematic as 
half the difference between including and not including these backgrounds in our model of the \sht\ and ANN output shape.

\end{itemize}

The systematic uncertainties are summarized in Table~\ref{tab:allsys_1d} for \sht\
and in Table~\ref{tab:allsys_ann} for ANN. When the same systematic effect 
has an impact on both the \ttbar\ acceptance and the shape of the \ttbar\ kinematic distributions, 
we treat the uncertainties as 100\% correlated and calculate the total uncertainty by adding the acceptance and shape 
systematic numbers linearly. 
For multiple component systematic uncertainties like those from PDFs, ISR and FSR, the acceptance and shape uncertainties for each
component are first combined linearly, then the components are added in quadrature. Finally, 
the overall systematic uncertainty is obtained by adding the total contributions from uncorrelated effects in quadrature.

\begin{table}[!hbtp]
\caption{\label{tab:allsys_1d}
Systematic uncertainties in \% on the cross section, for 
fits to the total transverse energy, \sht, in the $W+\geq$3 jets sample. 
The overall uncertainty is given by the sum in quadrature of the numbers in the last column.
} 
\begin{ruledtabular}
\begin{tabular}{lccc} 
Effect                      & Acceptance (\%) 	& Shape (\%)     & Total (\%)      \\ 
\hline
Jet $\et$ Scale             & 4.7	 	& 21.4    & 26.1     	\\
W+jets Q$^2$ Scale          & -       		& 24.6    & 24.6       	\\
QCD fraction		    & -     		& 2.4     & 2.4 	\\
QCD shape		    & -     		& 4.5     & 4.5 	\\
Other EWK		    & -			& 1.8	  & 1.8		\\
\ttbar\ PDF                 & 1.5		&  2.2    & 4.7		\\
\ttbar\ ISR		    & 2.1		&  1.1	  & 2.9 	\\	
\ttbar\ FSR		    & 1.7		&  1.5 	  & 3.7 	\\
\ttbar\ generator	    & 1.4		&  1.0    & 2.4		\\
Lepton ID/trigger           & 2.0		&  -      & 2.0		\\
Lepton Isolation	    & 5.0		& - 	  & 5.0 	\\
Luminosity       	    &  -          	& -       & 5.9     	\\
\hline
%Total                      &  7.9          	& 33.2        & 37.8    \\ 
Overall                     &			&	   &37.8        	        \\
\end{tabular} 
\end{ruledtabular}
\end{table}

\begin{table}[!btp]
\caption{\label{tab:allsys_ann}
Systematic uncertainties in \% on the cross section, for fits to the ANN output distribution
 in the $W+\geq$3 jets sample. The overall uncertainty is given by the sum in quadrature
of the numbers in the last column.
}
\begin{ruledtabular}
\begin{tabular}{lccc} 
Effect                      & Acceptance (\%) 	& Shape (\%)     & Total (\%)      \\ 
\hline
Jet $\et$ Scale             & 4.7	 	& 12.2    & 16.9     	\\
W+jets Q$^2$ Scale          & -       		& 10.2    & 10.2       	\\
QCD fraction		    & -     		& 0.6     & 0.6 	\\
QCD shape		    & -     		& 1.1     & 1.1 	\\
Other EWK		    & -			& 2.0	  & 2.0		\\
\ttbar\ PDF                 & 1.5		& 2.9     & 4.4		\\
\ttbar\ ISR		    & 2.1		& 1.9	  & 3.0 	\\	
\ttbar\ FSR		    & 1.7		& 1.0 	  & 2.7 	\\
\ttbar\ generator	    & 1.4		& 0.3     & 1.7		\\
Lepton ID/trigger           & 2.0		& -       & 2.0		\\
Lepton Isolation	    & 5.0		& - 	  & 5.0 	\\
Luminosity       	    &  -          	& -       & 5.9     	\\
\hline
%Total                       &  7.9          	& 16.5        & 22.3     	\\ 
Overall                      &   		& 	  & 22.3       	        \\
\end{tabular}
\end{ruledtabular}
\end{table}

%\begin{figure}[htbp]
%  \begin{center}
%\resizebox{3.2in}{!}{ \includegraphics{fig10.eps}}
%  \end{center}
%  \caption{\label{fig:jesys} Jet energy scale systematics as function of the signal cross section, 
%  $\sigma_{\ttbar}$ for \sht\ fits.
%The circles(squares) show the shifts in the mean fitted signal contribution for a positive(negative) shift in the jet 
%energy scale.  The triangles represent the average shift.
%  } 
%  \end{figure}

%------------------------------------------------------------------
\section{\label{sec:res} Results}

\begin{figure}[htbp]
  \begin{center}
\resizebox{3.0in}{!}{ \includegraphics{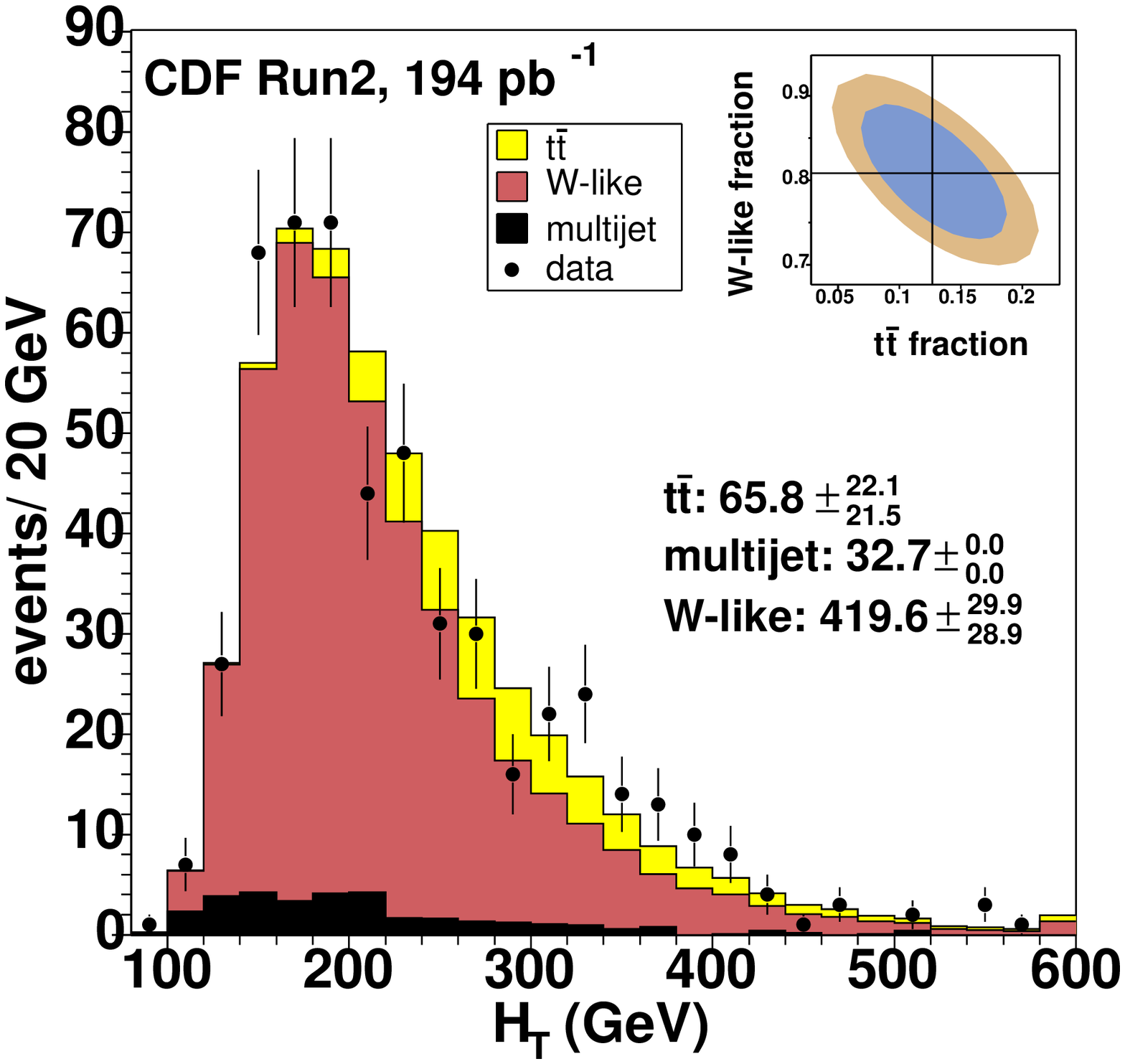}}
% \resizebox{3.0in}{!}{ \includegraphics{Win04fit-HtT3J.eps}}
  \end{center}
  \caption{\label{fig:result_ht_3j} Distribution of observed $\sht$ 
           in the $W+\geq$3 jets sample, compared with the
           result of the fit.  The inset shows the 1- and 
           2-standard-deviation contours of the free parameters in 
           the fit, normalized to the total number of observed events. 
The contribution of multi-jet background to the fit is fixed.}
\end{figure}

\begin{figure}[htbp]
  \begin{center}
\resizebox{3.0in}{!}{ \includegraphics{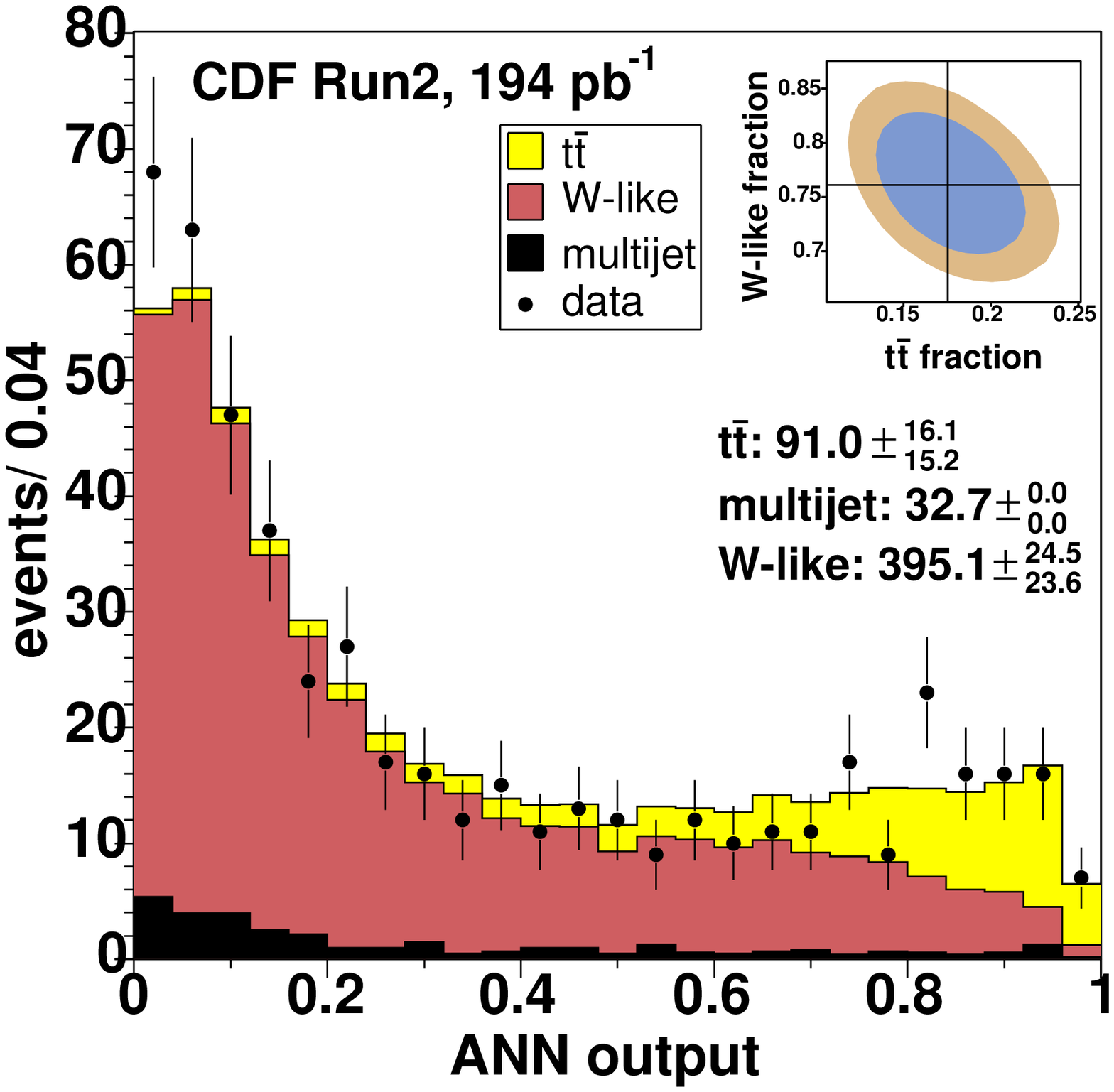}}
  \end{center}
  \caption{\label{fig:result_ann_3j} Distribution of observed ANN output 
           in the $W+\geq$3 jets sample, compared with the
           result of the fit.  The inset shows the 1- and 
           2-standard-deviation contours of the free parameters in 
           the fit, normalized to the total  number of observed events. The contribution of multi-jet background to the fit is fixed.}
\end{figure}

We have applied the method described in Section~\ref{sec:kinvar} to a dataset with an 
integrated luminosity of 194 pb$^{-1}$, where 519 events pass the $W+\geq$3\ jets selection criteria (Table~\ref{tab:ndata}).  
Figures~\ref{fig:result_ht_3j} and \ref{fig:result_ann_3j} show the distribution of data 
events for the single property, \sht\, and the output of an ANN respectively. We maximize the likelihood of 
Equation~\ref{eqn:lik} to extract the most probable number of \ttbar\ signal events:

\begin{center} 
   $\mu_{\ttbar} = 65.8 \pm  21.8$ ($H_{T}$),
\\ $\mu_{\ttbar} = 91.0 \pm  15.6$ (ANN), 

%%   $f_{\ttbar} = 0.124 \pm  0.042$ ($H_{T}$),  %% * 519
%%\\ $f_{\ttbar} = 0.176 \pm  0.030$ (ANN),      %% * 519
\end{center}

\noindent where the uncertainty is statistical only and we have assumed a top mass of 175 GeV/$c^{2}$. 
Using our estimate of 7.11$\pm$0.56\%\ for the \ttbar\ acceptance and 194$\pm$11~\pbi\ for the integrated luminosity
in Equation~\ref{eqn:llk}, we measure a top pair production cross section of:
 \begin{center}
 $\sigma_{\ttbar} = 4.8 \pm 1.6 \pm 1.8$~pb ($H_{T}$),
\\ $\sigma_{\ttbar} = 6.6 \pm 1.1 \pm 1.5$~pb (ANN),
 \end{center}
where the uncertainties are statistical and systematic, respectively. These results agree well with the 
theoretical prediction of $6.7\pm^{0.7}_{0.9}$~pb \cite{mlm} for a top mass of 175 GeV/$c^{2}$. 
From simulated experiments with this top mass, we estimate a probability of 10\% to find a difference equal to or larger 
than the observed difference between the results from the correlated \sht\ and ANN distributions.  
The observed 33\% statistical uncertainty for the \sht\ fit is slightly larger than we would 
expect in 68\% of simulated experiments.  However, the observed 17\% uncertainty for the 
ANN fit is close to the median from simulated experiments.

\begin{figure}[htbp]
  \begin{center}
  \resizebox{3.75in}{!}{ \includegraphics{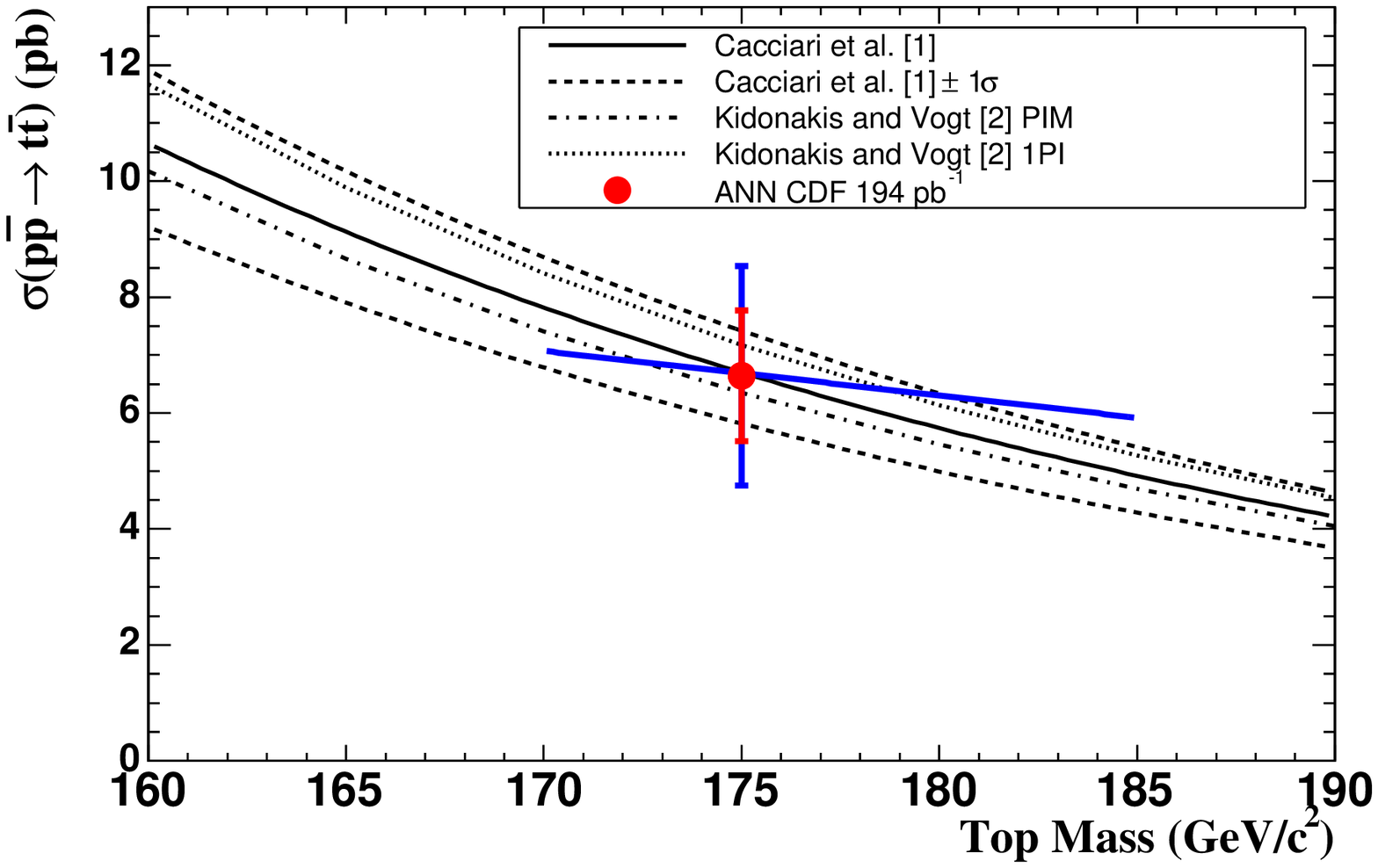}}
  \end{center}
  \caption{  \label{fig:sigtt} Theoretical predictions for the 
top quark pair production cross section\cite{mlm,kidonakis} compared to our measurement with the ANN 
in the $W+\geq$3 jets data sample. The nearly horizontal error bar shows how the central value of
our measurement evolves with top quark mass between 170 and 185~GeV/$c^{2}$.
	   %Points are shown slightly displaced along the $x$ axis in order to accomodate both the 
	   %\sht\ and the ANN fit results. For the 175 GeV/$c^{2}$ result, systematic errors are 
	   %shown with black error bars.
	   }
\end{figure}

We note that both the acceptance and the kinematic distributions for \ttbar\ depend on our assumed value 
for the top quark mass.  We quote the dependence of our result for the top pair production cross section
on the assumed top quark mass in Table~\ref{xs-topmass}.  Fig.~\ref{fig:sigtt} compares the ANN result with the 
theoretical predictions \cite{mlm,kidonakis}.

\begin{table}[htb]
 \caption{\label{xs-topmass} The \ttbar\ production cross section results (pb) in the 
  $W+\geq$3 jets sample at different top quark masses. The uncertainty is statistical only. }

  \begin{ruledtabular}
  \begin{tabular}{ccc}
Generated Top Mass &  $\sigma_{\ttbar}$\ from \sht\ & $\sigma_{\ttbar}$\ from ANN \\
\hline
160 & 5.2$\pm$2.1 & 7.9$\pm$1.3 \\
165 & 5.1$\pm$1.9 & 7.5$\pm$1.3 \\
170 & 4.9$\pm$1.7 & 7.0$\pm$1.2 \\
175 & 4.8$\pm$1.6 & 6.6$\pm$1.1 \\
180 & 4.5$\pm$1.4 & 6.3$\pm$1.1 \\
185 & 4.4$\pm$1.3 & 5.9$\pm$1.0 \\
190 & 4.2$\pm$1.2 & 5.7$\pm$1.0 \\
  \end{tabular}
  \end{ruledtabular}
  \end{table}

\section{\label{sec:xsec} Cross-Checks}
%------------------------------------------------------------------

\begin{figure}[!htbp]
  \begin{center}
\resizebox{3.0in}{!}{ \includegraphics{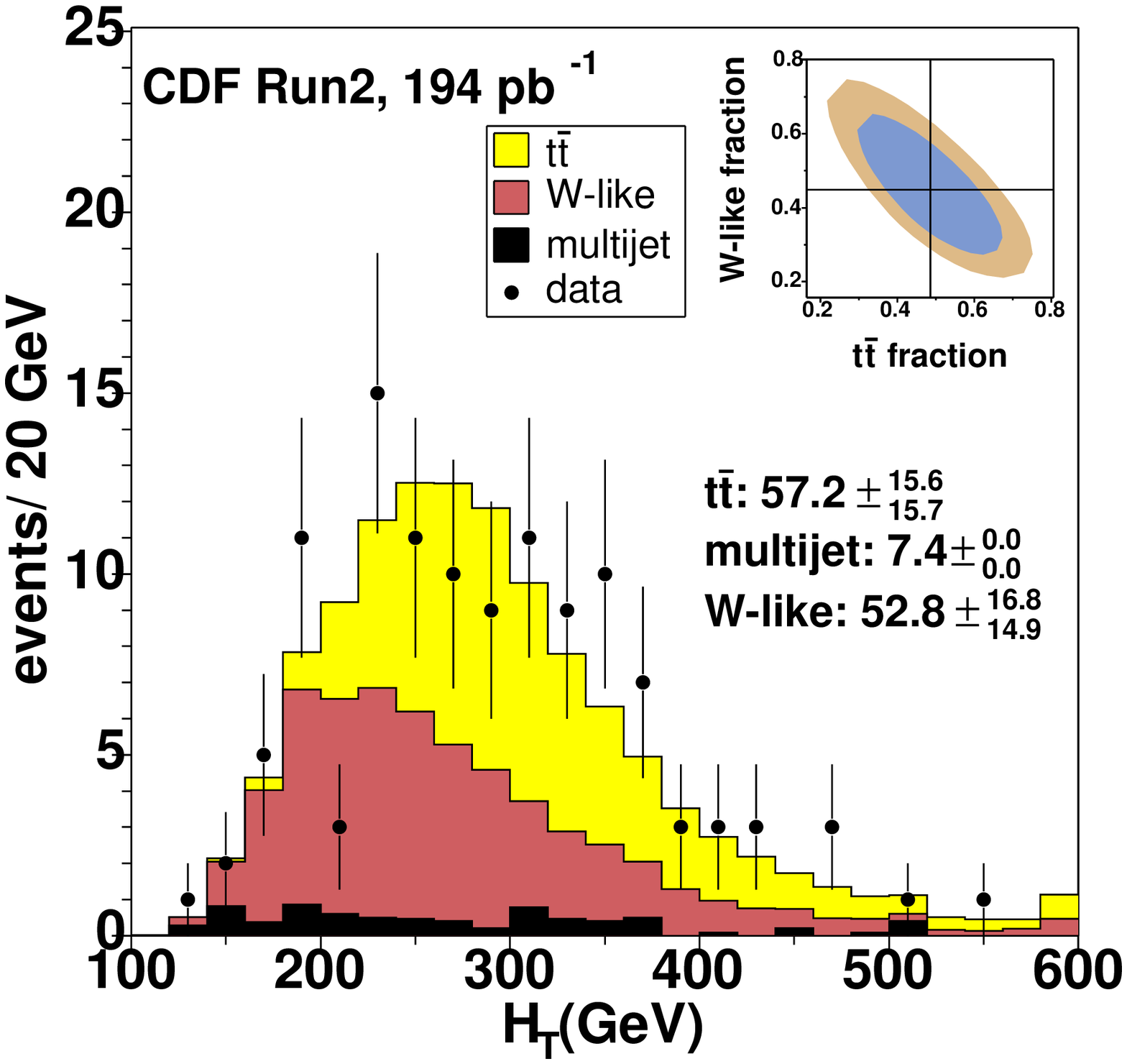}}
  \end{center}
  \caption{\label{fig:result_ht_4j} Distribution of observed $\sht$ 
           in the $W+\geq$4 jets sample, compared with the
           result of the fit.  The inset shows the 1- and 
           2-standard-deviation contours of the free parameters in 
           the fit,, normalized to the total number of observed events. The contribution of multi-jet background to the fit is fixed.}
\end{figure}

\begin{figure}[!hbp]
  \begin{center}
\resizebox{3.0in}{!}{ \includegraphics{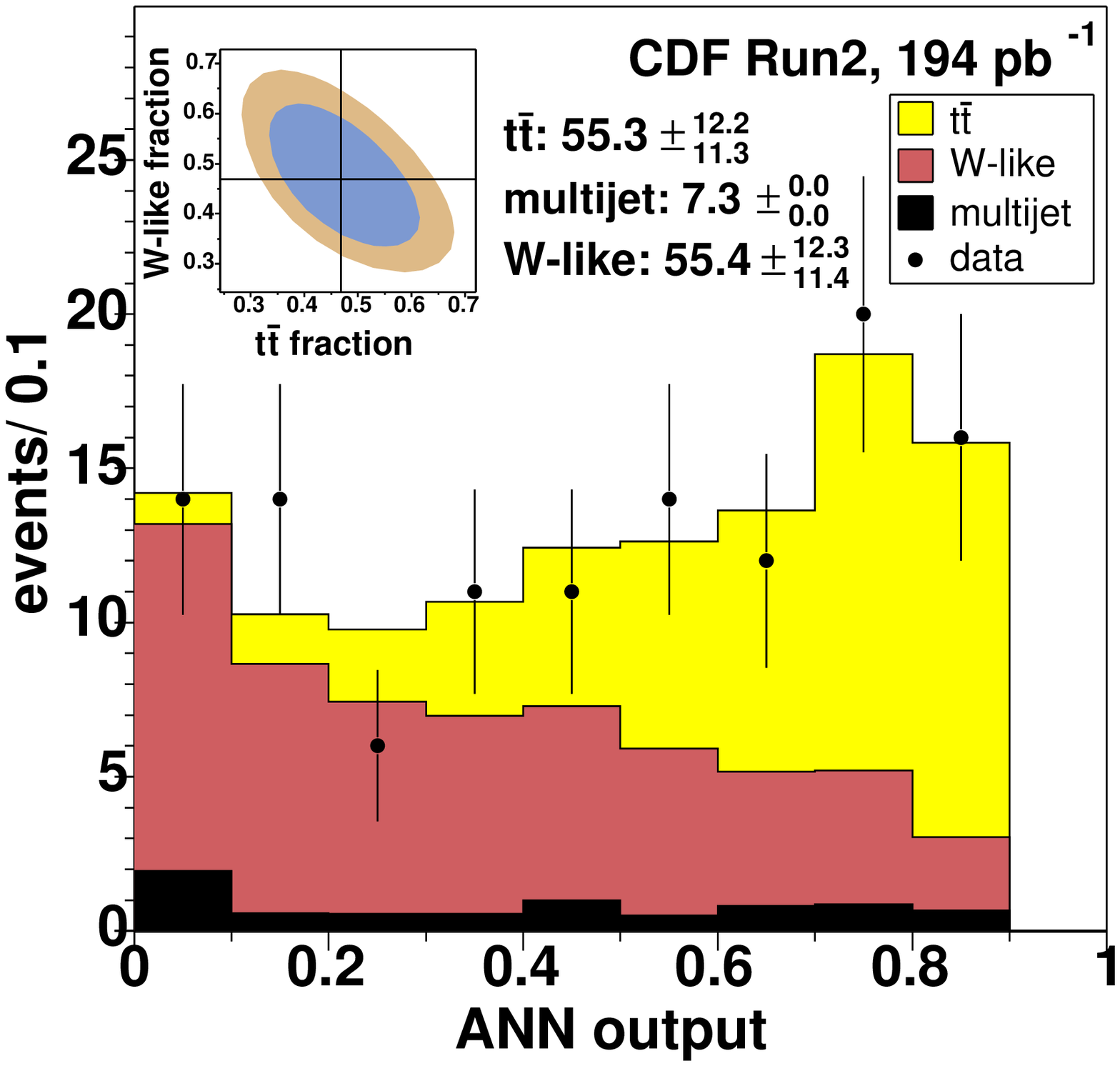}}
  \end{center}
  \caption{\label{fig:result_ann_4j} Distribution of observed ANN output 
           in the $W+\geq$4 jets sample, compared with the
           result of the fit.  The inset shows the 1- and 
           2-standard-deviation contours of the free parameters in 
           the fit, normalized to the total number of observed events. The contribution of multi-jet background to the fit is fixed.}
\end{figure}

We found the smallest expected statistical and systematic
uncertainties {\it a priori} for the ANN in the $W+\geq$3 jets data sample.  
As a cross-check, we repeat the analysis in the $W+\geq$4 jet sample, where there is a higher expected signal fraction of about 42\%.  We find 118 events pass the event selection criteria in our data sample with an integrated luminosity of 
194~\pbi. Figures~\ref{fig:result_ht_4j} 
and \ref{fig:result_ann_4j} show the distribution of data events for \sht\ and an ANN specially 
trained to obtain good separation in the $W+\geq$4 jet sample. 
We extract the most probable number of \ttbar\ signal events:

\begin{center}
   $\mu_{\ttbar}= 57.1 \pm 15.7$ ($H_{T}$), 
\\ $\mu_{\ttbar}= 55.3 \pm 11.7$ (ANN),

%%   $f_{\ttbar}= 0.503 \pm 0.128$ ($H_{T}$), %%*118
%%\\ $f_{\ttbar}= 0.473 \pm 0.100$ (ANN),  %%*118
\end{center}


\noindent where the uncertainty is statistical only and we have assumed a top mass of 175 GeV/$c^{2}$.
The requirement of a fourth jet with transverse energy above 15~GeV\ reduces the \ttbar\ acceptance 
to 3.85$\pm$0.47\%.  The measured top pair production cross section is then:
\begin{center}
$\sigma_{\ttbar} = 7.7\pm 2.1\pm 3.5$~pb ($H_{T}$),
\\ $\sigma_{\ttbar} = 7.4 \pm 1.6 \pm 2.0$~pb (ANN),
\end{center}
where the uncertainties are statistical and systematic, respectively. We observe good agreement
here between the results of the \sht\ and ANN fits.  
From simulated experiments, we estimate a probability of 13\% to find a difference equal to or larger 
than the observed difference between the results from the \sht\ distributions in the correlated $W+\geq$3 jets and
$W+\geq$4 jets samples.  
The observed 27\% statistical uncertainty for the \sht\ fit is now on the low edge
of what we would expect in 68\% of simulated experiments.  The observed 22\% uncertainty
for the ANN fit is close to the median from simulated experiments.

\begin{figure}[htbp]
  \begin{center}
\resizebox{3.6in}{!}{ \includegraphics{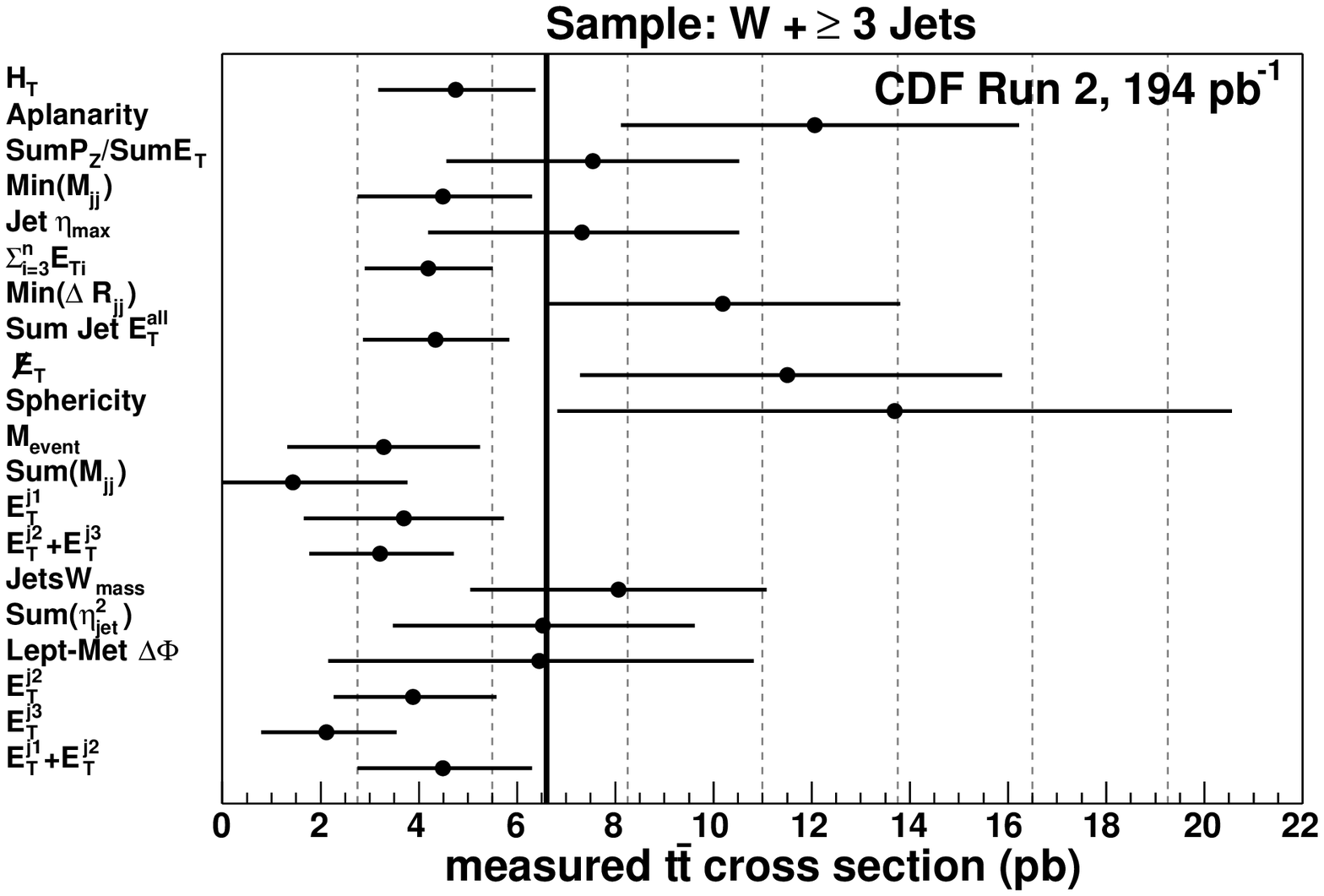}}
% \resizebox{3.0in}{!}{ \includegraphics{Win04fit-HtT4J.eps}}
  \end{center}
  \caption{\label{fig:result_single_3j} Measured \ttbar\ cross section in
	the $W+\geq$3 jets sample for all twenty kinematic and topological properties considered.  The uncertainty is statistical only, the
	vertical line shows the measured cross section with the ANN of 6.6~pb.}
\end{figure}

\begin{figure}[htbp]
  \begin{center}
\resizebox{3.6in}{!}{ \includegraphics{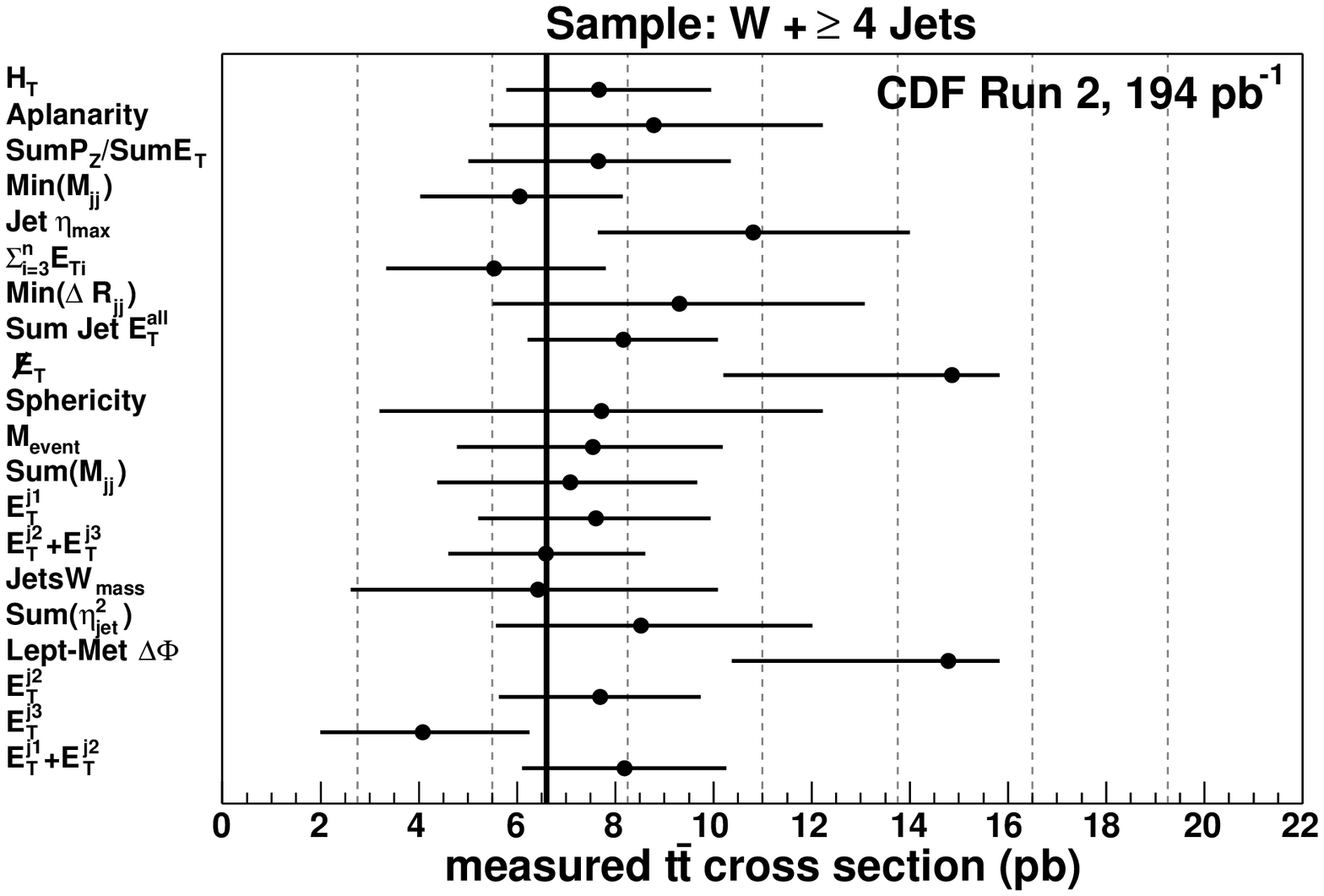}}
  \end{center}
  \caption{\label{fig:result_single_4j} Measured \ttbar\ cross section in
	the $W+\geq$4 jets sample for all twenty kinematic and topological properties.  The uncertainty is statistical only, the
	vertical line shows the measured cross section with the ANN of 6.6~pb.}
\end{figure}

We show the results of a fit to each of the twenty kinematic and topological properties
listed in Table~\ref{tab:vardef} for the $W+\geq$3 jet data sample in Fig.~\ref{fig:result_single_3j}
and the $W+\geq$4 jets data sample in Fig.~\ref{fig:result_single_4j}.  Some of these properties
are highly correlated with each other.  We do not observe any significant difference in the results for properties
used or not used by the ANN.

\begin{figure}[hbtp]
  \begin{center}
\resizebox{2.7in}{!}{ \includegraphics{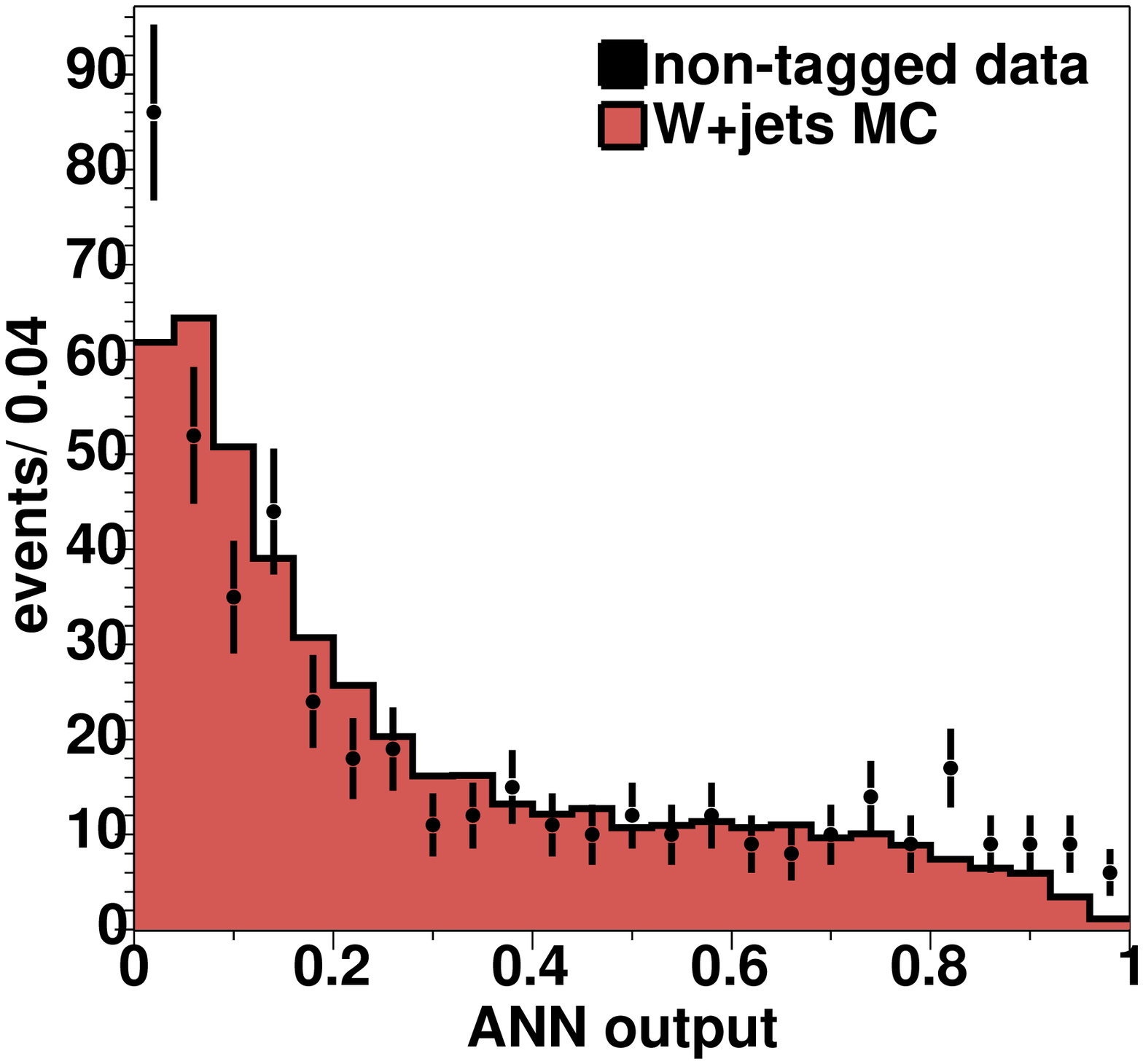}}
\resizebox{2.7in}{!}{ \includegraphics{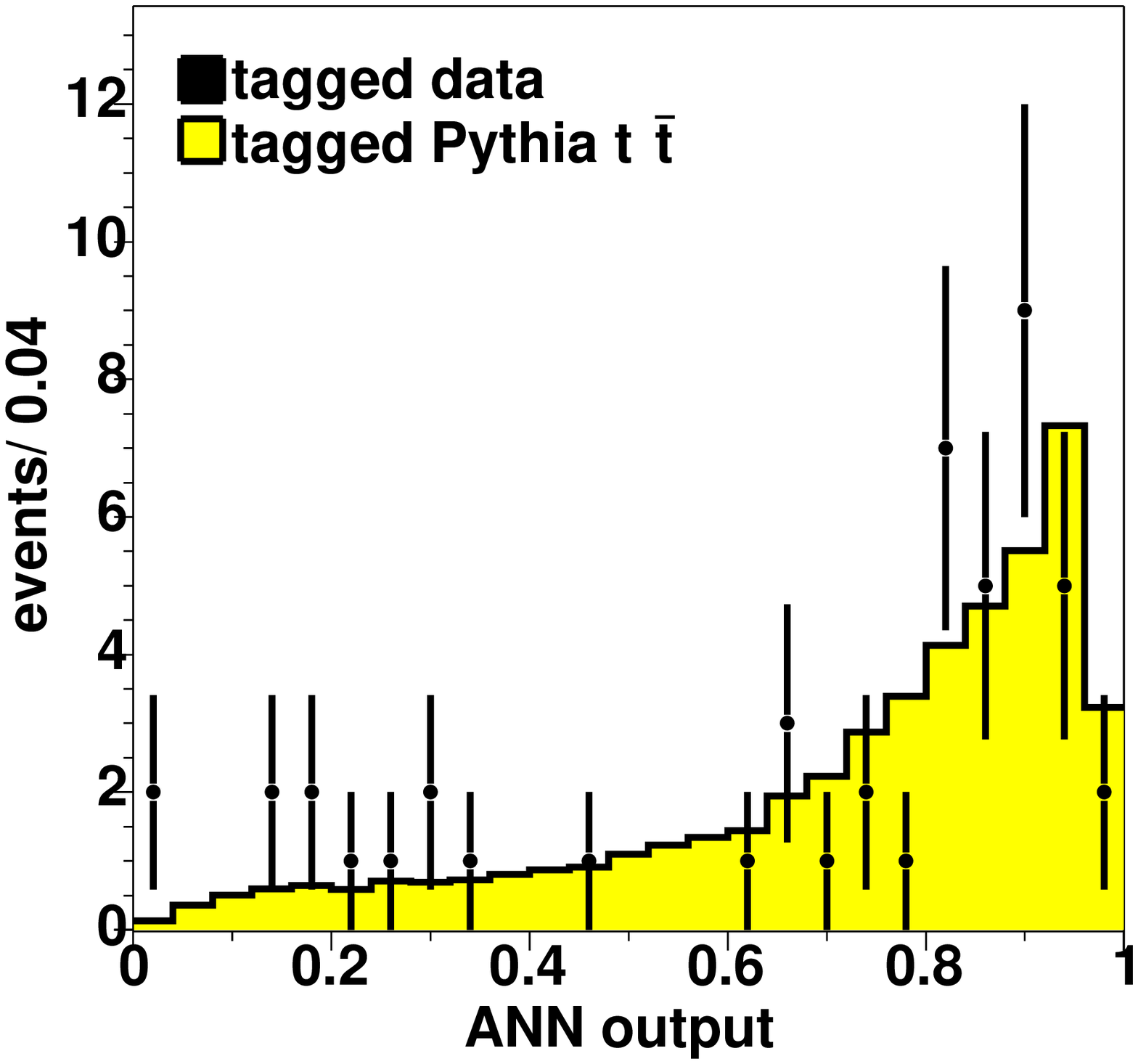}}
%\resizebox{2.7in}{!}{ \includegraphics{fig16_NonTagged.eps}}
%\resizebox{2.7in}{!}{ \includegraphics{fig16.eps}}
  \end{center}
  \caption{  \label{fig:ann_btag} The ANN output for events in the $W+\geq$3 jets data sample with (left) 
no identified $b$-jet  compared to $W$+3 parton ALPGEN+HERWIG Monte Carlo, 
and (right) at least one identified $b$-jet compared to PYTHIA \ttbar\ Monte Carlo.  The distributions are normalised
to equal area.}
\end{figure}

We note that two other CDF analyses \cite{svxruniipaper,Taka+Mel} select a top 
sample, and measure a top cross section in the lepton+jets channel, by using a displaced
secondary vertex to tag the presence of $b$ quarks from the $t\rightarrow W b$ decay. 
It is of interest to see how our neural net classifies this top sample. 
Fig.~\ref{fig:ann_btag} shows the ANN output for $W+\geq$3 jet data events with, and without, at least 
one $b$-jet identified using this secondary vertex algorithm. 
The output of the network is indeed close to 1 for many of the events with at least one identified $b$-jet. 
This provides verification that the kinematics of the $b$-tagged events are top-like, or, 
alternatively, the $b$-tag algorithm provides verification that the ANN efficiently isolates top events using kinematics only.

%------------------------------------------------------------------
\section{\label{sec:con} Conclusions}

We present a measurement of the top pair production cross section in $p\overline{p}$ collisions
at $\sqrt{s}=1.96$~TeV with an integrated luminosity of 194$\pm$11~pb$^{-1}$.  
We select events in the top lepton+jets channel by requiring one isolated lepton with $\et \geq 20$~GeV, missing 
transverse energy $\met \geq 20$~GeV, and three jets with $\et \geq 15$~GeV.
This selection accepts an estimated  7.11 $\pm$ 0.56\% of all \ttbar\ events. 
We develop an artificial neural network technique, which combines the information from seven kinematic and 
topological properties, to discriminate between \ttbar\ and background processes.  
Relative to the discrimination from only the total transverse energy, 
this artificial neural network technique reduces the expected statistical uncertainty by 30\% and
the estimated systematic uncertainty by 40\%. 
We perform a binned maximum likelihood fit to the artificial neural network output distribution 
observed in data, where we rely on Monte Carlo simulation to model the \ttbar\ and $W$+jets processes.
In a  data sample of 519 events, we find 91$\pm$16 \ttbar\ events, where the uncertainty is statistical only. 
We measure a top pair production cross section of $\sigma_{\ttbar}$=6.6$\pm$1.1 $\pm$1.5~pb,
where the uncertainties are statistical and systematic, respectively.  

%------------------------------------------------------------------

\begin{acknowledgments}

%%grabbed acknowledgements on March 8 from spokes page.
%http://www-cdf.fnal.gov/internal/physics/godparents/guidelines.html

We thank Michelangelo L.~Mangano and Claudio Ferretti for their assistance with the ALPGEN Monte Carlo generator. 
We thank the Fermilab staff and the technical staffs of the participating institutions for their vital contributions. This work was supported by the U.S. Department of Energy and National Science Foundation; the Italian Istituto Nazionale di Fisica Nucleare; the Ministry of Education, Culture, Sports, Science and Technology of Japan; the Natural Sciences and Engineering Research Council of Canada; the National Science Council of the Republic of China; the Swiss National Science Foundation; the A.P. Sloan Foundation; the Bundesministerium f\"ur Bildung und Forschung, Germany; the Korean Science and Engineering Foundation and the Korean Research Foundation; the Particle Physics and Astronomy Research Council and the Royal Society, UK; the Russian Foundation for Basic Research; the Comision Interministerial de Ciencia y Tecnologia, Spain; and in part by the European Community's Human Potential Programme under contract HPRN-CT-2002-00292, Probe for New Physics. 
\end{acknowledgments}


\begin{thebibliography}{99}
\bibitem{itsthegeometrystupid}{The detector has cylindrical  geometry with forward-backward and  axial symmetry around the beam 
 axis.  CDF uses a cylindrical coordinate system with the $z$ (longitudinal) axis along the 
 proton-beam direction and $\phi$ the azimuthal angle. The pseudo-rapidity $\eta$ of a particle 
 three-momentum is defined in terms of the polar angle $\theta$ of the momentum to the beam 
direction measured from the $z=0$ point at the center of the detector: 
$\eta = -\ln(tan({\theta}/{2}))$.}
\end{thebibliography}
\end{document}